\documentclass[useAMS,usenatbib]{mn2e}
\usepackage{natbib}
\pdfoutput=1 
\usepackage{journals}
\usepackage[pdftex]{graphicx,color}
\usepackage[latin1]{inputenc} 
\usepackage[T1]{fontenc}
\usepackage{graphics}
\usepackage{amsfonts}
\usepackage{amsmath}
\usepackage{multicol}
\usepackage{layout}
\usepackage{amssymb}
\newcommand{\mbf}{\boldsymbol}
\title[Halo properties from weak and strong lensing]
      {Mass and Concentration estimates from Weak and Strong Gravitational
        Lensing: a Systematic Study}
      \author[Giocoli et al. 2013] 
             {\parbox{\textwidth}{Carlo Giocoli$^{1,2,3}$\thanks{E-mail:
                   carlo.giocoli@unibo.it}, 
                        Massimo Meneghetti$^{2,3,4}$,
                        R. Benton Metcalf$^1$,  
                        Stefano Ettori$^{2,3}$, Lauro Moscardini$^{1,2,3}$}\\ \\
               $^1$ Dipartimento di Fisica e Astronomia, Alma Mater Studiorum Universit\`{a} di 
               Bologna, viale Berti Pichat, 6/2, 40127 Bologna, Italy\\ 
               $^{2}$ INAF - Osservatorio Astronomico di
               Bologna, via Ranzani 1, 40127, Bologna, Italy \\ 
               $^{3}$ INFN - Sezione di Bologna, viale Berti Pichat 6/2, 
               40127, Bologna, Italy \\
               $^4$ Jet Propulsion Laboratory, 4800 Oak Grove Dr. Pasadena, CA 91109, USA
             }
       
\begin{document}
\maketitle
\label{firstpage}
\pagerange{\pageref{firstpage}--\pageref{lastpage}} \pubyear{2014}
\begin{abstract}
  We study how well halo properties  of galaxy clusters, like mass and
  concentration,  are  recovered  using  lensing data.   In  order  to
  generate a large sample of systems at different redshifts we use the
  code MOKA. We measure halo mass and concentration using weak lensing
  data alone  (WL), fitting to  an NFW profile the  reduced tangential
  shear  profile, or  by combining  weak and  strong lensing  data, by
  adding information  about the size  of the Einstein  radius (WL+SL).
  For different redshifts,  we measure the mass  and the concentration
  biases  and  find  that  these  are  mainly  caused  by  the  random
  orientation   of   the   halo   ellipsoid  with   respect   to   the
  line-of-sight. Since our  simulations account for the  presence of a
  bright   central  galaxy,   we   perform   mass  and   concentration
  measurements using a generalized NFW profile which allows for a free
  inner  slope.  This  reduces  both the  mass  and the  concentration
  biases.  We discuss how the mass function and the concentration mass
  relation change when  using WL and WL+SL  estimates.  We investigate
  how  selection   effects  impact  the   measured  concentration-mass
  relation showing that strong lens  clusters may have a concentration
  $20-30\%$ higher than  the average, at fixed  mass, considering also
  the particular  case of strong  lensing selected samples  of relaxed
  clusters.  Finally, we  notice that  selecting a  sample of  relaxed
  galaxy clusters,  as is  done in some  cluster surveys,  explain the
  concentration-mass relation biases.
\end{abstract}

\section{Introduction}

Galaxy  clusters represent a  very important  cosmological laboratory.
Their  abundance and  evolution is  related to  important cosmological
parameters.   To first  order, the  cluster  counts as  a function  of
redshift  mainly  depend  on  the  matter  density  of  the  universe
$\Omega_m$, the  dark energy equation  of state parameter $w$  and the
normalization of  the initial power spectrum  of density fluctuations,
$\sigma_8$.  In the era of precision cosmology, and with the advent of
future wide  field surveys, new  and independent cosmological  probes are
necessary  for disentangling  degeneracies  between some  cosmological
parameters.

In  particular,  in order  to  use  the  cluster  mass function  as  a
cosmological probe it  is necessary to be able to  estimate their mass
with  very  high  accuracy.   Observations  in  different  wavelengths
provide different, indirect  measurements of cluster mass  that can be
compared  or  combined  with  each other.   Many  clusters  have  been
observed in both X-ray (CHANDRA and XMM) \citep{ettori09,ettori10} and
optical and  near-IR (HST, SUBRAU and  VLT) \citep{newman09,newman11}.
Using X-ray data  it is possible to extract the  cluster mass from the
observed brightness and gas temperature profiles by assuming spherical
symmetry and hydrostatic  equilibrium \citep{ettori13}.  However, some
biases might affect  the estimated mass due  to unresolved non-thermal
contribution to  the total gas pressure  \cite{rasia06,rasia12}.  In a
recent paper  \citep{planckxx} the Planck collaboration  has presented
the constraints on cosmological parameters ($\Omega_m$ and $\sigma_8$)
using number  counts as a function  of redshift for a  sample of $189$
galaxy  clusters selected  thanks to  their Sunyaev-Zeldovich  signal.
The authors  point out that  the measured cosmological  parameters are
degenerate with  the hydrostatic  mass bias  and that  only a  bias of
about $45\%$ can reconcile the measured $\sigma_8$ from cluster counts
with the one measured from  the primary CMB anisotropies (although see
\citet{hajian13}).

Optical  and near-IR data  allow one  to measure  the (weak  and strong)
lensing signal of background  galaxies.  The mass and the  critical lines can
be  recovered from  elongated  arcs and  multiple images.   Conversely,
estimating the cluster mass is  also very important for predicting the
redshift of  the arcs and  to identify highly magnified  high redshift
galaxies \citep{coe13} --  recall that the size of  the critical curve
depends on the redshift of the source \citep{zieser12,zitrin13b}.

These  estimates  can be  turned  into  indirect measurements  of  the
projected mass  distribution of the  cluster along the line  of sight,
and  then  into  an  estimate  of   the  virial  mass  of  the  system
\citep{hoekstra03,meneghetti10b,hoekstra13}.  Many  studies have shown
inconsistencies between  the mass  estimated from $X$-ray  and lensing
observables \citep{meneghetti10b,rasia12}, mainly because clusters are
not spherical or  relaxed system and contain  substructures -- cluster
members.  It is  also reasonable to suspect that  clusters with strong
lensing features  preferentially have their major  axis oriented along
the line of  sight \citep{sereno12} or very elongated in  the plane of
the sky \citep{zitrin13a,zitrin13b} -- also because in merging or post
merging phase,  which breaks  down the  usual assumption  of spherical
symmetry when deprojecting the mass model from two to three dimensions
\citep{limousin13}.

By combining the estimated mass and concentration of observed clusters
we can measure the mass-concentration relation and compare it with the
predictions  extracted from  numerical  simulations.   However, it  is
worth  noting  that  many  studies are  finding  a  mass-concentration
relation  that tends  to be  higher than  what is  expected from  dark
matter only $N$-body simulations  \citep{rasia13}.  However, it should
be emphasized that  the cooling of baryons is expected  to make haloes
more concentrated through adiabatic contraction while at the same time
an uncertain  amount of  feedback from the  central AGN  can partially
counteract this contraction \citep{killedar12,fedeli12}.

Several  systematic errors  contribute to  a scatter  and bias  in the
estimated  cluster   masses  from  lensing.   The   presence  of  such
systematics can be  seen directly in the data.  For  example, the mass
obtained for  a single cluster  using different source  galaxy samples
differs  by $\sim  10\%$.  The  estimated  masses also  depend on  the
radial  range over  which the  fit is  performed.  \citet{applegate12}
suggest that  a range  at least  out to $2  \times R_{500}$  should be
used.  What mass model is used in  the fit can also affect the result.
\citet{applegate12}  fix   the  concentration   while  \citet{okabe13}
consider it as a parameter to be estimated from the fit.  In addition,
there  are systematics  arising  from the  assumed  mass model,  shear
calibration    and    background   galaxy    redshift    distribution.
\citet{mahdavi13},  estimating the  hydrostatic and  the weak  lensing
mass of  a sample  of $50$ clusters,  noticed that  hydrostatic masses
underestimate  weak  lensing  masses  by  $10\%$  on  average,  within
$R_{500}$.

The mass estimates  of strong lensing selected  clusters are typically
uncertain   to   within   $30\%$  \citep{bartelmann96b}   because   of
substructures  and  projection  effects.    More  recent  analyses  by
\citet{kneib11,meneghetti10b} show that the  estimates of cluster core
masses  from  strong  lensing  are accurate  to  within  $10\%$.   For
comparison, mass estimates from X-ray  obsrvations are biased low with
respect  to  lensing  masses  by around  $25\%$  because  they  assume
hydrostatic equilibrium \citep{meneghetti10b,rasia12}.   However it is
important to note  that X-ray and lensing (weak  plus strong) analyses
can be  combined in such  a way as  to resolve the  degeneracy between
mass and elongation \citep{morandi10}.

In  this work  we will  address  where systematic  effects arise  when
trying to estimate  the mass and the concentration  from lensing data,
considering  the contributions  coming from  the halo  triaxiality and
orientation, and from the presence  of substructures and/or the bright
central galaxy.

Another  complication when  estimating the  mass and  concentration of
halos from lensing  comes from the presence of objects  along the line
of sight that  are unrelated to the halo being  considered.  This is a
relatively unexplored  source of systematic  error.  In this  paper we
will consider  only substructures in  the main halo.  In  a subsequent
paper we will address the effects of line of sight structures.

\begin{figure*}
\includegraphics[width=\hsize]{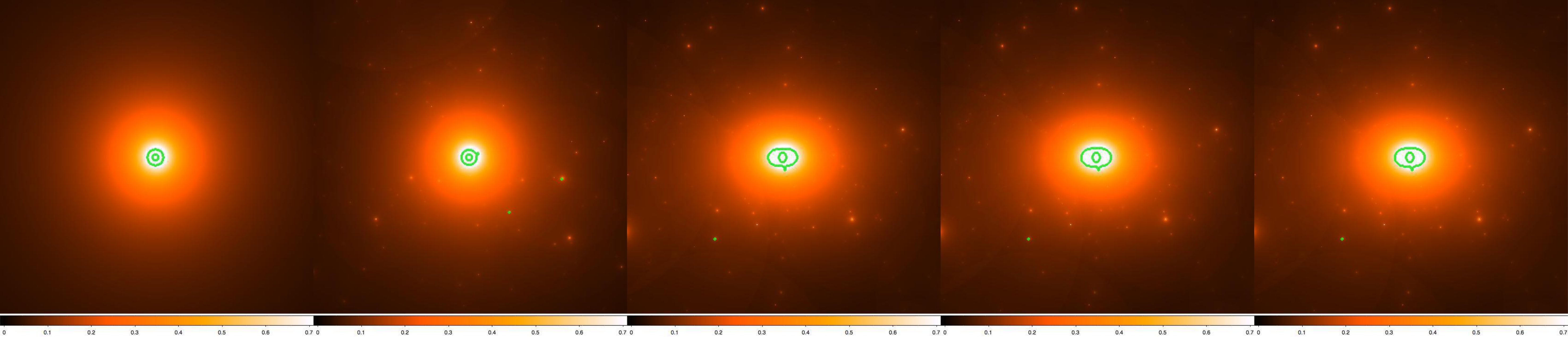}
\caption{Convergence   map  of   a   cluster  with   virial  mass   of
  $M_{vir}=5\times  10^{14}M_{\odot}/h$ and  concentration $c_{vir}=6$
  located  at  redshift  $z=0.288$  and  sources  placed  at  redshift
  $z_s=2$. In the  first panel on the left we  show the spherical halo
  without substructures, while they are  included in the second panel.
  In  the third  panel we  assign a  3D ellipticity  to the  main halo
  ellipsoid and randomly rotate it with  respect to the line of sight;
  in the forth and fifth panel we  consider also the presence of a BCG
  located at the centre of the cluster with the difference that in the
  latter  we  account  also  for adiabatic  contraction  of  the  dark
  matter. In each panel the green  curve in the central region defines
  the location of  the critical curves, where  the magnification $\mu$
  is infinite.
  \label{figMOKA}}
\end{figure*}

The paper is organized as follows. In Section~\ref{secmod} we describe
our lens model and explain how  we extract from each simulated cluster
weak  and  strong  lensing  information.   Section~\ref{secmcNFWf}  is
dedicated to  present how  well mass  and concentration  are recovered
using  as a  reference  a  \citet{navarro96} (NFW)  model  to fit  the
lensing  data, while  in Section~\ref{secmcgNFWf}  we investigate  the
case  of  a  generalized  NFW model.   In  Section~\ref{seccosmos}  we
present the results for a  cosmological sample of clusters and discuss
how mass  and concentration  uncertainties are  reflected in  the halo
mass   function   and   in  the   concentration   mass-relation.    In
Section~\ref{secsc} we summarize our finding.

\section{The Method} \label{secmod}

In  order to  create  a  large sample  of  triaxial and  substructured
convergence maps of galaxy-cluster size haloes we make use of the code
MOKA\footnote{http://cgiocoli.wordpress.com/research-interests/moka}
\citep{giocoli12}.  MOKA builds up the convergence map of haloes in an
analytical way, treating them as made  up of three components: (i) the
main halo --  assumed to be smooth, triaxial, with  an NFW profile (in
the  code  has  been  also   implemented  the  possibility  to  use  a
generalized-NFW  profile),  (ii)  the cluster  members  --  subhaloes,
distributed to follow  the main halo and to have  a truncated Singular
Isothermal Sphere profile \citep{metcalf01} -- and (iii) the brightest
cluster galaxy  (BCG) modelled with a  \citet{hernquist90} profile (in
MOKA a \citet{jaffe83}  model for the BCG has  also been implemented).
The axial  ratios, $a/b$  and $a/c$,  of the  main halo  ellipsoid are
randomly  drawn   from  the  \citet{jing02}   distributions  requiring
$abc=1$.  The halo ellipsoid is  randomly oriented choosing a point on
a sphere identified by its  azimuthal and elevation angles.  We assign
the same projected ellipticity to the smooth component, to the stellar
density and to the subhalo spatial distribution.  This is motivated by
the   hierarchical  clustering   scenario  where   the  BCG   and  the
substructures are related to the cluster  as a whole and retain memory
of  the  directions  of  the  accretion  of  repeated  merging  events
\citep{kazantzidis04,kazantzidis08,kazandzidis09,fasano10}.    In  our
simulations, we also account for the adiabatic contraction of the dark
matter  caused  by  the  BCG.    We  have  implemented  the  adiabatic
contraction   as   described   by  \citet{keeton01}   both   for   the
\citet{hernquist90} and the \citet{jaffe83} profiles. For more details
about  the MOKA  code we  refer to  \citet{giocoli12,giocoli12c}.  The
code also takes into account  the correlation between assembly history
and different halo properties: (i)  less massive haloes typically tend
to be more concentrated than the  more massive ones, and (ii) at fixed
mass,  earlier   forming  haloes   are  more  concentrated   and  less
substructured.  These  recipes have  been implemented  considering the
recent results  from numerical simulations.  In  particular, we assume
the \citet{zhao09} relation to link  the concentration to mass and the
\citet{giocoli10}   relation   for   the  subhalo   abundance.    When
substructures are included we define  smooth mass as $M_{\rm smooth} =
M_{vir} - \sum_i m_{\mathrm{sub},i}$ and its concentration $c_{\rm s}$
is  set in  such a  way that  the total  (smooth+clumps) mass  density
profile has a concentration $c_{vir}$, equal to the original one.

Throughout the  paper we will  denote with $M_{vir}$ (or  $M_{3D}$) the
cluster   mass   and   with    $c_{vir}$  (or   $c_{3D}$)   the   halo
concentration. For these definitions we  assume the one adopted for the
spherical collapse model:
\begin{equation}
M_{vir} = \dfrac{4 \pi}{3} R_{vir}^3 \dfrac{\Delta_{vir}}{\Omega_m(z)}
\Omega_0 \rho_c\,,
\end{equation}
where  $\rho_c=2.77  \times  10^{11}\,  h^2\,  M_{\odot}/\mathrm{Mpc}$
represents     the    critical     density     of    the     Universe,
$\Omega_0=\Omega_m(0)$ is  the matter  density parameter  at  the present
time, $\Delta_{vir}$  is the virial  overdensity \citep{eke96,bryan98}
and $R_{vir}$ symbolizes the virial  radius of the halo, i.e.
the distance  from the halo  centre that encloses the  desired density
contrast; and:
\begin{equation}
c_{vir}(M_{vir}, z) \equiv \dfrac{R_{vir}}{r_s} = 4 \left\{ 1 + \left[
    \dfrac{t(z)}{3.75 t_{4\%}} \right]^{8.4} \right\}^{1/8}\,,
\end{equation}
with  $r_s$  the  radius  at   which  the  NFW  profile  approaches  a
logarithmic  slope of $-2$,  $t(z)$ is the  cosmic time  corresponding at
redshift $z$ and  $t_{4\%}$ the one at which  the main halo progenitor
assembles $4\%$ of its mass \citep{zhao09,giocoli12b}. 

The characterization of galaxy cluster  properties done in this way is
simplified,  but on  average  resembles  -- in  the  best  way --  the
properties  measured   from  numerical   simulations.  Since   we  are
interested  in discussing  only the  average lensing  properties of  a
large  sample  of  systems,  we  do not  include  in  this  work  more
particular asymmetries present in galaxy clusters.

The  whole  halo  catalogue  is  made  up  by  galaxy  clusters  above
$10^{14}M_{\odot}/h$  at six  different redshifts  ($z=0.187$, $0.288$,
$0.352$, $0.450$, $0.548$ and $0.890$) with sources located at $z_2=2$.
To make our results easily comparable to recent observational data, 
these  redshifts have  been chosen  to  match those  of some  clusters
observed during the CLASH  program \citep{postman12}: Abell 383, Abell
611,    MACS1115.9+0129,     RXJ1347.5-1145,    MACS0717.5+3745    and
CLJ1226.9+3332.   For  each redshift,  we  generate  in total  $12288$
cluster maps from $10^{14}$ to $3.16 \times 10^{15} M_{\odot}/h$, with
a constant  bin size of  $\mathrm{d} \log(M) = 0.25$,  creating $2048$
maps for each considered bin.

\subsection{WL and SL signals from the simulated clusters}
Because  of their  mass density  distribution, triaxiality  and baryon
content, galaxy  clusters represent interesting  gravitational lenses,
deflecting  the light  rays  from background  galaxies. Observing  the
source  images, we  can  typically distinguish  two different  lensing
regimes:  one  in the  outer regions, where  the images  of background
galaxies  appear   slightly  distorted  and   magnified,  called  weak
gravitational  lensing (WL) \citep{bartelmann01},  and one  in  the central
part,  where background  sources are  highly magnified,  distorted and
multiply     imaged,    called     strong     gravitational    lensing (SL)
\citep{kneib11,meneghetti13}.

In Fig.~\ref{figMOKA}, we  show, as an example,  the projected density
map  of one  of our  clusters  with virial  mass $M_{vir}  = 5  \times
10^{14} M_{\odot}/h$ and concentration $c_{vir}=6$ located at redshift
$z=0.288$.  In  the first  panel on  the left we  present the  case in
which the halo is perfectly smooth  and spherical. In the second panel
the halo contain  substructures whose mass function  resembles the one
obtained   by   \citet{giocoli10}   from  a   cosmological   numerical
simulation.   In the  third  panel, we  introduce  triaxiality to  the
smooth  component  and  to  the satellite  spatial  distribution,  and
randomly orient the major axis with respect to the line of sight.  The
forth panel shows  the projected density map of the  cluster where the
BCG is included, finally in  the fifth panel the adiabatic contraction
of the  dark matter is  also considered.   The median distance  of the
tangential critical points $\theta_E$ --  that we will introduce later
in the  text, increases from left  to right and assumes  the following
values:
\begin{equation}
\theta_E = 8.2,\, 7.9, \, 11.4, \, 11.7, \, 12.4 \, \mathrm{arcsec}\,.
\end{equation} 
The small changes of the Einstein radius when substructures are included
is due to the redistribution of the  virial mass between the smooth and the
clump components. 

From the  projected density $\Sigma(r) =  \int \rho(r,z) \mathrm{d}z$,
we can define the convergence as:
\[
\kappa(r) = \dfrac{\Sigma(r)}{\Sigma_{\rm crit}}\,,\;\;\;\mathrm{with} \;\;\;
\Sigma_{\rm crit} = \dfrac{c^2}{4 \pi G} \dfrac{D_l}{D_s D_{ls}}\,.
\]
where  $c$  represents  the  speed  of  light  and  $G$  the  universal
gravitational  constant;  $D_l$ $D_s$  and  $D_{ls}$  are the  angular
diameter  distances  observer-lens, oberver-source  and  source-lens,
respectively.   Using the  convergence,  we can  define the  effective
potential $\Phi(x,y)$, the scaled deflection angle $\mbf{\alpha}(x,y)$
and introduce the  pseudo-vector field of the shear  using the complex
notation, $\mbf{\gamma} = \gamma_1 + i \gamma_2$ by its components:
\begin{equation}
\gamma_1(x,y) = \dfrac{1}{2} \left( \Phi_{11} - \Phi_{22} \right)
\end{equation}
and  
\begin{equation}
\gamma_2(x,y) = \Phi_{12} = \Phi_{21}\,;
\end{equation}
from which we can define the tangential and the cross components of the shear
\begin{equation}
\gamma_t = - \left[ \gamma_1 \cos(2 \phi) + \gamma_2 \sin(2 \phi) \right]
\end{equation}
\begin{equation}
\gamma_{\times} = - \gamma_1 \sin(2 \phi) + \gamma_2 \cos(2 \phi) \,,
\end{equation}
respectively.
These components  are  respectively  perpendicular and parallel
to the radius vector; $\phi$  specifies the angles with respect to the
centre of  the coordinate frame; in  what follows we  will denote with
$g$ the reduced shear as:
\begin{equation}
g \equiv \dfrac{\gamma}{1 - \kappa}\,.
\end{equation}
The  reduced  shear, and  not  the  actual  shear, is  the  observable
quantity  from image  ellipticities  \citep{bartelmann01,viola11}.  We
recall that in the weak  lensing regime the tangential
shear  is   related  to  the   mass  density  through   the  relation:
$\gamma_t(\theta) = \bar{\kappa}(<\theta) - \kappa(\theta)$.
From the  definitions of  the tangential and  cross components  of the
shear we can compute  the corresponding reduced shear quantities $g_t$
and $g_{\times}$ that are related  to the corresponding components of the
image ellipticity  $\epsilon_t$ and $\epsilon_{\times}$  of background
sources \citep{bartelmann01}. In the  absence of higher order effects,
the  azimuthal average  of the  cross component  ($\gamma_\times$) is
expected to  vanish. In practice, the  presence of cross  modes can be
used to check for systematic errors.

\begin{figure}
\includegraphics[width=\hsize]{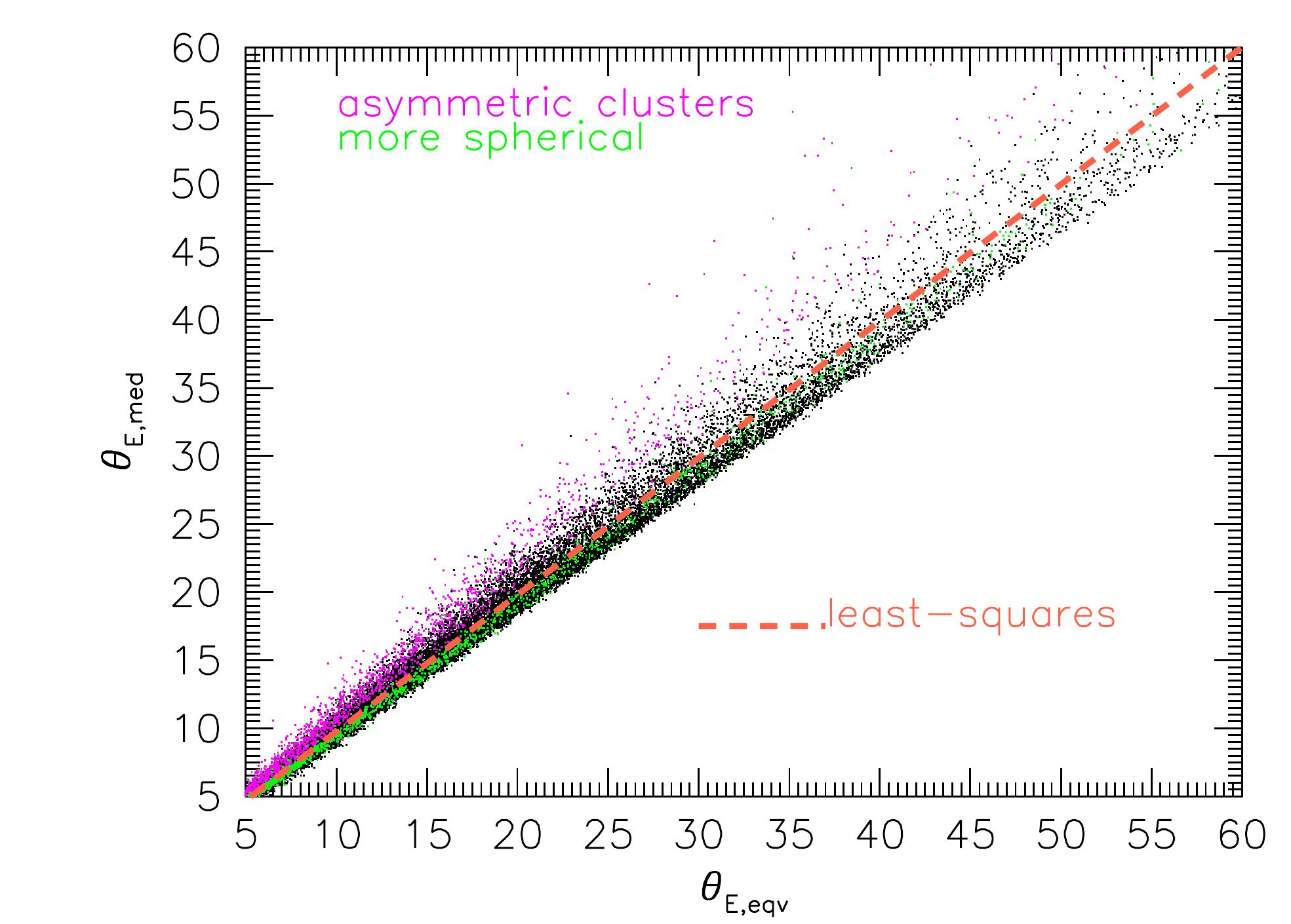}
\caption{Correlation between  the median  and the  equivalent Einstein
  radius   definitions  for   the   whole  sample   of  haloes   above
  $10^{14}M_{\odot}/h$ at  the six  considered redshifts.   The dashed
  line  shows  the  least-squares  fit  to the  data,  and  is  almost
  equivalent with the bisector. The magenta points show the relation 
  for clusters with a convergence ellipticity within $R_{2500}$ $\epsilon_{\kappa,2500}  > 0.7$, 
  while the green points for clusters with $\epsilon_{\kappa,2500}<0.1$.
  \label{figmedequiv}}
\end{figure}

The  differential deflection  of  light bundles  propagating from  the
source to the observer are given by to the Jacobian matrix:
\begin{equation}
 A =   \left( \delta_{ij} - \Phi_{ij} \right) = \begin{pmatrix}
  1 - \kappa - \gamma_1 & -\gamma_2 \\
  -\gamma_2  & 1 - \gamma - \gamma_1
 \end{pmatrix}\,,
\end{equation}
with eigenvalues
\begin{equation}
\lambda_t = 1 - \kappa - \gamma
\label{eqlambdat}
\end{equation}
and
\begin{equation}
\lambda_r = 1 - \kappa + \gamma\,.
\end{equation}
The  cases  $\lambda_r=0$ and  $\lambda_t=0$  define  the location  of
radial  and tangential  critical lines  in the  lens plane;  where the
magnification $\mu$  is infinite (green lines  in Fig.~\ref{figMOKA}).
In order  to define the Einstein  radius of the lens,  we consider all
the points in  the lens plane with $\lambda_t=0$ and  connect the ones
that    enclose     the    cluster    centre.     As     adopted    by
\citet{meneghetti10a,meneghetti11} we will  define the Einstein radius
$\theta_{E,med}$  as the  median  distance from  the  centre of  these
points. To  be compatible  with other  definitions \citep{reidlich12},
for  each  system  we  also  compute  the  effective  Einstein  radius
$\theta_{E,eqv}$, defined  as the radius  of the circle  enclosing the
same area  as central  critical points.  In  Fig.~\ref{figmedequiv} we
show  the correlation  between the  two different  definitions of  the
Einstein radius  for the total  sample of $12,288$ clusters  with mass
between $10^{14}M_{\odot}/h$  and $3.16 \times  10^{15}M_{\odot}/h$ at
six different  redshifts.  From the  figure we notice that  the median
Einstein radius  definition better captures the  presence asymmetry of
the matter distribution  towards the cluster centre  since many points
lie     above    the     least-squares     fit     to    the     data:
$\theta_{\mathrm{E,med}}=\theta_{\mathrm{E,eqv}}  -0.3$. To  emphasize
this point  we have  colored the points  referring to  more asymmetric
clusters (presenting  a convergence  ellipticity within  $R_{2500}$ --
the  radius which  encloses 2500  times  the critical  density of  the
universe   --   $\epsilon_{\kappa,2500}>0.7$)  magenta   while   those
referring      to     the      more      spherical     ones      (with
$\epsilon_{\kappa,2500}<0.1$) are green.  We also note that the sample
by  \citet{meneghetti11} possesses  a larger  scatter in  the relation
mainly because of  the large number of  asymmetrical objects presented
in their sample of strong lensing selected clusters.

While the  halo mass  and the concentration  are two  derived quantities
from the weak and strong lensing signals, the size of the Einstein radius
is more  directly estimated  from the position  of multiple  images of
background sources.  In the literature there are many clusters in which
tens of  multiple images  have been identified,  allowing a  very good
determination   of   the   size   of   the   strong   lensing   region
\citep{broadhurst05a,broadhurst05b,zitrin11b}.

In  the MOKA  code,  all  $\gamma$  and $\alpha$  are computed  in
Fourier space where derivatives are easily and efficiently calculated.
The Fourier transform of the convergence $\hat{\kappa}(\mbf{l})$, is:
\begin{equation}
\hat{\kappa}(\mbf{l}) = \int_{\mathbb{R}^2} \mathrm{d}^2 \theta
  \kappa(\mbf{\theta}) \exp \left( i \mbf{l}  \cdot \mbf{\theta} \right)\,.
\end{equation}
This is  computed on a  map of $1024 \times  1024$ pixels with  a zero
padding region of  $512$ pixels to avoid  artificial boundary effects.
For  each system,  from the  potential  and the  convergence maps,  we
compute the ellipticity $\epsilon_{\Phi,500}$  of the potential and of
the convergence $\epsilon_{\kappa,500}$ at  $R_{500}$, at the distance
where the enclosed density drops below $500$ times the critical value.
We    are   interested    in   the    potential   ellipticity    since
$\epsilon_{\Phi,500}$ is a quantity that can be directly compared with
the X-ray morphology of observed galaxy clusters: systems that possess
a  regular  and  smooth X-ray  map  tend  to  have  a small  value  of
$\epsilon_{\Phi,500}$.

Combining the  convergence and shear  maps, for each MOKA  cluster, we
compute the reduced tangential shear  profile and the error associated
to each  radial bin  as follows.   We assume  a background  density of
sources of $n_g = 30$ gal/arcmin$^2$ (which is a reasonable number for
current and future space-based observations;  see also the ESA mission
EUCLID \citep{euclidredbook}),  and locate on the  map $N_g=n_g \times
A_{map}$ random  points, where  $A_{map}$ represents  the area  of the
map.  We measure  the reduced tangential shear at each  of these $N_g$
points and  build a logarithmically azimuthally  averaged profile from
$0.01\,\mathrm{Mpc}/h$ up to the virial radius with bins equispaced by
$\mathrm{d}\log(r)=0.05$.  This ensures a good  $S/N$ in each bin both
for low and  high-redshift clusters.  To each radial bin  we assign an
error given by the sum of two components:
\begin{equation}
 \sigma_{g_+}^2 = \sigma_{\rm int}^2 + \sigma^2_{g,\epsilon}\,,
\label{eqsgt}
\end{equation}
where $\sigma_{\rm  int}$ represents the  rms of the  measured reduced
tangential shear from the map, and $\sigma_{g,\epsilon}$ is the intrinsic shape
of  the background  galaxies, that  depends on  the  considered number
density of  the sources and  on the intrinsic scatter in  the ellipticity
$\sigma_{\epsilon}=0.3$ \citep{hoekstra03,hoekstra13}:
\begin{equation}
\sigma_{g,\epsilon}^2 = \dfrac{\sigma_{\epsilon}^2}{\pi
  \left(\theta^2_2 - \theta^2_1   \right) n_g}\,;
\end{equation}
here  $\theta_1$ and  $\theta_2$  the  two extrema  of  the bin.   The
intrinsic  scatter  of  the  ellipticity of  the  background  galaxies
provides a  noise that limits  the accuracy of the  shear measurements
\citep{hoekstra03,hirata04}. We recall that in our analysis we did not
consider  any  uncertainty on  the  photometric  redshift which  would
dilute the weak lensing signal.

\begin{figure}
\includegraphics[width=\hsize]{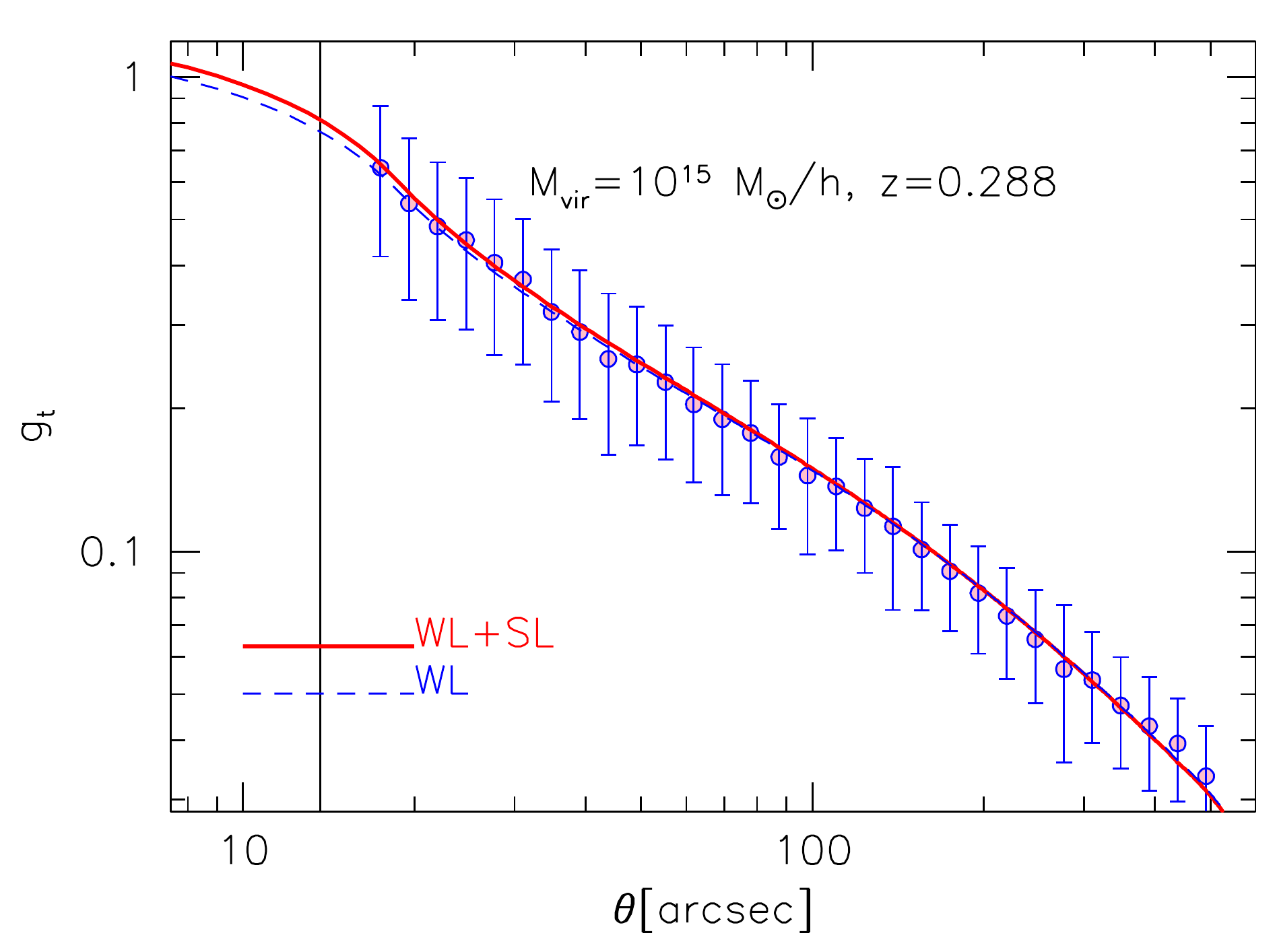}
\caption{The reduced tangential  shear profile of a  galaxy cluster at
  redshift $z=0.288$ with sources at  redshift $z_s=2$. The error bars
  show     the     error     associated    to     $g_t$,     as     in
  equation~(\ref{eqsgt}). The  vertical line  defines the size  of the
  Einstein radius of the cluster.  The solid curve represents the best
  fit NFW profile obtained by taking into account both weak and strong
  lensing data,  while the dashed  one is the  best fit when  only the
  reduced tangential shear profile is considered.\label{figprof}}
\end{figure}

An example  is shown in  Fig.~\ref{figprof} where the red  circle with
the  error bars  shows the  average reduced  tangential shear  profile
measured for a  cluster at redshift $z=0.288$ with  sources located at
redshift $z_s=2$; the vertical line  represents the median distance of
the  critical points  from the  cluster  centre.  The  average of  the
reduced tangential  shear in each  annulus is calculated  by averaging
the  corresponding values.  Here we  prefer  not to  use the  weighted
average, as done  by \citet{umetsu11}, because we  assume the variance
for  the shear  estimate to  be  zero since  the value  of $g_{+}$  is
directly  computed  using  the   corresponding  map.  To  compute  the
spherical averaged profile we take the centre of the cluster to be the
position of the BCG.  In fact the  location of the BCG can sometime be
offset  from the  mass centroid  of the  corresponding matter  density
distribution \citep{oguri10,oguri11}. However, in analyzing the strong
lensing  mass model  of five  clusters, \citet{umetsu11}  find only  a
small offset  (of the order  of $20\,\mathrm{kpc}/h$) between  the BCG
and the centre of mass.

\begin{figure*}
\includegraphics[height=5.6cm]{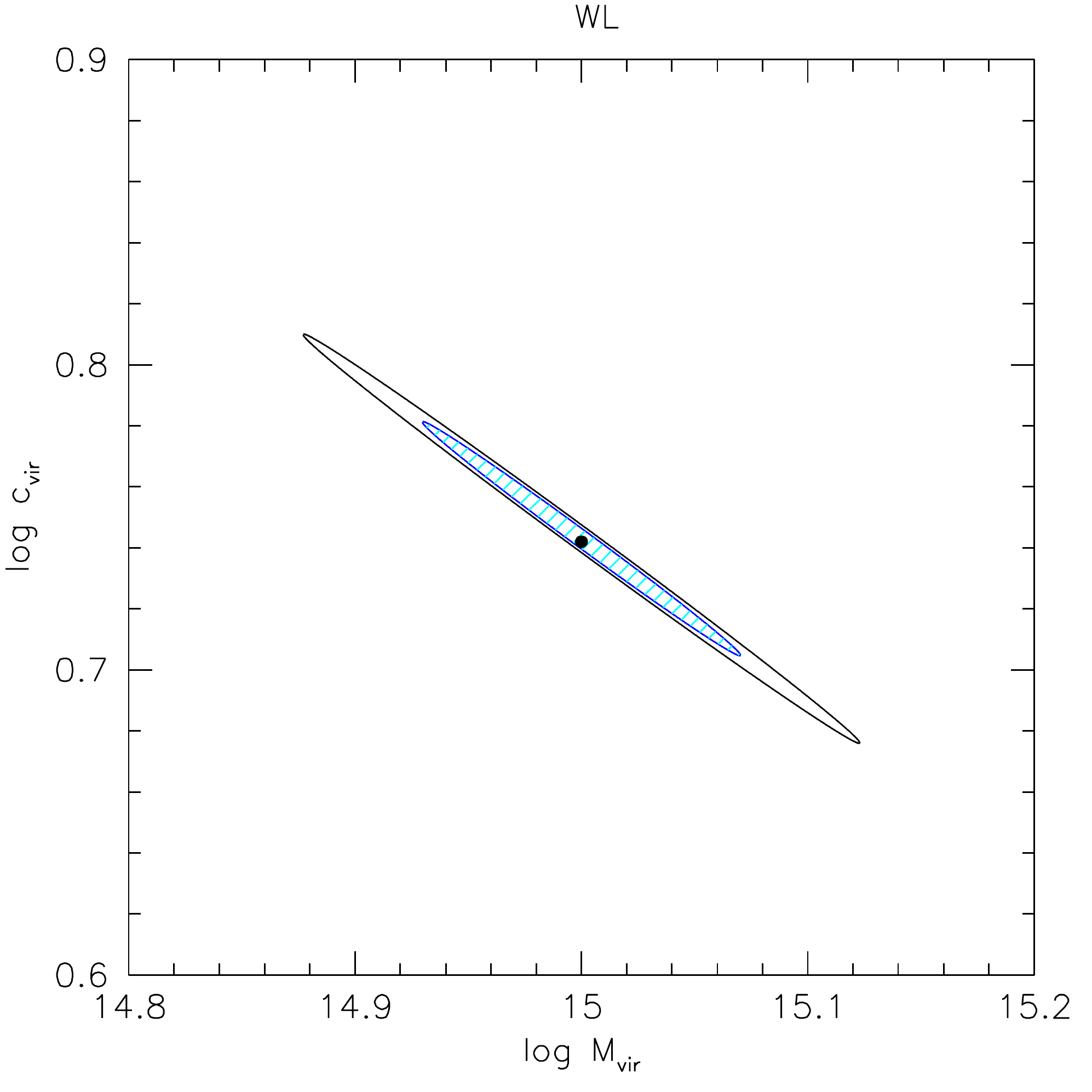}
\includegraphics[height=5.6cm]{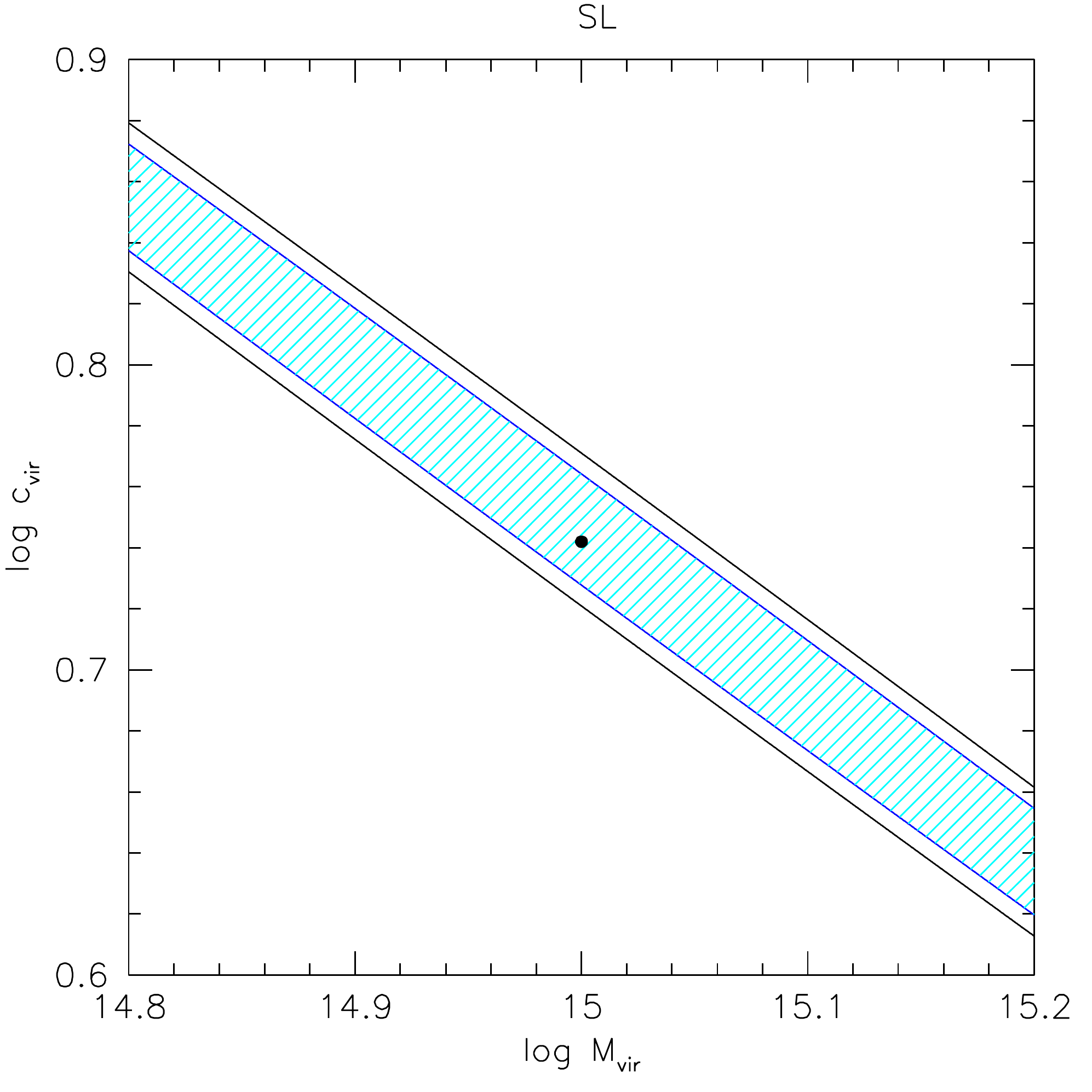}
\includegraphics[height=5.6cm]{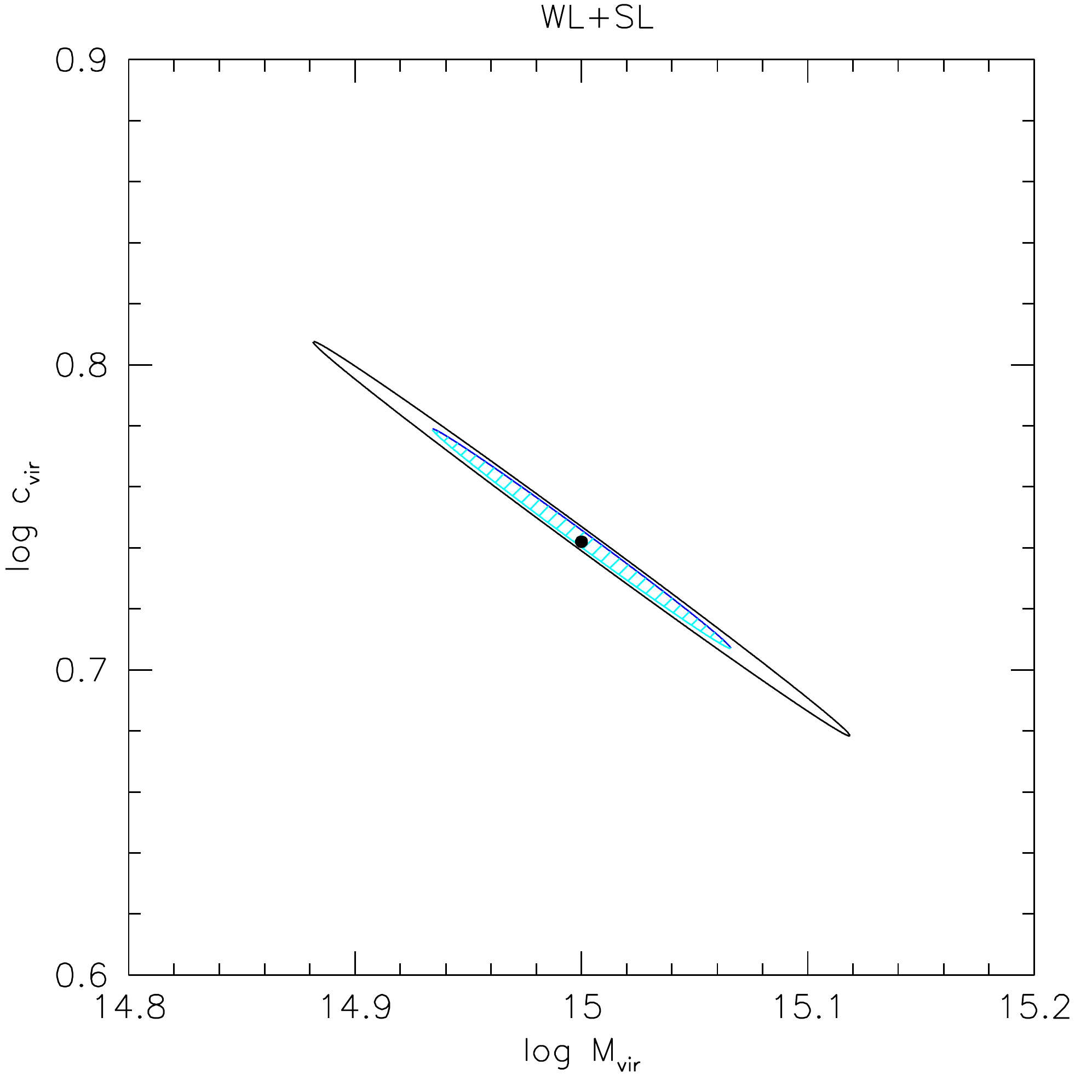}
\caption{Weak lensing (left),  strong lensing (centre) and weak+strong
  lensing (right) constraints estimating  mass and concentration for
  a smooth, spherical NFW  halo. The black  dot 
   represents the  input halo  mass and  concentration.  In all
  three cases the  minimum of the $\chi^2$ corresponds  with the black
  dot. \label{figconstrains}}
\end{figure*}

For each cluster, we estimate the weak-lensing (WL) mass $M_{est}$ and
concentration  $c_{est}$  using an  NFW  profile  fit  to the  reduced
tangential shear profile by minimizing the quantity:
\begin{equation}
\chi^2_{WL}(M,c) = \sum_{i=1}^{N} \dfrac{\left[ g_{NFW}(\theta_i|M,c)
      - g_t(\theta_i) \right]^2}{\sigma^2_{g_+,i}}\,,
\end{equation}
where  the index $i$ runs  on  the  number of  bins  and $g_{NFW}(\theta_i)$  is
computed using the  relations for the convergence and  the shear valid for a
NFW  halo  \citep{bartelmann96}.  In  Fig.~\ref{figprof},  the
dashed blue  curve represents the  best fit to the  reduced tangential
shear profile.

The information about  the Einstein radius of the  cluster allows also
to define a  strong lensing constraint \citep{sereno12,zitrin11a}.  For
a spherical NFW halo we use  analytic formula for the convergence and the
shear \citep{bartelmann96} and compute from equation~(\ref{eqlambdat})
the   corresponding  Einstein   radius  given   a  certain   mass  and
concentration.   For the  strong  lensing constraint  on  the mass  and
concentration estimates we minimize the following quantity:
\begin{equation}
\chi^2_{SL}(M,c) = \dfrac{\left[\theta_E - \theta_{E,NFW}(M,c)\right]^2}{\sigma_E^2}\,,;
\end{equation}
we               assume              $\sigma_E=1$               arcsec
\citep{jullo10,zitrin11a,zitrin11c,host12}   which    represents   the
measurements error associated  with the number of  multiple images and
from  the positions  of tangentially  distorted galaxies  available to
reconstruct the  critical curve of  the cluster.  In this  approach we
are comparing the measured size of  the Einstein radius of the cluster
with a spherical symmetric model. It is worth to notice that with this
method  we tend  to overestimate  the  mass, since  real clusters  are
triaxial  and  present  substructures  --  so  they  are  not  axially
symmetric.  As  described by \citet{bartelmann95}, a  smaller and more
realistic mass  could be obtained  modeling the strong lensing  with a
triaxial shape.  In Fig.~\ref{figprof}  the solid red curve represents
the  tangential   shear  profile  of   the  cluster  where   mass  and
concentration have been  estimated minimizing the combined  WL and SL:
$\chi^2 = \chi^2_{WL} + \eta \chi^2_{SL}$.  The parameter $\eta$ is an
integer $(1\leq \eta < N)$ equal to the bins of $g_{+}$ containing the
value  of the  Einstein  radius $\theta_E$.   This  ensures that  when
combining weak and strong lensing $\chi^2$ we are putting together the
same  information,   taking  into  account  that   the  shear  profile
represents a differential quantity of the cluster matter distribution,
while the size  of the Einstein radius is a  cumulative quantity since
it     is      related     to     the     total      enclosed     mass
\citep{narayan96,bartelmann10}    (for   the    case   presented    in
Fig.~\ref{figprof} we have  $\eta=1$). We recall the  reader that even
if we are using only the  Einstein radius as strong lensing constraint
we will label it as SL in all the figures.

In  Fig.~\ref{figconstrains}, we  show the  constraints obtained  when
recovering mass and concentration for  the reference case of a smooth,
spherical  NFW halo.   The left,  center and  right panels  show weak,
strong and weak+strong lensing  constraints, respectively.  In all the
panels the black dot represents the  mass and the concentration of the
input  halo which  corresponds in  all cases  to the  location of  the
minimum of the corresponding $\chi^2$.

\begin{figure*}
\includegraphics[width=5.8cm]{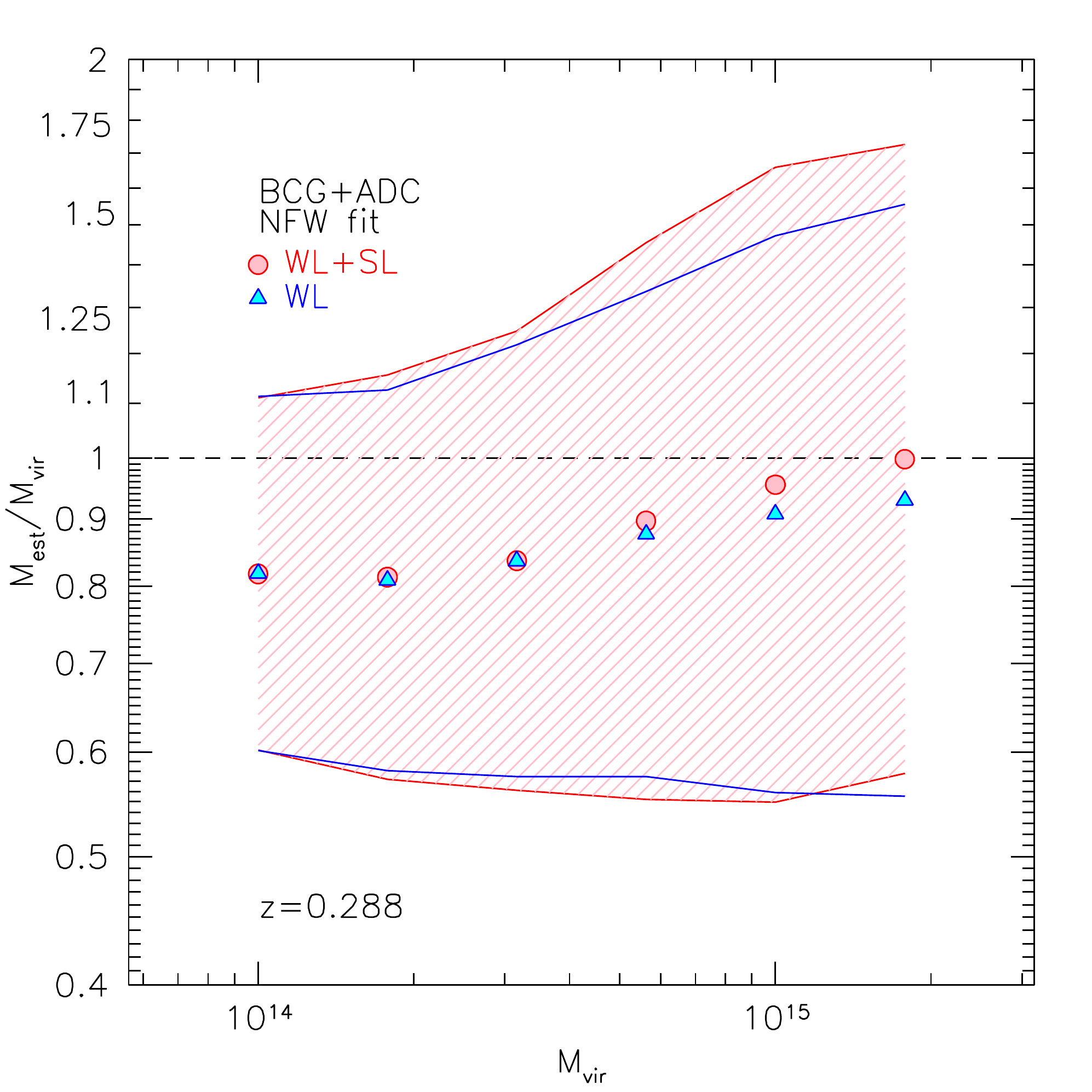}
\includegraphics[width=5.8cm]{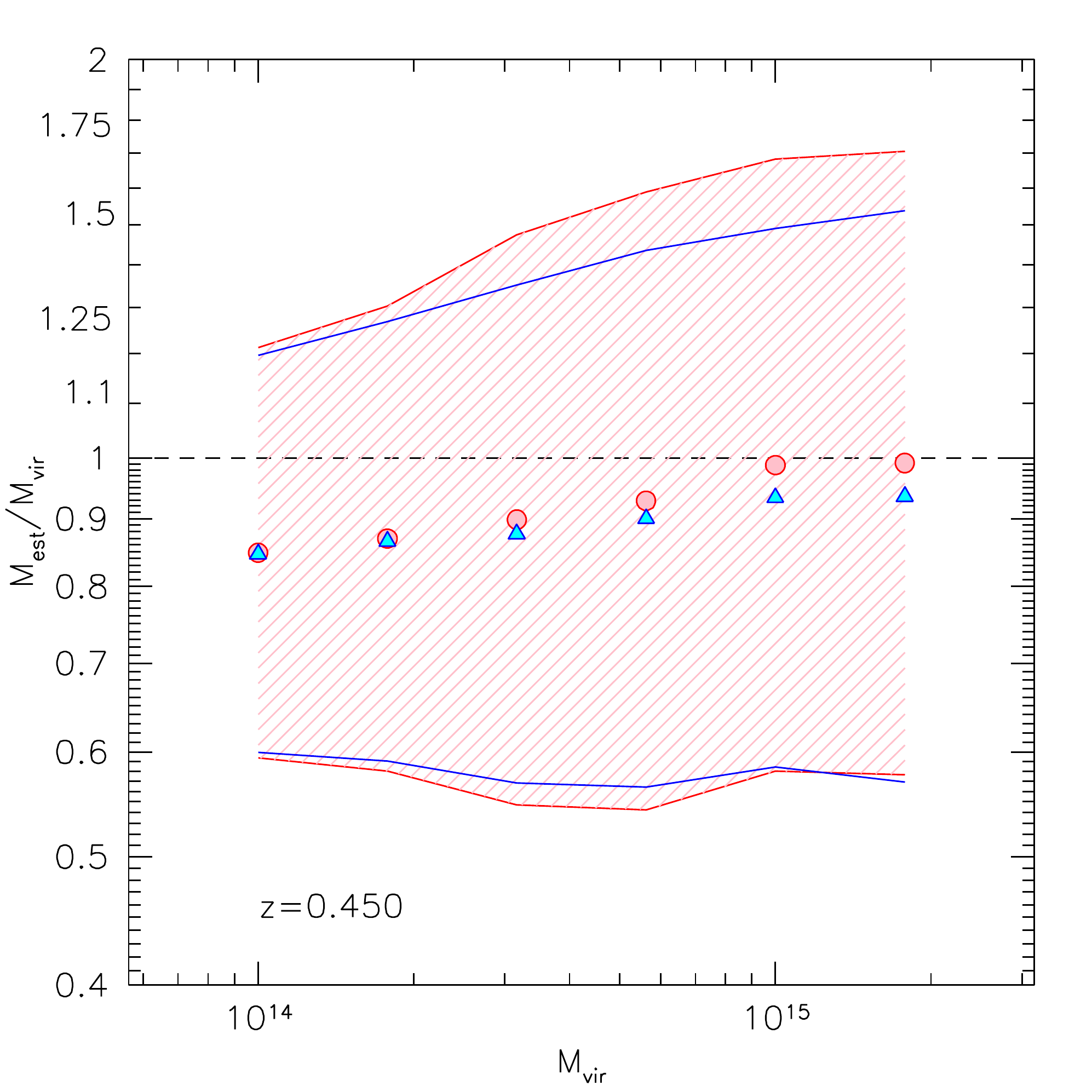}
\includegraphics[width=5.8cm]{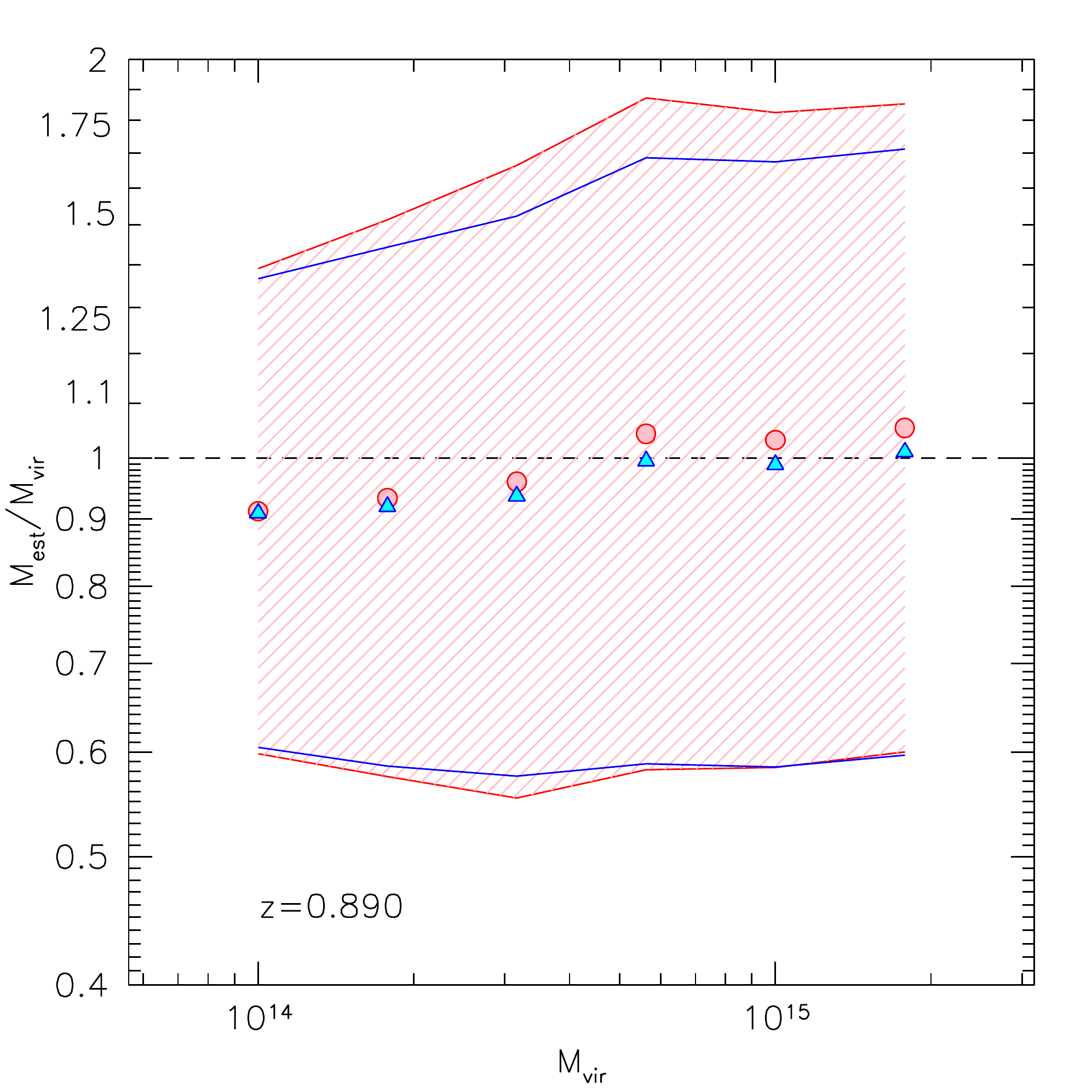}
\caption{Average of the  rescaled estimated mass as a  function of the
  true cluster mass.  We show the  case in which the mass is estimated
  using  WL+SL  (circles)  and  only WL  (triangles)  information  for
  systems at  three different redshifts.   The upper and  lower curves
  show the $1\sigma$ scatter  of the corresponding distributions, red
  for  WL+SL and  blue for  WL only.   Each mass  bin contains  $2048$
  cluster realizations. \label{figmassest}}
\end{figure*}

\section{Masses and Concentrations from NFW fitting} \label{secmcNFWf}

In  this section  we will  present  the results  on the  mass and  the
concentration  estimates  obtained  from the  whole  constructed  MOKA
cluster sample.  Since average  halo structural properties  depends on
mass and  redshift we will study  how the mass and  concentration bias
depends on these quantities.

In Fig.~\ref{figmassest}  we show the average  rescaled estimated mass
as a  function of  the cluster  mass for three  of the  six considered
redshifts.   The results  of  the three  redshifts  not displayed  are
consistent with those here presented.   The rescaled mass is the ratio
between the  estimated and  the true (3D)  cluster mass  (see Appendix
\ref{app2D}  for  the  comparison  between the  true  and  2D  cluster
masses).  We recall that the  simulations of the MOKA clusters include
the presence  of a BCG,  the adiabatic  contraction (ADC) of  the dark
matter  component,  triaxiality  and subhaloes.   Filled  circles  and
triangles show  the masses estimated  considering SL+WL and  WL alone,
respectively; the shaded region encloses  the $1\sigma$ scatter of the
distribution.  From  the figure  we notice that  for groups  and small
clusters the  mass is  typically underestimated  by about  $15\%$, for
massive clusters the mass has a bias  ranging from $5\%$ down to a few
precent,    consistently    with    what    has    been    found    by
\citet{becker11,meneghetti10b,rasia12}.  The  higher bias in  the mass
estimate for the  smallest systems is due to the  triaxiality model by
\citet{jing02} implemented in MOKA and  extended down to these masses.
In this model, typically smaller systems  tend to be more prolate than
the more  massive ones in agreement  with the fact that  they are more
stretched by the gravitational field of the surrounding matter density
distribution during their collapse  \citep{sheth01b}.  The small trend
of        the        normalization       of        the        relation
$M_{\textrm{est}}/M_{\textrm{vir}}-M_{\textrm{vir}}$ as  a function of
redshift reflects the fact that MOKA clusters at higher redshifts tend
on       average       to       possess       more       substructures
\citep{vandenbosch05,giocoli10} and  to have  a larger  3D ellipticity
$\epsilon$ \citep{shaw06,shaw07,giocoli10,despali12,limousin13}.  From
the  figure we  notice that  when the  constraint on  the size  of the
Einstein radius  is included  in the  mass estimate  the corresponding
mass bias  tends to be  reduced: the  modeling of the  Einstein radius
size is  done using  a spherical model  and so on  average we  tend to
measure  a higher  mass with  the SL  constrain \citep{bartelmann96b}.
The absence  of difference in  the rescaled estimated mass  between WL
and WL+SL for the  smallest mass bins is due to the  fact that most of
those clusters are not strong lenses  or have small Einstein radii: in
those cases $\chi^2_{SL}$  has a negligible contribution  on the total
$\chi^2$.   \citet{meneghetti10b}  perform  a similar  analysis.  They
study  the lensing  signals  of three  projections  of three  clusters
extracted  from a  numerical simulation.  In particular,  in the  left
panel of their  Fig.~16 they show the ratio between  the estimated and
true  3D  mass obtained  best  fitting  the reduced  tangential  shear
profile of  each cluster using,  as we  have done, an  NFW functional.
For the  case $M_{200}$,  the authors  find a  negative bias  of about
$15\%-5\%$, consistent with what we  have found in this work. However,
the  flexibility and  the  speed of  our algorithm  MOKA  allow us  to
generate and  analyze a  sample of  clusters which  is more  than five
orders    of   magnitude    larger   than    the   one    studied   by
\citet{meneghetti10b}.

\begin{figure}
\includegraphics[width=\hsize]{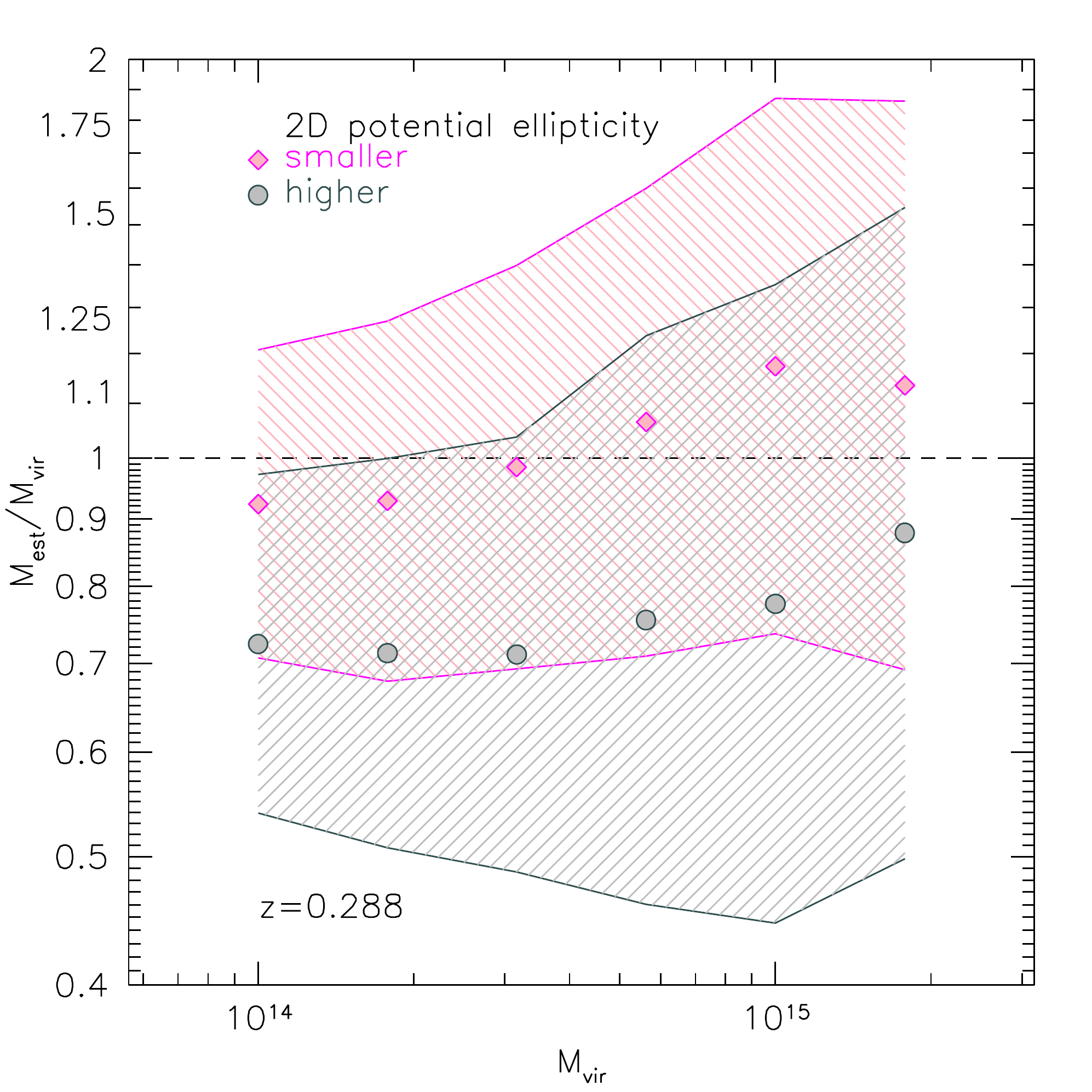}
\caption{Average  of the rescaled estimated  mass as a  function of the
  true cluster mass for haloes  with higher or smaller ellipticity
  the plane of the sky at redshift $z=0.288$. Masses and concentrations have
  been evaluated using both weak and strong lensing constraints. \label{figmestfsubPell}}
\end{figure}

We have also  investigated what effect the cluster  ellipticity on the
plane of  the sky have  on the estimated mass.   For each mass  bin we
have computed  the median potential  ellipticity $\epsilon_{\Phi,500}$
--  measured within  $R_{500}$ --  and  divided the  halo sample  into
haloes     with     smaller     or     higher     ellipticity.      In
Fig.~\ref{figmestfsubPell} we present the  average estimated mass as a
function of the  cluster mass for these two  samples, considering only
the  clusters  at  redshift  $z=0.288$.  The  results  for  the  other
redshifts are quantitatively consistent.  From  the figure we see that
the orientation of  the main halo ellipsoid is an  important source of
bias   in  the   measured   halo  mass,   as   already  discussed   by
\citet{meneghetti10b}. We consider in this case the situation in which
mass and concentration have been estimated from the WL+SL constraints.
Typically we see that for clusters  whose major axis is oriented along
the line of sight the mass tends  to be overestimated -- that are more
spherical  in  the  plane  of the  sky  \citep{morandi10},  while  the
opposite  occurs for  clusters elongated  in perpendicular  direction:
most spherical clusters are less biased than the most elliptical ones.
An analogous  result is presented in  \citet{meneghetti10b} (see their
fig.  17), where the cluster  masses tend to be under (over)-estimated
when large (small) angles between the line-of-sight and the major axis
of the halo ellipsoid are present.

\begin{figure*}
\includegraphics[width=5.8cm]{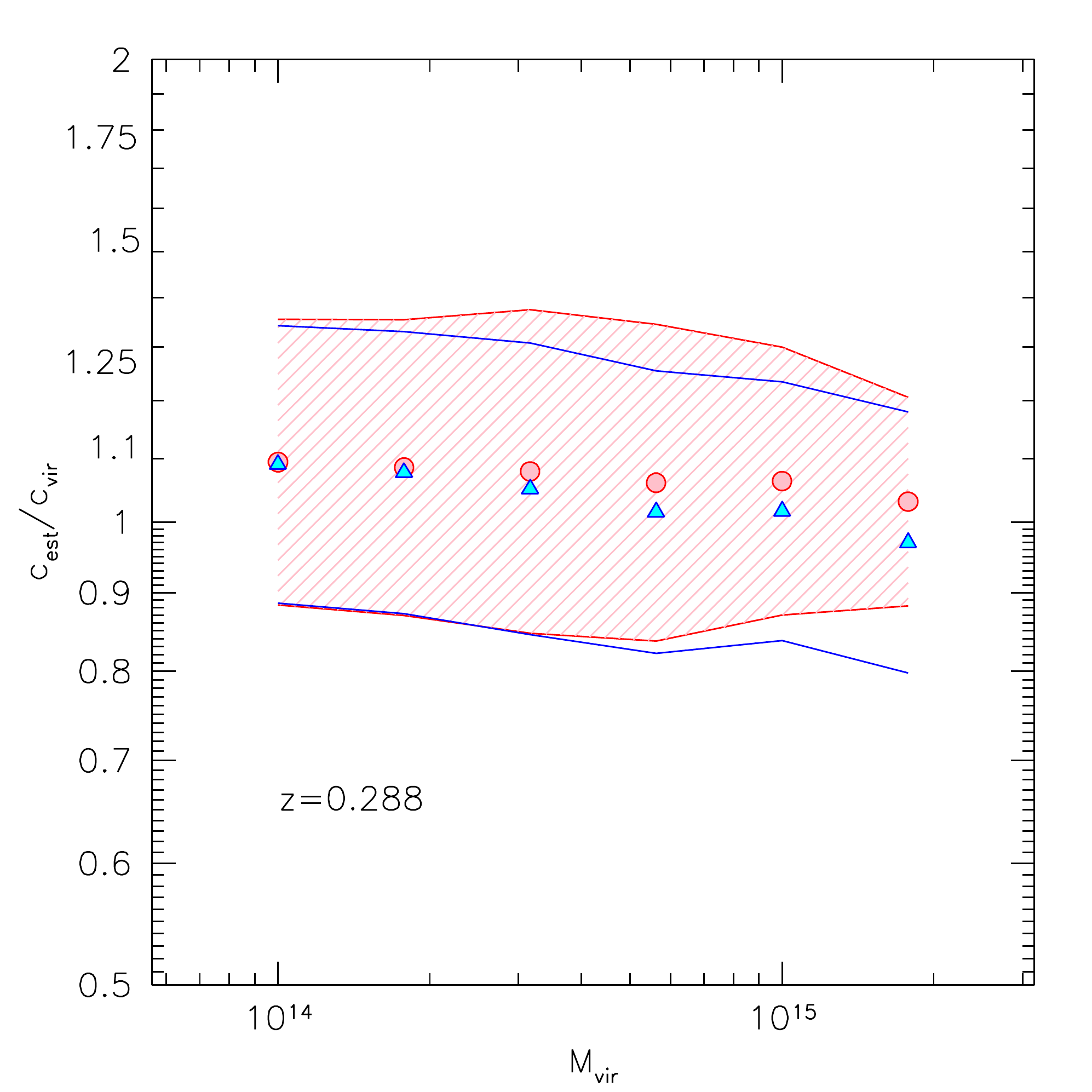}
\includegraphics[width=5.8cm]{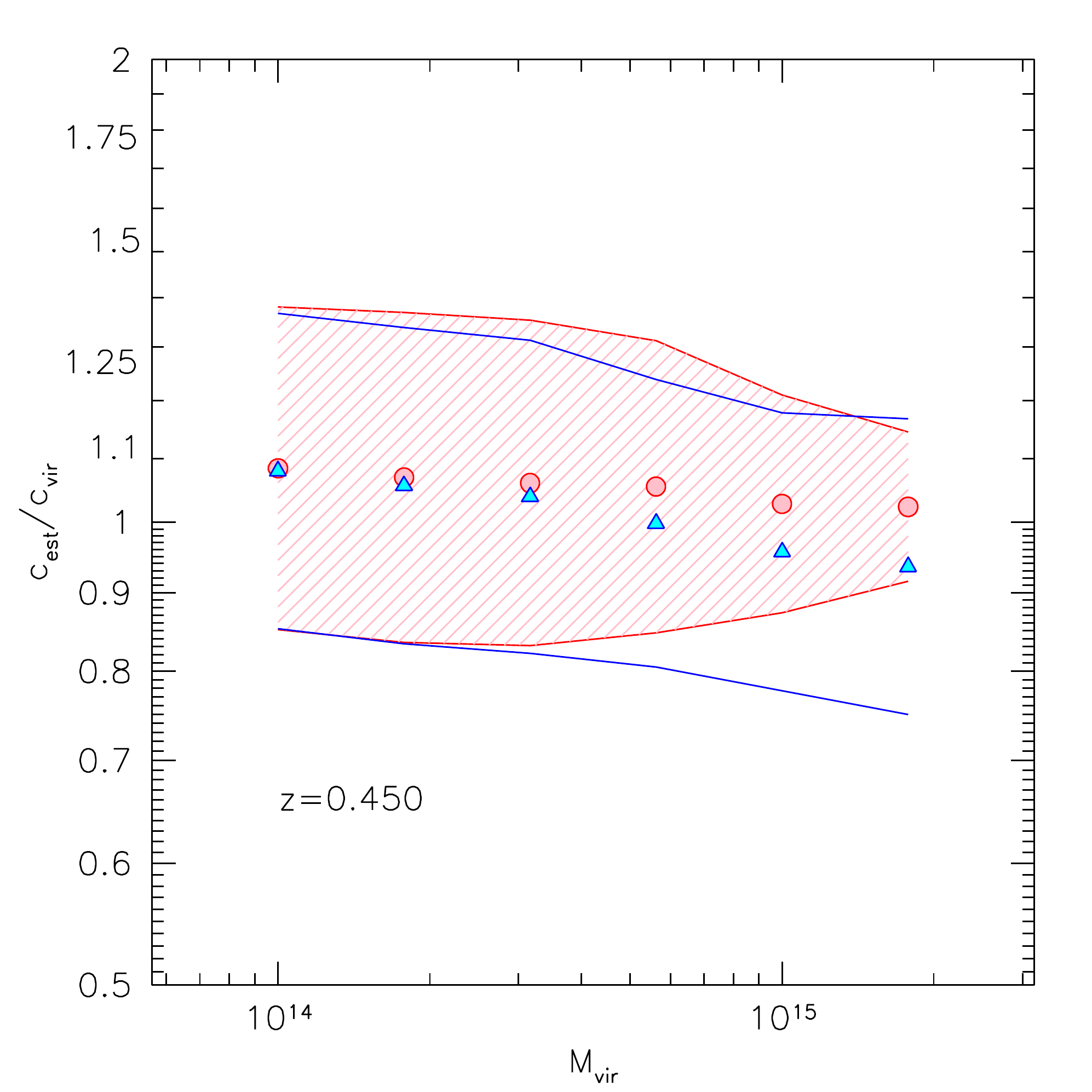}
\includegraphics[width=5.8cm]{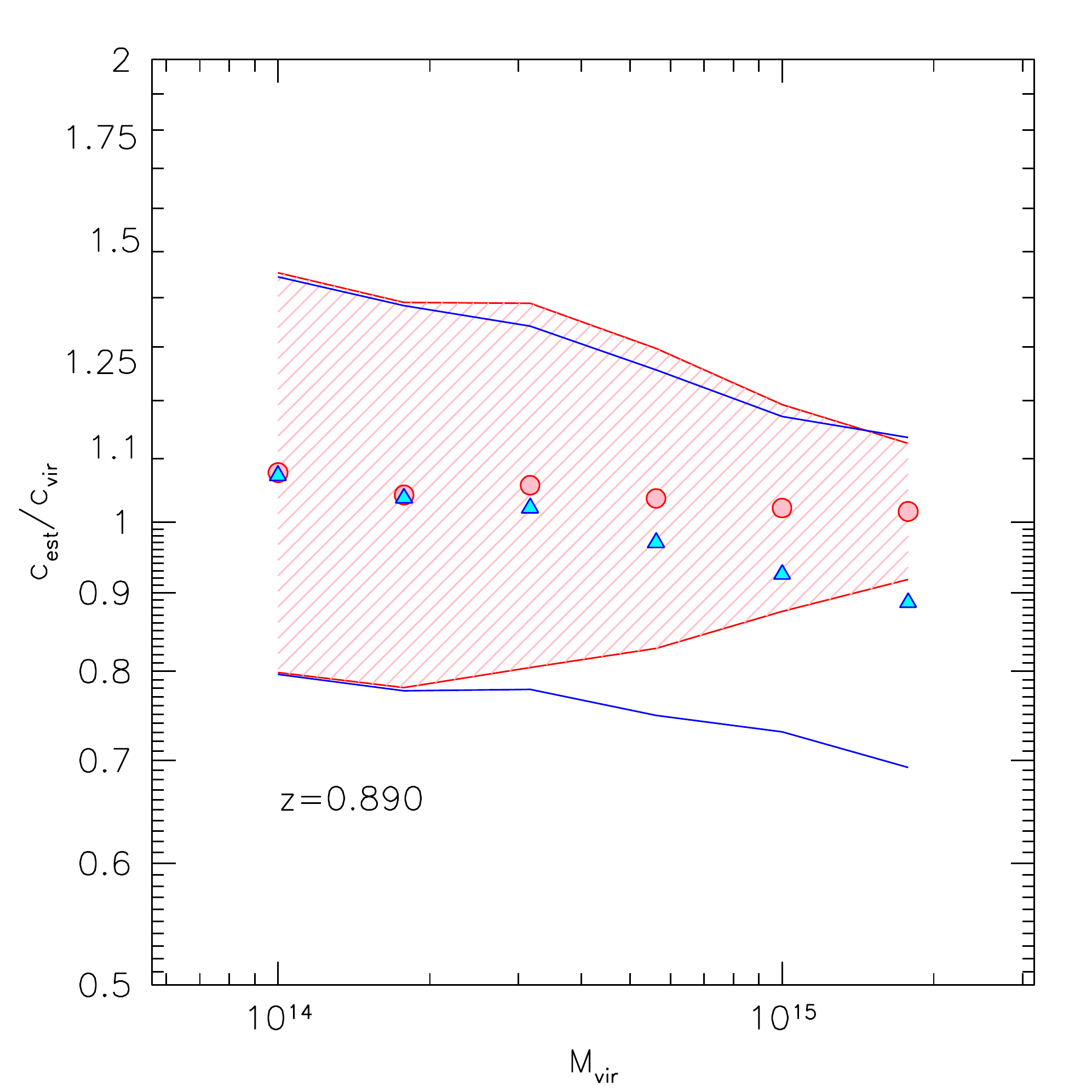}
\caption{Average of the rescaled estimated concentration as a function
  of  the  true  cluster  mass.   We   show  the  case  in  which  the
  concentration is  estimated using  WL+SL (red  circles) and  only WL
  blue  (triangles)   information  for  systems  at   three  different
  redshifts.  For  each considered  case, the corresponding  upper and
  lower curves  enclose the $1\sigma$  scatter of the  distribution at
  fixed mass.  \label{figcest6p}}
\end{figure*}

In  Fig.~\ref{figcest6p},  we  show  the  average  ratio  between  the
estimated concentration and the true one  as a function of the cluster
mass.  The different panels refer  to different redshifts and the data
points have the same meaning as in Fig.~\ref{figmassest}.  The results
show that adding to the fit the constraint on the size of the Einstein
radius  does  not  change  significantly  the  average  value  of  the
distribution,  while  it reduces  the  scatter  for the  most  massive
systems.  The  different panels show  the presence of a  positive bias
for the  smallest systems  by less than  $\sim 10\%$,  which decreases
with the lens redshift.

\begin{figure*}
\includegraphics[width=8.6cm]{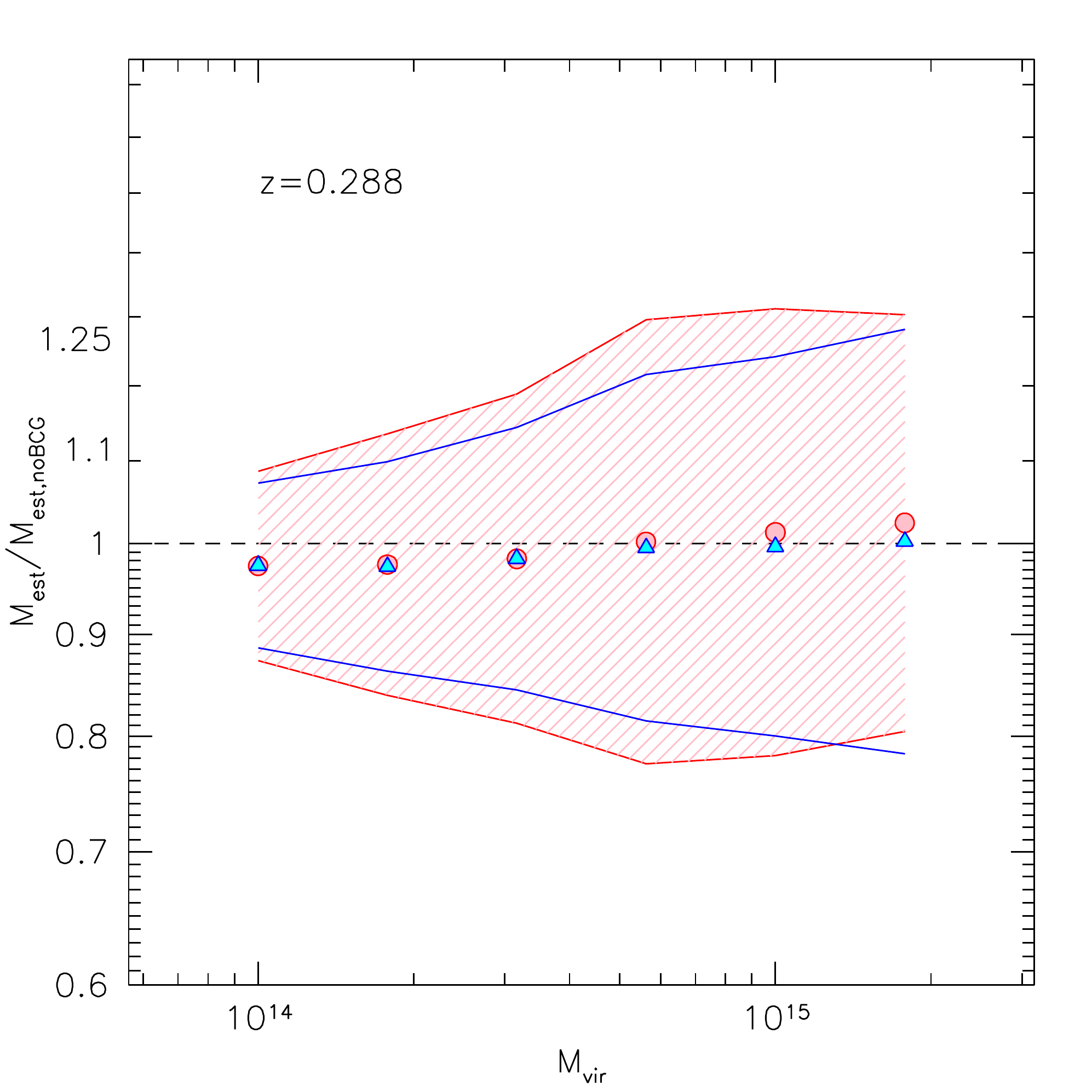}
\includegraphics[width=8.6cm]{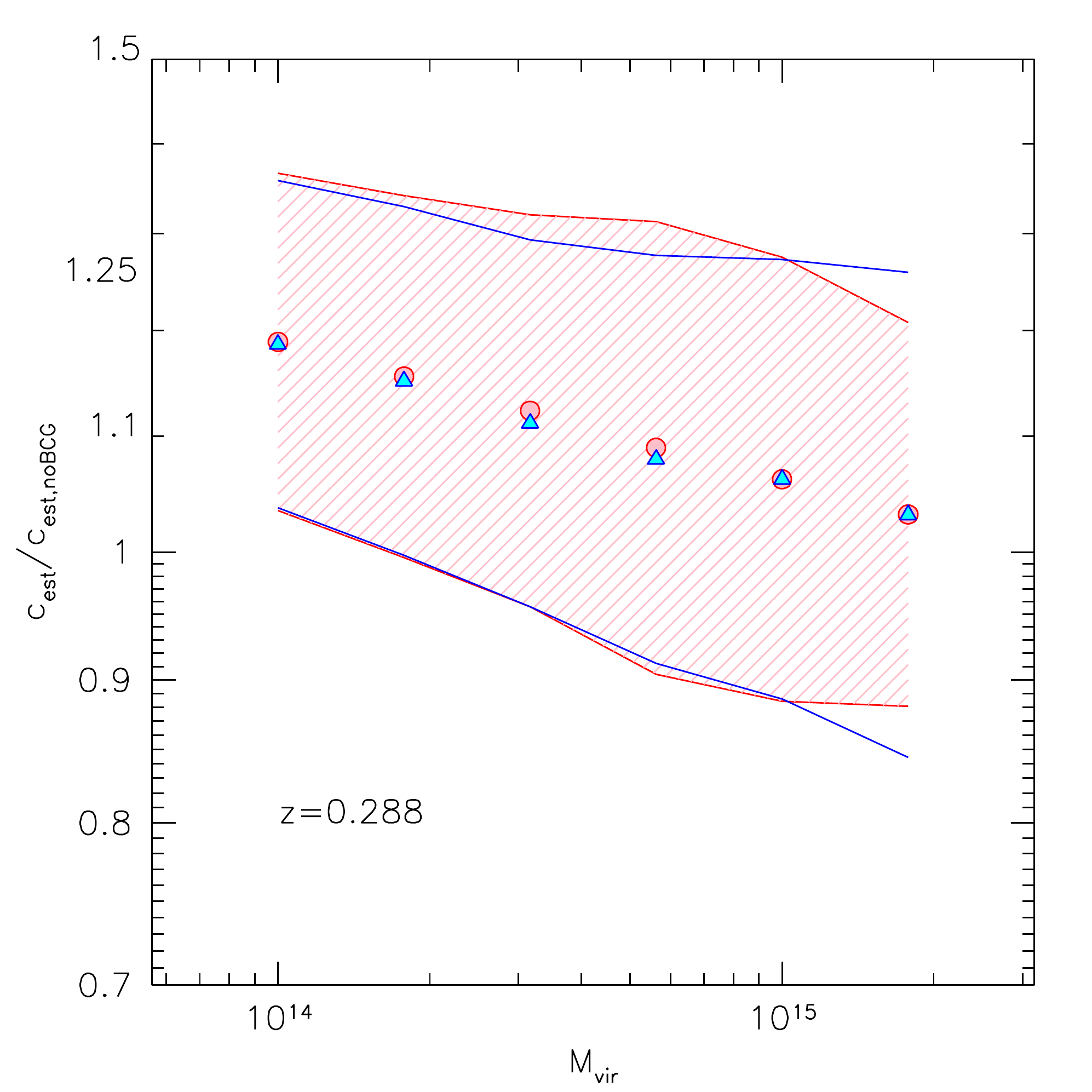}
\caption{Ratio  between different quantities estimated in simulations  
with BCG+ADC and without BCG  as a function of the  cluster mass.  Left and right
panels refer to mass and concentration, respectively. Blue triangles and  red circles refer  to the
  case where  mass and  concentration are estimated  using only  WL or
  WL+SL constrains, respectively. \label{figMandCnoBCG}}
\end{figure*}

Our cluster  lensing simulations include the  presence of a BCG  and a
recipe for  the adiabatic  contraction. To  understand how  much these
ingredients affect our mass and concentration measurements, we present
in  the  left  (right)  panel of  Fig.~\ref{figMandCnoBCG}  the  ratio
between the mass (concentration)  derived in simulations including BCG
and ADC and that estimated  in simulations without the central galaxy.
Blue triangles  and red  circles refer  to the case  where only  WL or
WL+SL constrains are considered, respectively.   We recall that in the
first case  holds $M_{\rm  vir} = M_{\rm  smooth}+ M_{\rm  in\,subs} +
M_{\rm   BCG}$  while   for  the   second  $M_{\rm   vir}  =   M'_{\rm
  smooth}+M_{\rm in\,subs}$.  We notice that fitting the whole profile
with an NFW function we have an  underestimate of the mass by only few
percent with respect to the case in which the BCG is not present.  The
trend  is different  for the  concentration which  for small  systems,
where  the  cold baryon  contribution  is  more important  since  star
formation is  fractionally more  efficient in  low-mass objects  it is
overestimated by $10-15\%$, while for  the more massive clusters it is
only few  percent. The  two data  points refer again  to the  cases in
which the  fit is  performed considering SL+WL  (circles) and  WL only
(triangles).  The shaded region encloses  the $1\sigma$ scatter of the
distribution for a fixed value of the true cluster mass.

\section{Masses and Concentrations from generalized NFW fitting} \label{secmcgNFWf}

Many studies analyzing haloes  from DM-only numerical simulations have
shown  that  the halo  density  profile  \citep{neto07,gao12} is  well
described by the  NFW relation. However, in our case,  we need to take
into account the  fact that real galaxy clusters are  not only made up
by dark matter -- that accounts for more than $85\%$ of the total mass
-- but  also by baryons, divided  into cold and hot  components. While
the hot component  is more evenly spread in the  potential well of the
cluster -- possessing a scale  radius of the density profile $r_{s,h}$
of the  order of hundreds  kpc/$h$ --  the cold component,  that turns
into the  presence of  a bright central  galaxy, is  more concentrated
toward  the center  with a  scale radius  $r_{s,c}$ much  smaller than
$r_{s,h}$.   This translates  into a  total density  profile which  is
different from an NFW relation and typically has an inner slope larger
than unity.   In order  to better  model the  increase of  the density
distribution  towards  the cluster  centre  we  can introduce  a  free
parameter $\beta$ in  the NFW equation which allows  the central slope
to freely vary \citep{zhao96,jing00}:
\begin{equation}
\rho_{gNFW}(r,\beta|M_{vir},c_{vir}) = \dfrac{\rho_s}{\left( r/r_s\right)^{\beta} 
\left(1+r/r_s\right)^{3-\beta}}\,,
\label{eqgNFW}
\end{equation}
where  $\rho_s$  represents  the   density  within  the  scale  radius
$r_s$. In order to define  the concentration it is useful to introduce
the quantity $r_{-2}$  as  the radius  at  which  the  logarithmic density  profile
is $-2$. This allows  us to write $r_{-2}=(2-\beta) r_s$ and
the  concentration $c_{-2} \equiv  R_{vir}/r_{-2}=c_{vir}/(2-\beta)$.  The
profile and the corresponding  definitions match the NFW function when
$\beta=1$.   The  generalized  NFW  convergence  can  be  obtained  by
integrating the  profile in equation~(\ref{eqgNFW}) along  the line of
sight:
\begin{equation}
\kappa_{gNFW}(r') = \int_{-\infty}^{\infty} \rho_{gNFW}(r,\zeta) \mathrm{d}\zeta\,,
\end{equation}
with $r^2  \rightarrow r'^2 + \zeta^2$  and $r'$ the  radius vector on
the plane of the sky. We can now define the dimensionless mass
$m(x)=\int_0^{x} x \kappa(x)\mathrm{d}x$ and the shear:
\begin{equation}
\gamma_{gNFW}(x) = \dfrac{m_{gNFW}(x)}{x^2} - \kappa_{gNFW}(x)\,,
\end{equation}
with     $x=r/r_s$.     By    setting     $\lambda_t    =     0$    in
equation~(\ref{eqlambdat}),  we can  find  the Einstein  radius for  the
generalized  NFW   profile,  that  we   use  as  reference   model  in
constructing  $\chi^2_{SL}$.   The   $\chi^2$  is  in  this  case
minimized  in  order to  obtain  three  parameters: the virial  mass,
the concentration  and    the  inner  slope $\beta$ of  the  total  density
profile.

\begin{figure*}
\includegraphics[width=5.8cm]{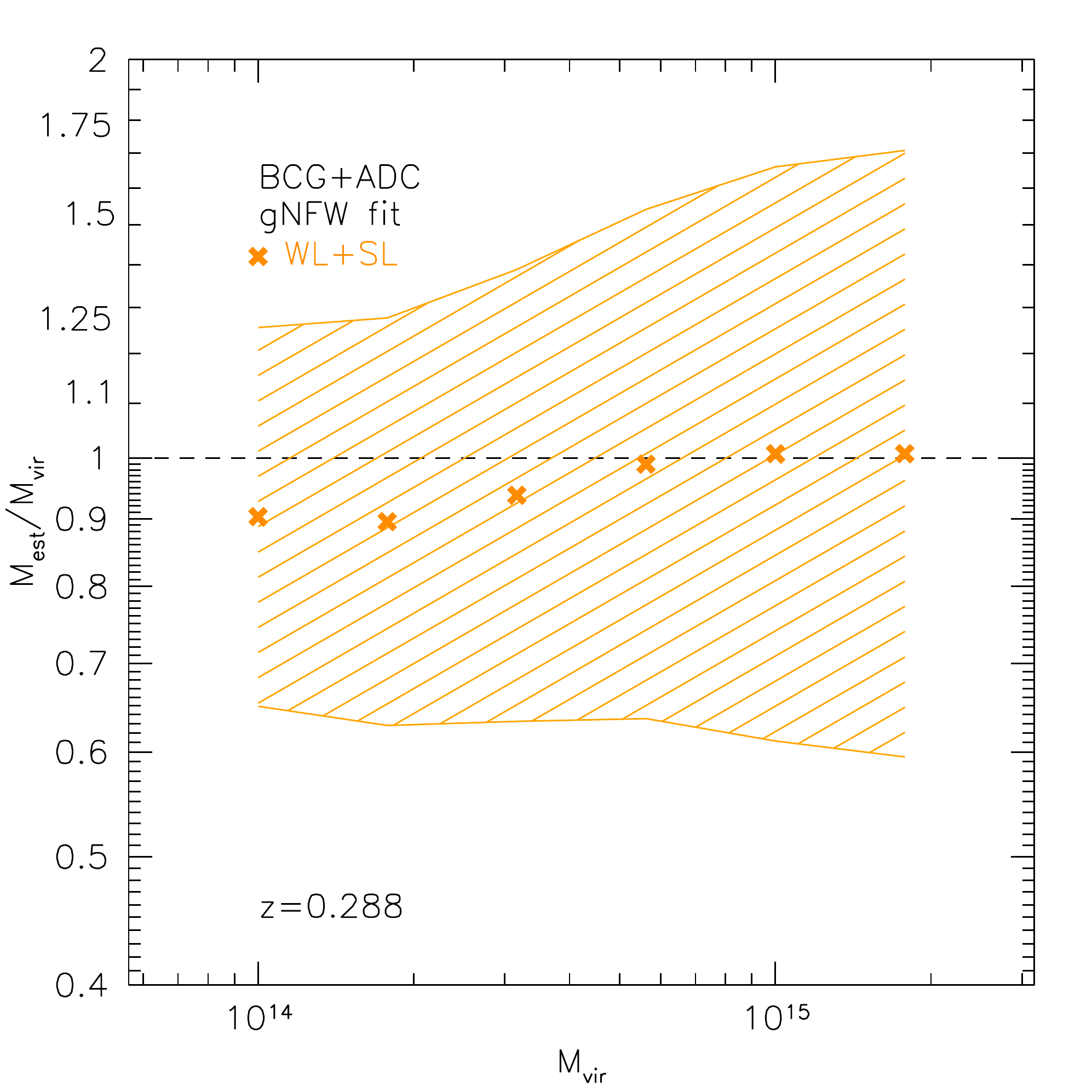}
\includegraphics[width=5.8cm]{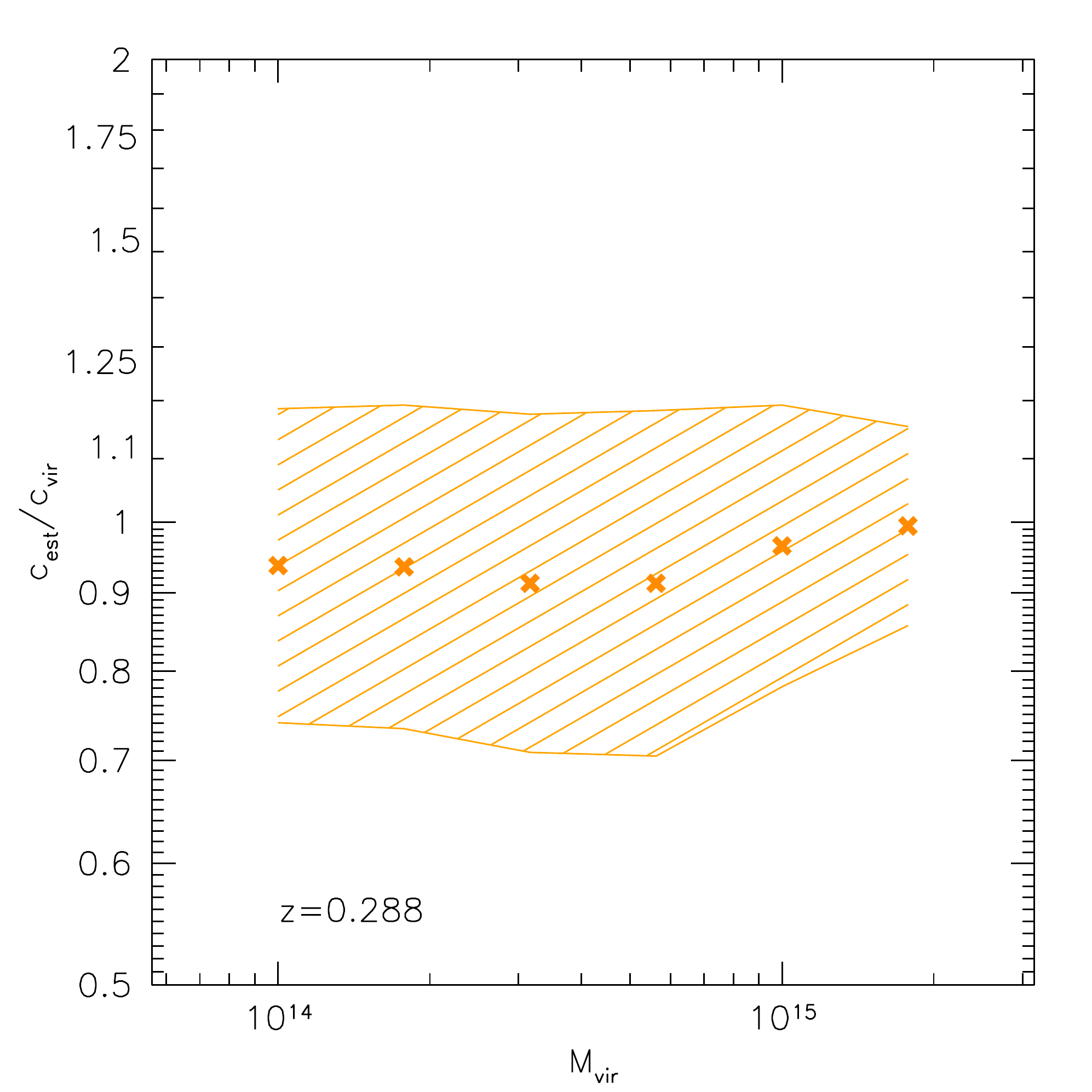}
\includegraphics[width=5.8cm]{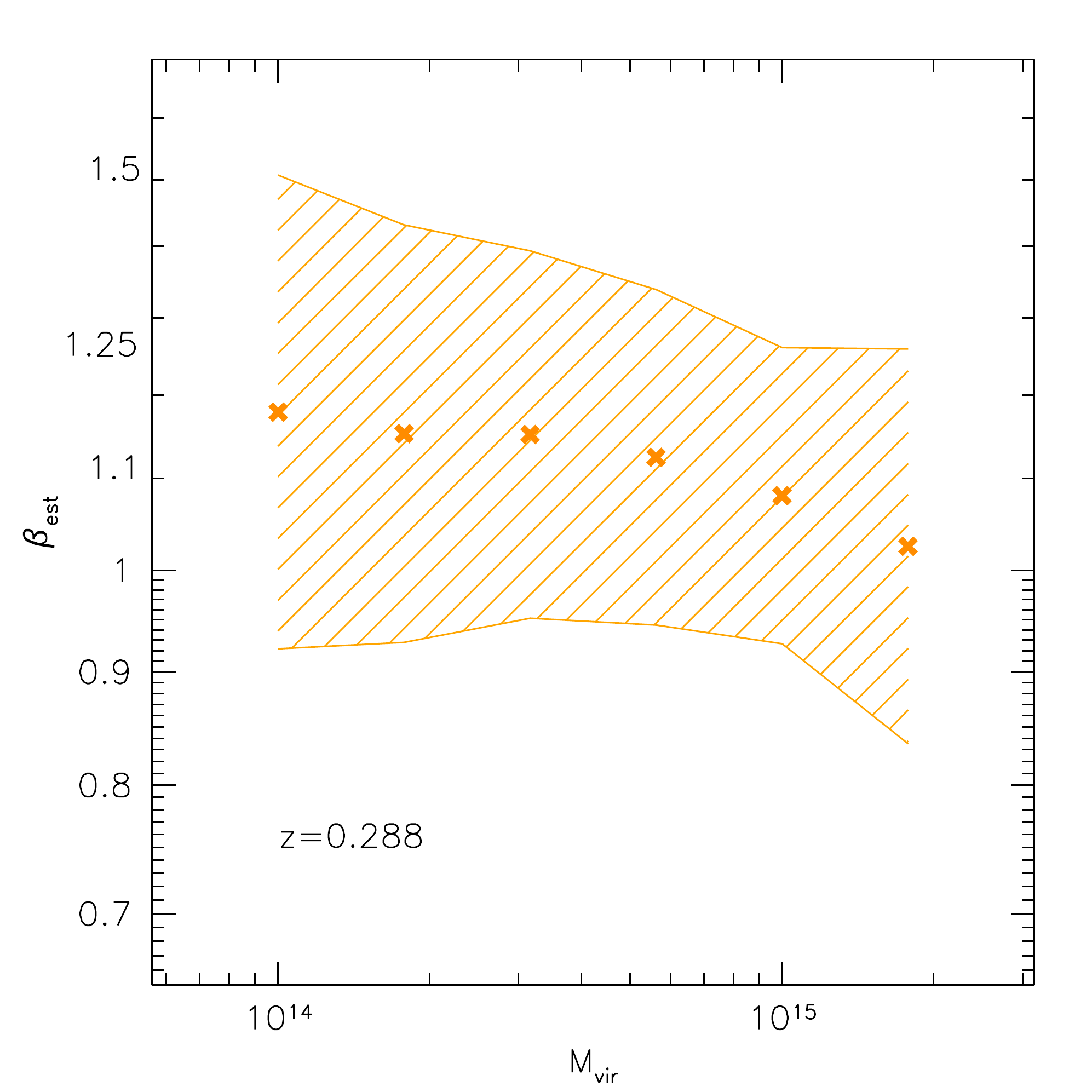}
\caption{Average mass  (left panel), concentration (center  panel) and
  inner  slope   (right  panel)   estimated  best  fitting   with  the
  generalized NFW  profile the reduced  shear profile and the  size of
  the Einstein radius, as a function  of the cluster mass.  The shaded
  region     encloses     the     $1\sigma$     scatter     of     the
  distribution.\label{figgNFW}}
\end{figure*}

In Fig.~\ref{figgNFW},  we show  as a  function of  the halo  mass the
average rescaled mass and concentration, and the inner slope estimated
by fitting the  tangential shear profile and the size  of the Einstein
radius using  a generalized  NFW profile (the  relations at  the other
redshifts are consistent with what  presented here at $z=0.288$).  The
shaded region encloses  the $1\sigma$ scatter of  the distributions at
fixed halo mass,  while the points represent the  average value.  From
the figure we notice that using the  gNFW profile the bias in the mass
is reduced by  about $10\%$ for the smallest systems,  but the scatter
remains  as large  as in  the NFW  case.  The  average measure  of the
concentration is almost unbiased -- of the order of few percent.  From
the right panel we notice that on  average we tend to measure an inner
slope larger  than unity for  the smallest systems.  This  behavior is
related to  the fact that we  are fitting the total  2D matter density
distribution including  the contribution both  from DM and  the bright
central galaxy.  Since  the BCG steepens the profile  and affects more
significantly the core of the smallest  systems, we tend to measure on
average an inner slope that is even $20\%$ larger than one.  Even when
the BCG is  removed, a bias of $5-7\%$ still  remains for the smallest
clusters due to the prolateness of their ellipsoids.

\section{Cosmological Halo Sample} \label{seccosmos}

Until now  many numerical simulations and  analytical predictions have
been developed to interpret the  number of collapsed objects and their
concentration-mass relation at  a given redshift.  With  the advent of
the era  of precision cosmology  it is possible  to use the  halo mass
function  and   the  concentration-mass  relation  as   an  additional
cosmological  probe.    Currently  the  relatively  small   number  of
available data  combined with  the bias and  the scatter  in estimated
cluster   properties   limits    the   constraints   on   cosmological
parameters. However, recently the  Planck team \citep{planckxx}, using
a sample of 189 galaxy clusters from the Planck SZ catalogue, was able
to  put  good  constraints  on  $\Omega_m$  and  $\sigma_8$  the  also
emphasize the fact that the  values of the cosmological parameters are
degenerate  with the  hydrostatic  mass bias  and  that the  agreement
between the cluster counts and  the primordial CMB anisotropies can be
reached only  assuming a  mass bias  of about  $45\%$.  However  it is
worth mentioning that the  tension between the cosmological parameters
($\Omega_m$ and $\sigma_8$)  derived from cluster counts  and the ones
derived from the Planck CMB temperature maps is alleviated when the SZ
clusters are cross-correlated  with the X-ray cluster  maps from ROSAT
\citep{hajian13}.    At  the   same  time   the  CLASH   collaboration
\citep{postman12}, reconstructing the mass distribution of a sample of
$25$ galaxy  clusters using weak  and strong lensing  measurements, is
exploring the  possibility of  eventually measuring deviations  of the
concentration-mass    relation    from    the    one    measured    in
$\mathrm{\Lambda}$CDM  numerical  simulations.   However  these  works
require precise knowledge of bias and scatter when comparing estimated
and true cluster properties, like mass and concentration.

\begin{figure*}
\includegraphics[width=8.6cm]{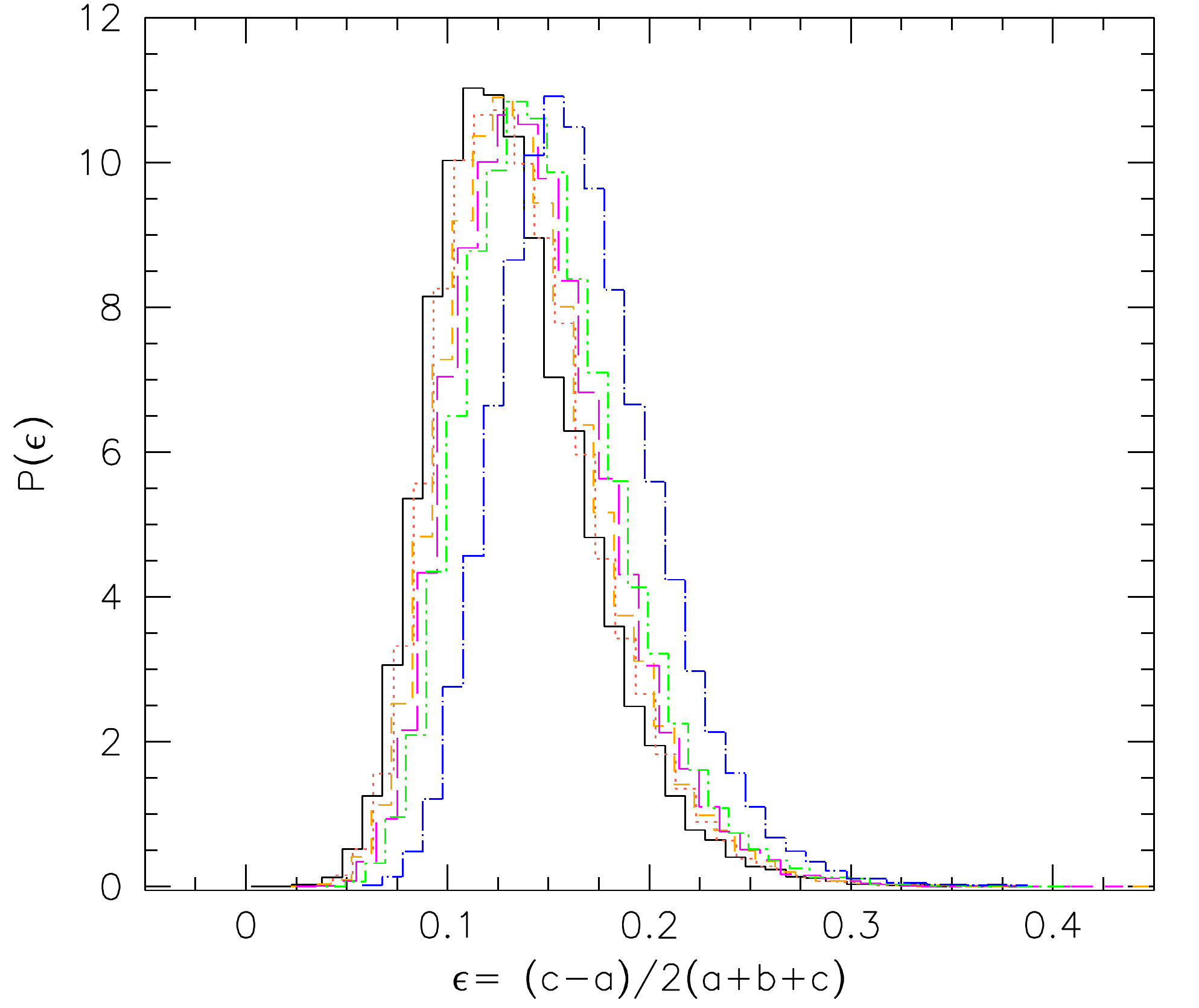}
\includegraphics[width=8.6cm]{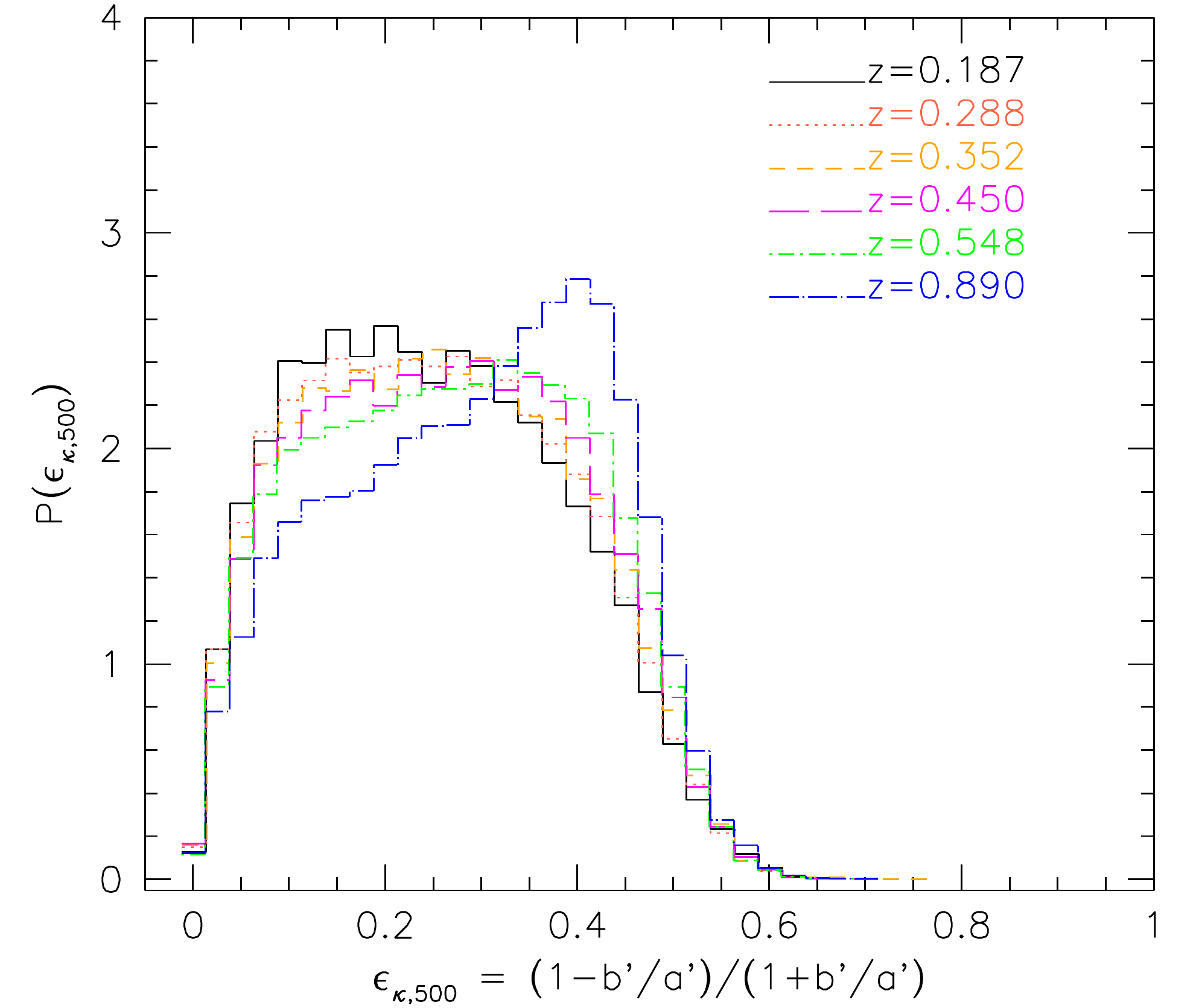}
\caption{Left panel: intrinsic 3D ellipticity distribution of the main
  halo component  for the samples  of galaxy cluster at  six different
  redshifts.  Right panel: measured  2D ellipticity of the convergence
  at the radius at which the  enclosed density reaches $500$ times the
  critical  value.   Different  line  styles  refer   to  the  various
  considered redshifts, as labelled.
  \label{figellipticity}}
\end{figure*}

In this  section, we will discuss  how the mass and  the concentration
measured by fitting the tangential  shear profile and constraining the
size Einstein  radius tend to modify  the intrinsic concentration-mass
relation  for galaxy  clusters.  We will  also  discuss how  selection
effects on  the cluster  sample could  modify the  slope and  the zero
point of the relation.

As done  previously, we consider  the case of six  different redshifts
but   with   the   halo   sample   extracted   from   the   analytical
\citet{sheth99b} mass function. We consider haloes with a mass larger than
$10^{14}M_{\odot}/h$.   At each  redshift,  the number  of
haloes is created to match  the number of collapsed objects present on
the whole sky between $z-\Delta  z/2$ and $z+\Delta z/2$ (with $\Delta
z = 0.01$).  To increase  the statistical sample, for each redshift we
perform $8$ different realizations.

We  investigate the  ellipticity  distribution  of these  cosmological
samples.  In the  left panel of Fig~\ref{figellipticity},  we show the
distribution of  the 2D ellipticities,  for each of the  six redshifts
considered.  The  parameters $a$,  $b$ and $c$  are the  smallest, the
intermediate and  largest axes  of the  halo ellipsoid  describing the
dark  matter  halo,  obtained  from  the  \citet{jing02}  model.   The
corresponding 2D  ellipticity distributions (on  the plane of  the sky
with $a'$ and  $b'$ representing the smallest and the  longest axis of
the  ellipse)  measured  from  the  cluster  convergence  maps  within
$R_{500}$ are shown  in the right panel.  It is  interesting to notice
the  agreement  with  the results  obtained  by  \citet{meneghetti10a}
analyzing  the  clusters  extracted   from  the  MARENOSTRUM  UNIVERSE
simulation \citep{gottlober06,gottlober07}:  our distributions  lie in
between  with respect  to the  their measurements  of the  convergence
ellipticity measured  at $R_{vir}$  and at  $0.1 \times  R_{vir}$.  As
expected from numerical  simulations \citep{jing02,despali12} and also
from  analytic predictions  of collapsing  ellipsoids \citep{rossi11},
high-redshift clusters tend to be  more elliptical, since more ongoing
merging events make  them typically unrelaxed. Notice  that the higher
ellipticities of high  redshift clusters tend to  enhance their strong
lensing  efficiency by  stretching  and increasing  the critical  area
\citet{zitrin13a,zitrin13b}.  In Appendix~\ref{app2D}  we discuss  the
correlation between convergence  and potential ellipticities comparing
our finding with the model proposed by \citet{golse02}.

\subsection{The cluster mass function}

As described previously in the paper, for each cluster we estimate the
mass and the concentration in two ways.  In the first case we use only
the tangential shear profile (WL), while  in the second one we combine this
with the  measurement of the size  of the Einstein radius  adopting as a
reference  a NFW  model (WL+SL);  for the  case WL+SL  we perform  the
measurement also considering a generalized NFW model (gNFW WL+SL).

\begin{figure*}
\includegraphics[width=5.8cm]{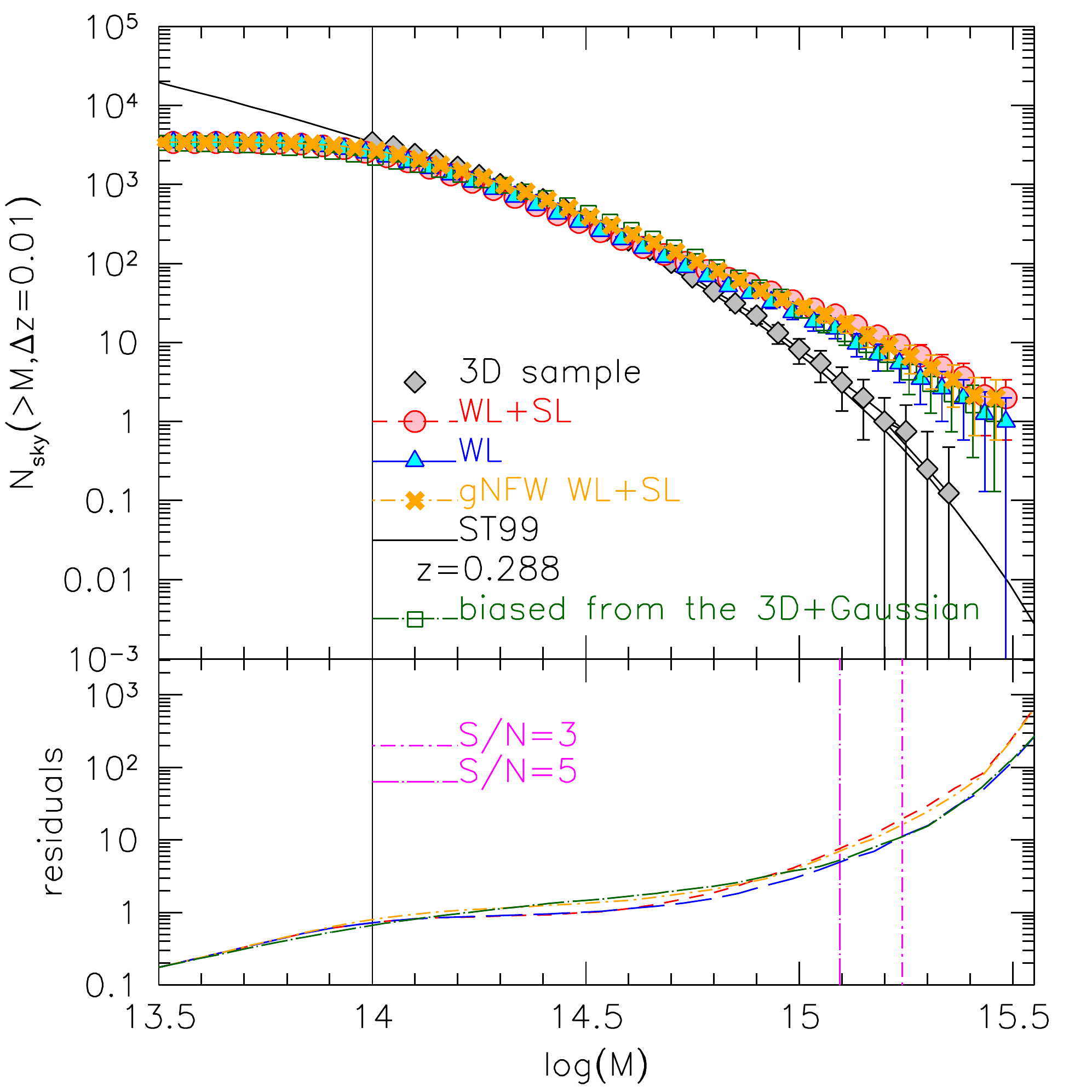}
\includegraphics[width=5.8cm]{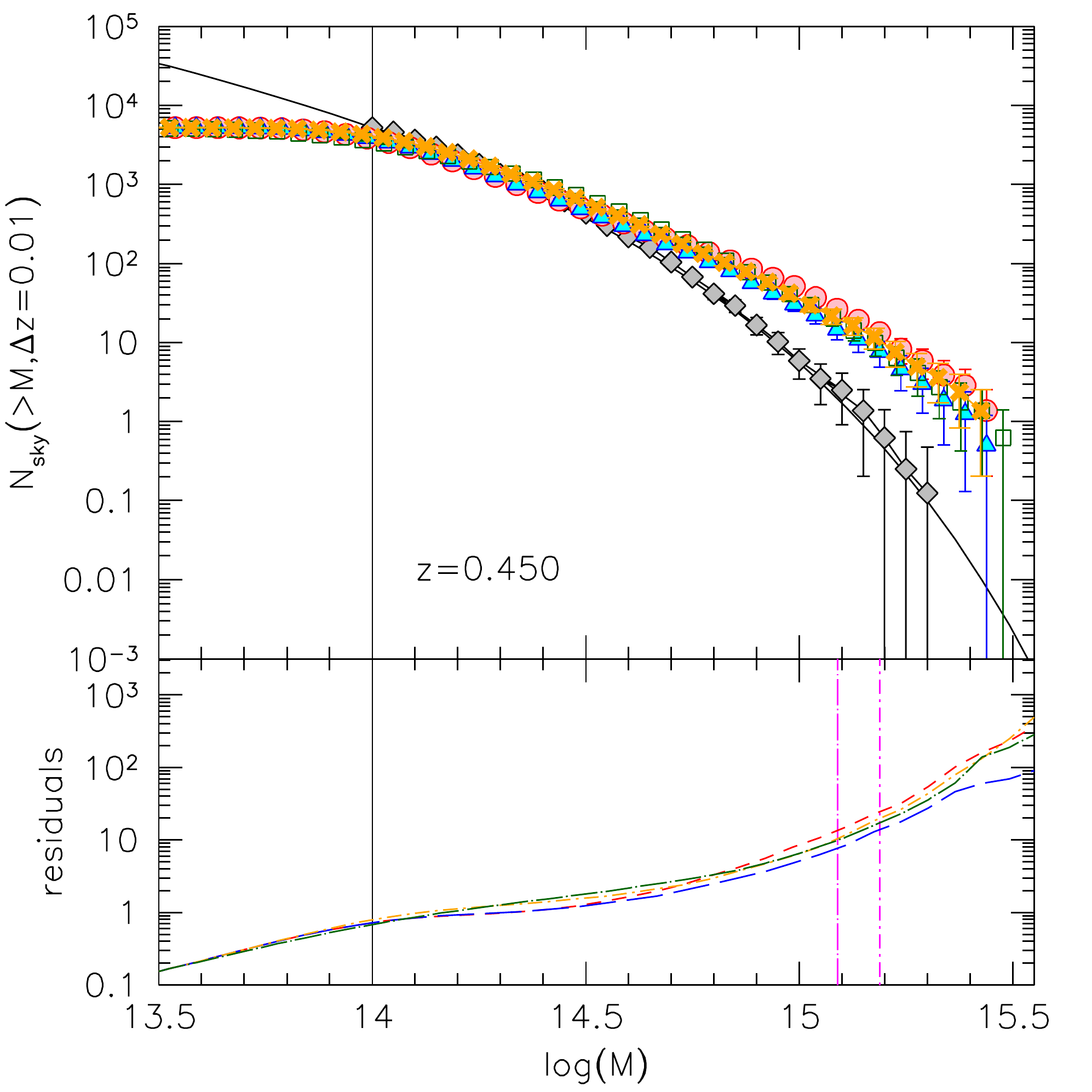}
\includegraphics[width=5.8cm]{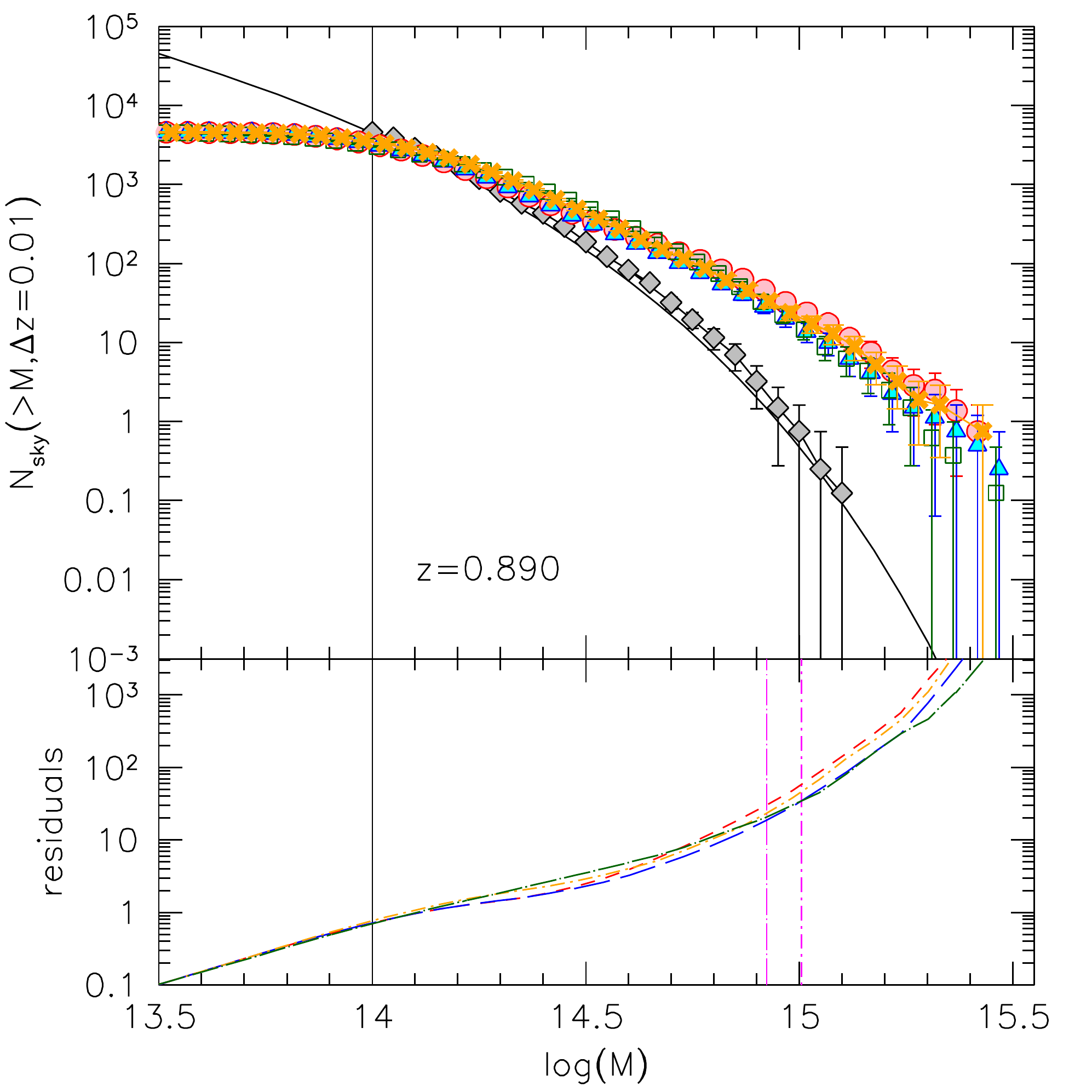}
\caption{Cumulative all sky cluster  mass functions at redshift z=0.288 (left panel), 
z=0.450 (central panel) and z=0.890 (right panel). In  each panel  panel, the  solid line  in the  top frame
  represents the  analytical halo  mass function  \citep{sheth99b},
  while  the black  filled  diamonds  show the  mass  function of  the
  cluster above $10^{14}M_{\odot}/h$  obtained sampling this function.
  The filled blue  triangles and the red filled  circles represent the
  cluster  mass   function  obtained   using  WL  and   WL+SL  masses,
  respectively.   For both cases, we  have considered  as a  reference
  model  a  NFW  halo.  The orange cross show the WL+SL mass function, where a
  generalized NFW model has been used as a reference. The open  green squares represent
  the mass  function obtained  biasing the halo  mass of  each cluster
  using the best fit relations reported in Table~\ref{tablelsqrelation}, and
  assuming a gaussian scatter of  $\sigma_{\log M}=0.25$.  For all data
  the   error    bars   represent   their    corresponding   Poisson
  uncertainties.   In  each panel,  the  bottom  frame represents  the
  residuals as  a function of  the cluster mass  of the WL,  WL+SL and
  WL+SL   using   gNFW   model   with  respect   to   the   analytical
  \citet{sheth99b} mass function. \label{figmassfunctions}}
\end{figure*}

In Fig.~\ref{figmassfunctions}, we show the cumulative all sky cluster
mass function at three different  redshifts.  In each panel, the solid
curve  represents the  analytical mass  function of  \citet{sheth99b},
while  the  black  filled  diamonds represent  a  catalogue  extracted
sampling this relation.  Red filled  circles and blue filled triangles
show   the  mass   function   from  weak   and  weak+strong   lensing,
respectively.     In   all    panels   the    vertical   line    marks
$10^{14}M_{\odot}/h$:  the  minimum  mass  of  the  extracted  cluster
sample.   We  notice that  while  for  small  masses the  behavior  is
dominated by the  difference between the true and  the estimated mass,
for large  masses -- since  there are less  haloes -- it  is dominated
mainly by  the scatter.  We  have tested this statement  modifying the
mass of  the cosmological  sample using the  least-squares fit  to the
WL+SL  masses  performed  to  the data  in  Fig.~\ref{figmassest},  as
expressed    in   Table~\ref{tablelsqrelation},    with   a    scatter
$\sigma_{\log M}  = 0.25$ (open  green squares points in  the figure).
From  the bottom  frame presented  in each  panel, where  we show  the
difference  of  the  measurements   with  respect  to  the  analytical
prediction, we  notice the biased  Gaussian sample matches  quite well
the WL+SL  distribution up  to the  mass bin where  the $S/N=5$  -- we
define  the signal  to noise  for each  bin as  the ratio  between the
number  of clusters  and the  corresponding Poisson  error.u The  mass
function obtained  using a gNFW model  as reference is similar  to the
others even if the mass bias is  reduced in this case.  This is due to
the  scatter  still  present  in  the  rescaled  estimated  mass:  the
discrepancy  with respect  to  the theoretical  prediction is  further
reduced if $\sigma_{\log  M} = 0.1$. If these  deviations with respect
to the theoretical  mass function are not reduced, they  will bias the
estimated  cosmological  parameters in  favor  of  models with  higher
$\sigma_8$ and lower $\Omega_m$.
\begin{table}
\caption{Least-squares fit to the estimated mass as a function of 
the true mass: $\log \mathrm{M_{set}/M_{3D}} = a \log \mathrm{M_{3D}} +
b$ ($M_{3D}$ is in unit of $M_{\odot}/h$). On the left for the case WL+SL, while on the right for the WL alone.}
\label{tablelsqrelation}
\begin{tabular}{@{}lccccc}
\hline
redshift & a & b &|& a & b  \\ \hline
0.187 & 0.062 & -0.970 &|& 0.045 & -0.731 \\
0.288 & 0.077 & -1.179 &|& 0.051 & -0.811 \\
0.352 & 0.073 & -1.121 &|& 0.046 & -0.727 \\
0.450 & 0.059 & -0.905 &|& 0.037 & -0.597 \\
0.548 & 0.059 & -0.900 &|& 0.040 & -0.627 \\
0.890 & 0.055 & -0.813 &|& 0.040 & -0.608 \\
\end{tabular}
\end{table}

\subsection{The concentration-mass relation}

\begin{figure*}
\includegraphics[width=5.8cm]{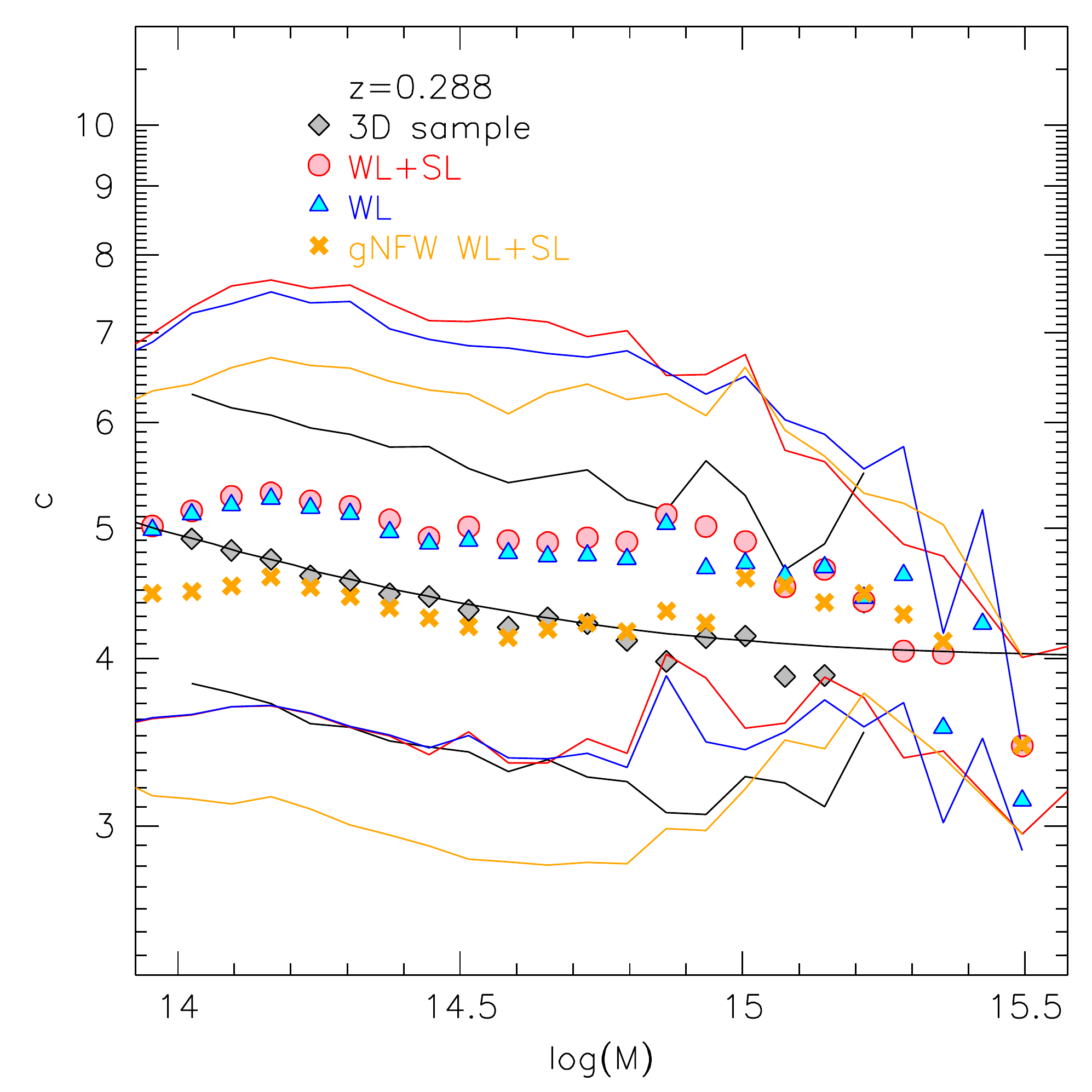}
\includegraphics[width=5.8cm]{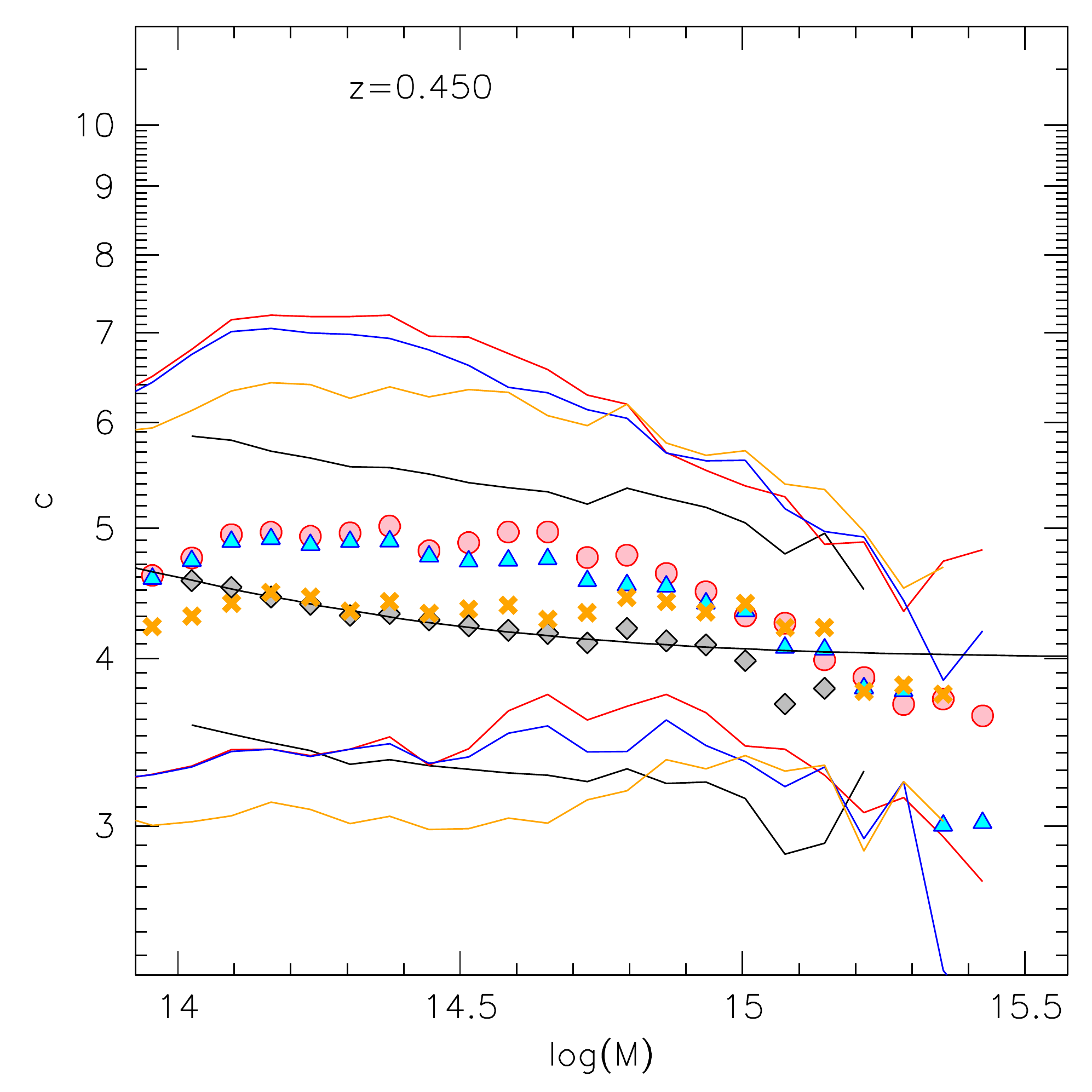}
\includegraphics[width=5.8cm]{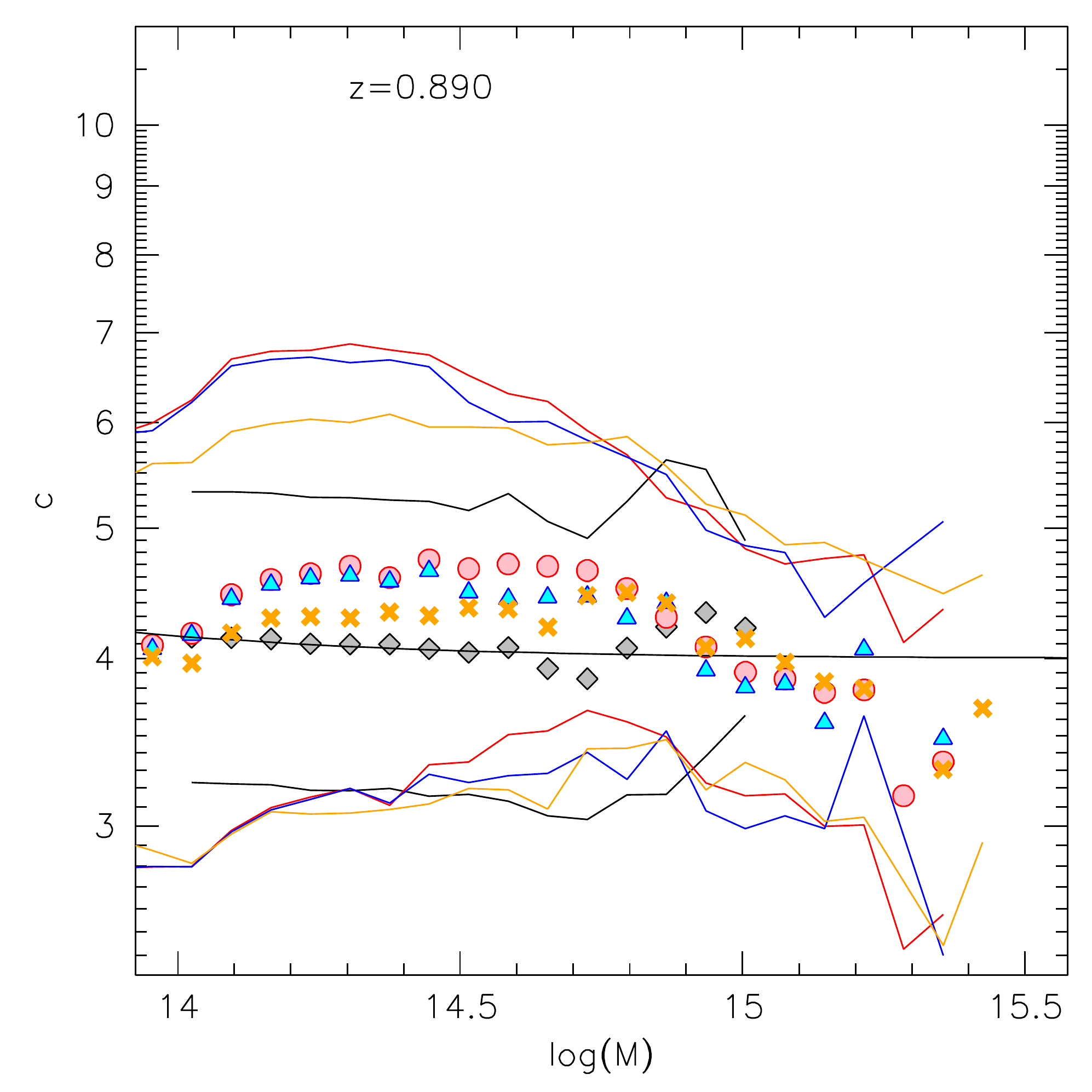}
\caption{Concentration-mass relation at  three different redshifts. In
  each panel the solid curve represents the \citet{zhao09} model, which
  we use as a reference when we assign the concentration to each halo.
  The  black filled  diamonds  represent the  intrinsic measured $c-M$  relation,
  while the blue  filled triangles and the red filled  circles correspond to the $c-M$
  relations  obtained  for  WL  and WL+SL,  respectively.   The  orange
  crosses  show  the  relation  for   WL+SL  when  the  mass  and  the
  concentration of  the halo are  obtained using  the gNFW model  as a
  reference. For each data, the curves with the same color enclose the
  first  and the  third quartiles  of the  distribution at  fixed halo
  mass. \label{figintcmrel}}
\end{figure*}

Biases  and   scatter  of  mass   and  concentration  impact   on  the
concentration-mass relation.   In Fig.~\ref{figintcmrel}, we  show the
concentration-mass relation at three  considered redshifts.  The solid
curve in  each panel represents  the $c-M$ model by  \citet{zhao09} at
the corresponding  redshift, while  the black  filled diamonds  is the
median intrinsic  relation for the  clusters obtained from  this model
assuming a  log-normal scatter of  $0.25$.  The blue  filled triangles
and the  red filled circles  show the median $c-M$  relations obtained
using  WL and  WL+SL estimates.   In both  cases, we  notice that  the
concentration-mass relation for clusters  tends to be overestimated by
$20\%$.  On average the difference with respect to the intrinsic $c-M$
relation  is  reduced using  gNFW  masses  and concentrations  (orange
crosses) and  the median  points are  on average  well inside  the two
central quartiles  of the intrinsic  distribution. For each  data, the
curves with  the corresponding color  enclose the first and  the third
quartiles of the distribution at fixed  halo mass.  From the figure we
notice that describing  the mass distribution as  a triaxial ellipsoid
tends to increase the normalization of the concentration-mass relation
\citep{comerford07} .  As discussed by \citet{oguri05} in the analysis
of A1689, this may eventually  reduce the apparent discrepancy between
theory and observations.

Different   observational  campaigns   \citep{okabe10b,postman12}  are
trying to use  the concentration-mass relation to  test the agreements
of  observations with  the  predictions of  structure  formation in  a
$\mathrm{\Lambda}$CDM cosmology.  However it  is reasonable to  ask if
the  selection  function of  the  observed  clusters is  important  in
reconstructing the $c-M$ relation and, if so, how does it reflects, in
both  the  true   and  reconstructed  samples.   For   example  it  is
interesting to point out the work by \citet{sereno12}, where the $c-M$
relation  of the  MACS cluster  sample has  been studied  using strong
lensing mass  model reconstructions.   This work first  underlines the
importance of  triaxiality of  the clusters when  reconstructing their
properties  and second  the  selection  function since  high-redshift,
unrelaxed  clusters may  form a  different class  of prominent  strong
gravitational lenses.

\begin{figure*}
\includegraphics[width=5.8cm]{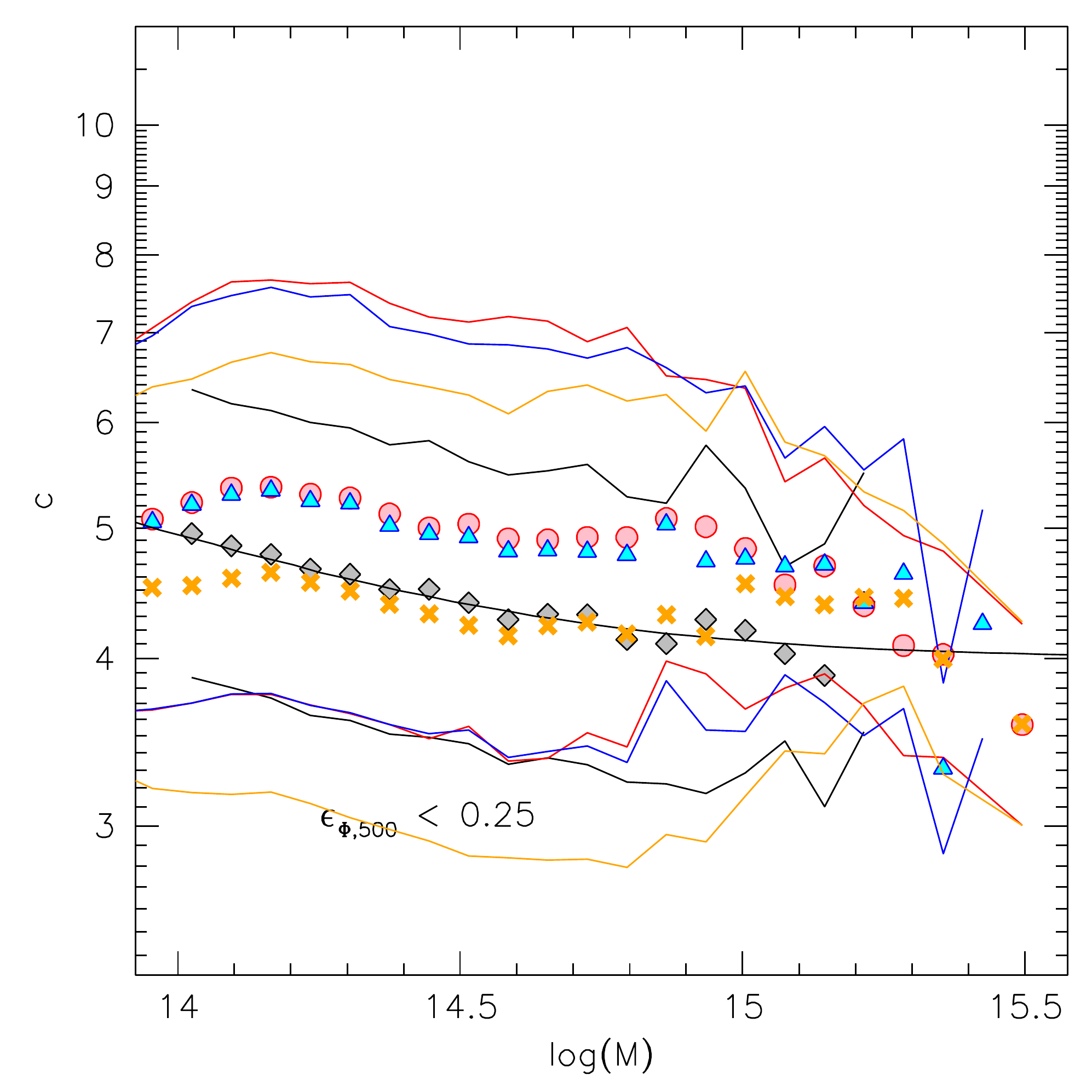}
\includegraphics[width=5.8cm]{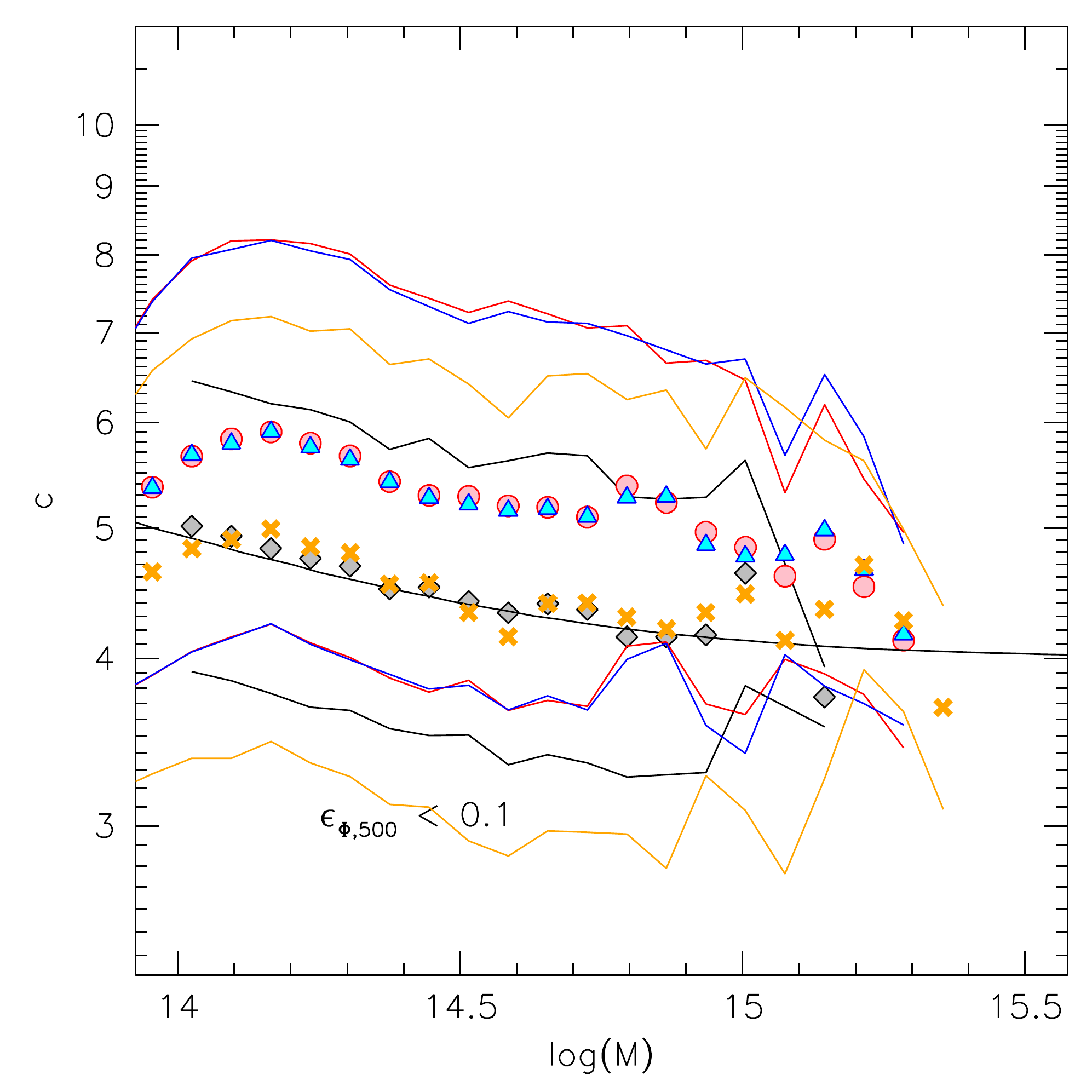}
\includegraphics[width=5.8cm]{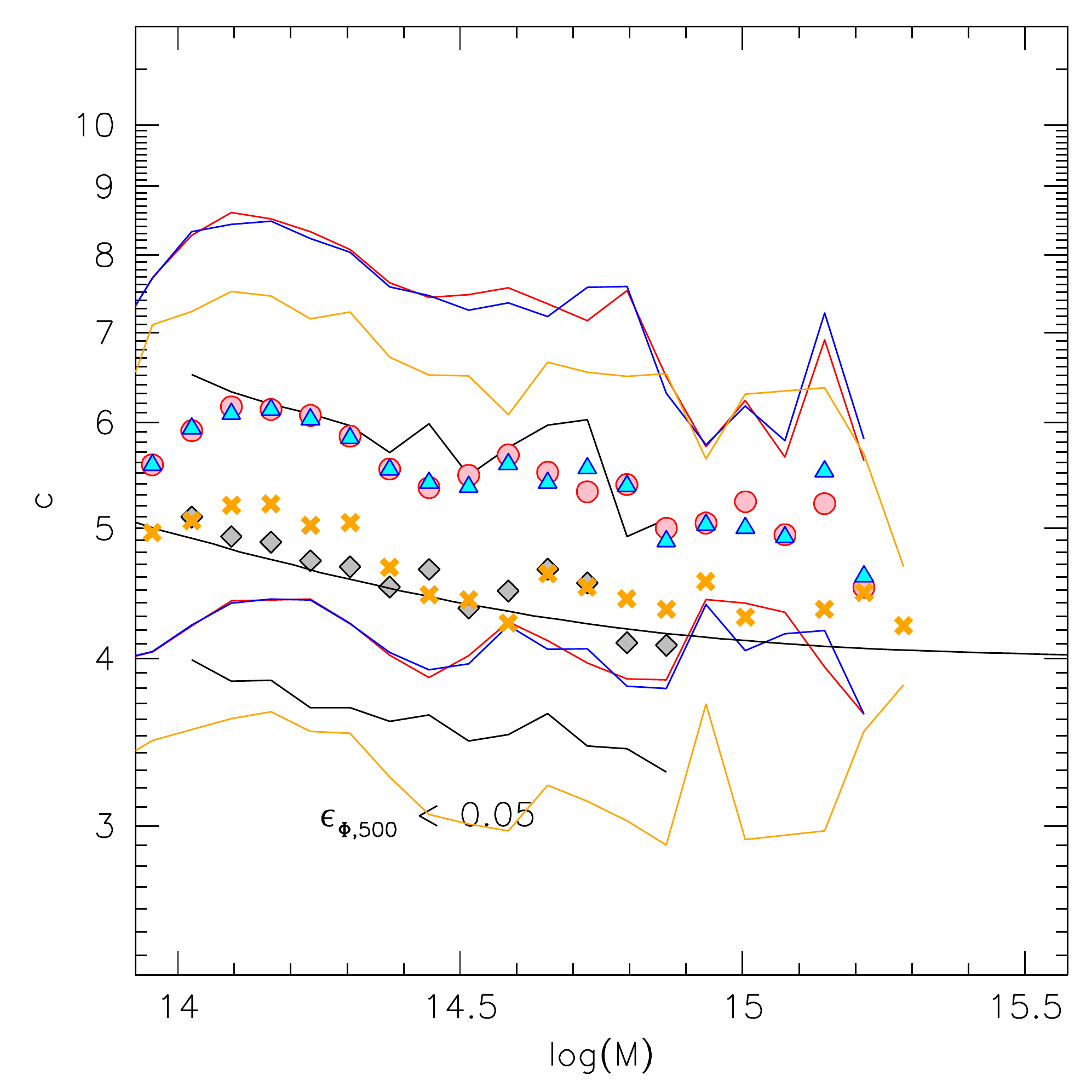}
\caption{Same as  Fig.~\ref{figintcmrel} but  for clusters  selected with
different limits in the potential ellipticity: $\epsilon_{\phi,500}<0.25$ (left panel), 
 $\epsilon_{\phi,500}<0.1$ (central panel) and $\epsilon_{\phi,500}<0.05$ (right panel).
 The results refer to $z=0.288$ only.\label{figcmrelX}}
\end{figure*}

In order  to understand how  different selection functions  change the
$c-M$   relation,   in  Fig.~\ref{figcmrelX}   we   show  the   median
relation  for  the  cluster  sample  at  redshift
$z=0.288$  (the results  at other redshifts  are  consistent with
these)  when  the objects  are   selected  to  have  different  projected
potential   ellipticities,  $\epsilon_{\Phi,500}$. Selecting clusters in 
ellipticity implies also a selection in the shape of the X-ray emission: 
the ones with smaller ellipticity will also present a more spherical and 
relaxed X-ray morphology.  Going  from the  left to the  right panel, we  
show the $c-M$ relations for clusters with $\epsilon_{\phi,500}<0.25$, $\epsilon_{\phi,500}<0.1$
and  $\epsilon_{\phi,500}<0.05$,  respectively;  the  adopted symbols are 
are the  same  as in  Fig.~\ref{figintcmrel}.   Looking at  the
filled black diamonds  we notice that no particular  bias appears
in the  intrinsic $c-M$ relation, while the  ones reconstructed using
the concentrations and  the masses derived from WL  and WL+SL tends to
move up for the more spherical systems.  This behavior reflects the
fact  that, since haloes  are in general prolate  ellipsoids, the  more spherical
they are in the plane of the sky, the more elongated they must be along the
line-of-sight -- or they could be intrinsically spherical. So,
even if the intrinsic $c-M$ relations lie on the theoretical ones, the 
selection in ellipticity could introduce a bias in the estimated masses 
and concentrations. This effect  is also seen in the $c-M$ relation of
clusters when their properties are  estimated using a gNFW model as a
reference: going  from left to right, this is shown by the orange crosses,
which tend to move toward higher values of concentration.

\begin{figure*}
\includegraphics[width=5.8cm]{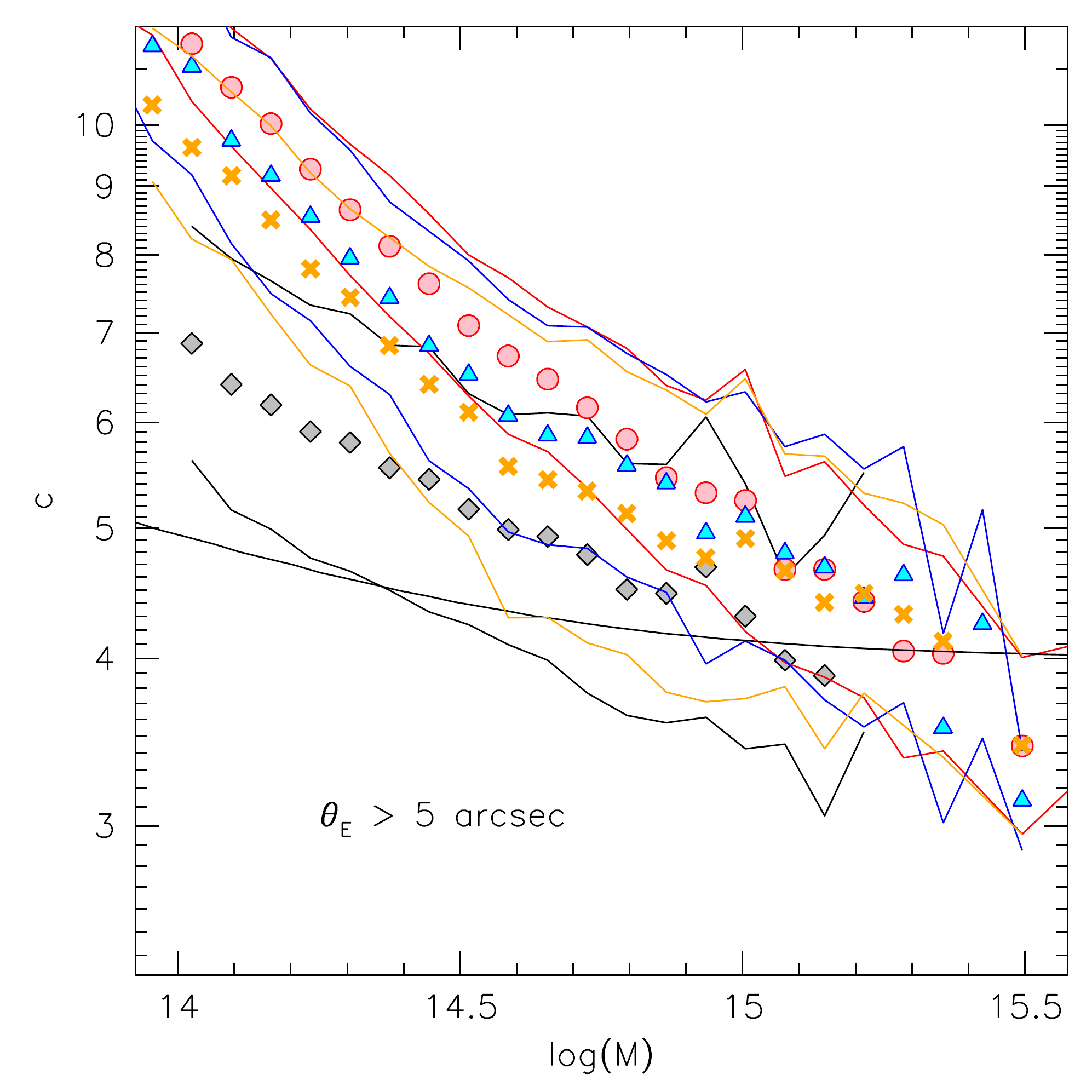}
\includegraphics[width=5.8cm]{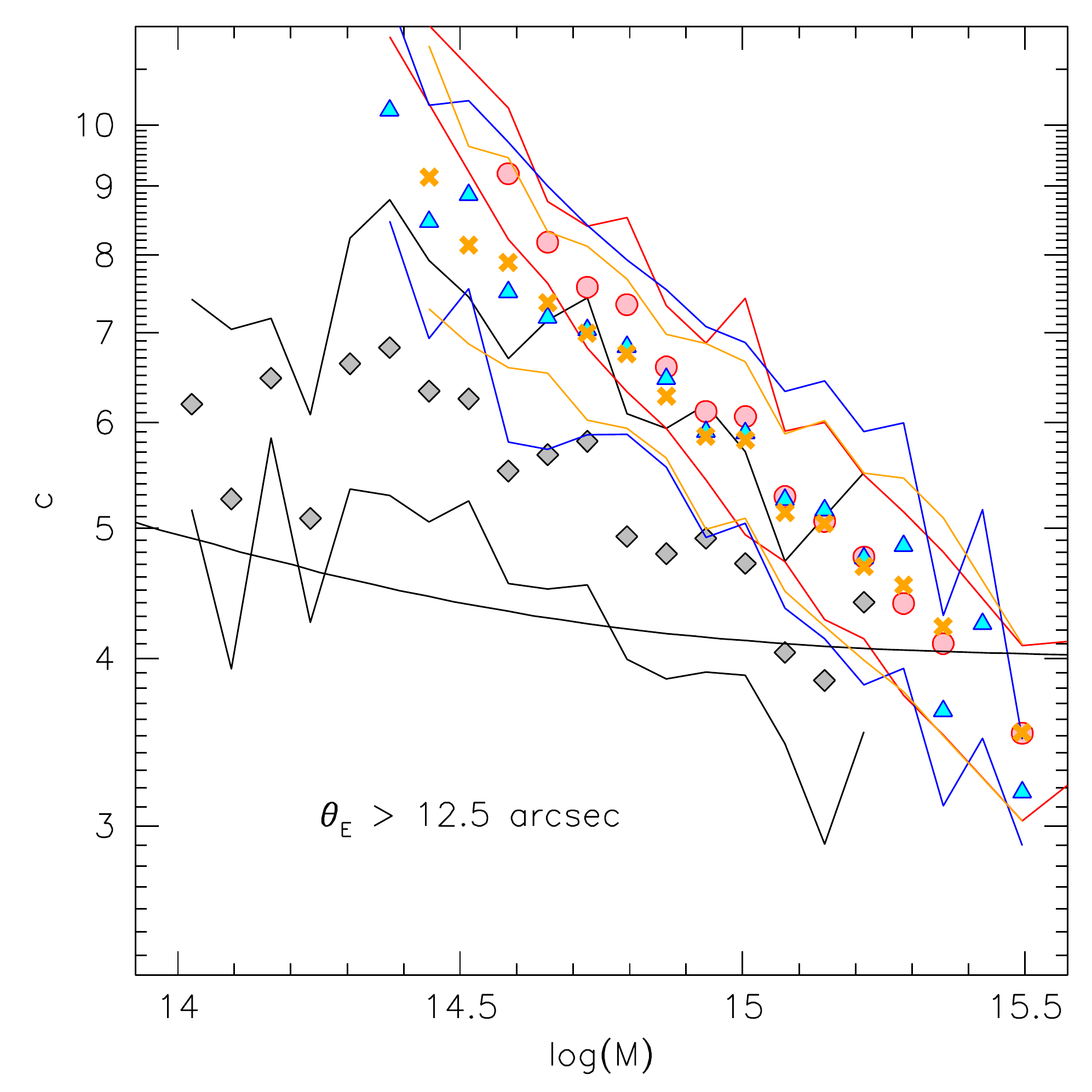}
\includegraphics[width=5.8cm]{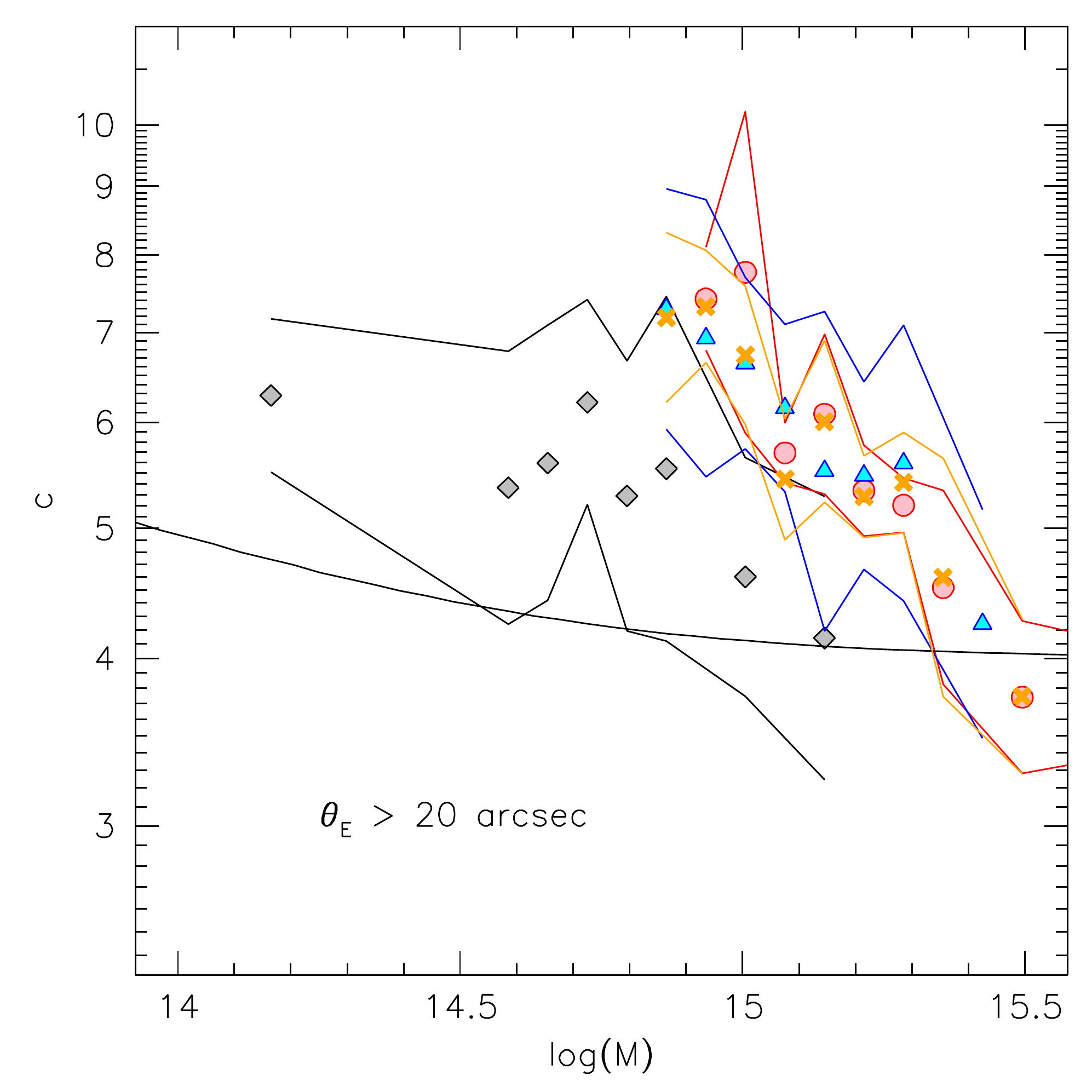}
\caption{Same as  Fig.~\ref{figintcmrel} but  for clusters  selected with
different limits in the size of the Einstein radius: $\theta_{E}>5$ (left panel), 
 $\theta_{E}>12.5$ (central panel) and $\theta_E>20$ arcsec (right panel).
 Also in this case the results refer to $z=0.288$ only.\label{figcmrelE}}
\end{figure*}

The effect  is more drastic  when we  select clusters by  their strong
lensing features.  In Fig.~\ref{figcmrelE}  we show the $c-M$ relation
selecting clusters to have an  Einstein radius larger than $5$, $12.5$
and $20$ arcsec, respectively.  In this case, we notice again that the
intrinsic $c-M$  relation --  showed by the  black filled  diamonds --
tends to  move up with  respect to  the solid curve,  representing the
reference  model: intrinsically  more concentrated  haloes tend  to be
selected   \citep{meneghetti10a}   .   Lensing   reconstructed   $c-M$
relations  are  usually  above  the solid  curve  even  by  $20-30\%$.
Selecting clusters by their strong  lensing features not only picks up
the ones that are more elongated along  the line of sight but also the
ones   that  are   more   elliptical   in  the   plane   of  the   sky
\citep{zitrin13b}.

\subsection{c-M relation for relaxed lenses}

As  a  first  case,  we  analyze the  situation  of  building  up  the
concentration-mass  relation  using  a   sample  of  relaxed  clusters
selected in a similar way to the CLASH sample.  Since our cosmological
cluster catalogues  have been created  for six discrete  redshifts, we
group the CLASH clusters in these bins. This corresponds to create six
sub-samples,  for the  $25$  clusters,  with a  temporal  bin size  of
approximately $1$  Gyr which is of  the order of the  cluster relaxing
time. Each realization is constructed  looping through the 25 clusters
and  then  randomly selecting,  from  the  corresponding catalogue,  a
cluster with a  true mass of at least  $5 \times 10^{14}\,M_{\odot}/h$
and  with a  2D ellipticity  of the  potential as  in the  Table 4  of
\citet{postman12}, when available.  We  repeat this procedure creating
10,000  different   realizations  of   the  CLASH-like   sample.   The
ellipticity distribution of the sample  nicely shows that the stronger
the lens the  larger is the elongation, and that  the strongest lenses
are  very elongated  either along  the line  of sight  or in  the lens
plane.

\begin{figure*}
\includegraphics[width=8.cm]{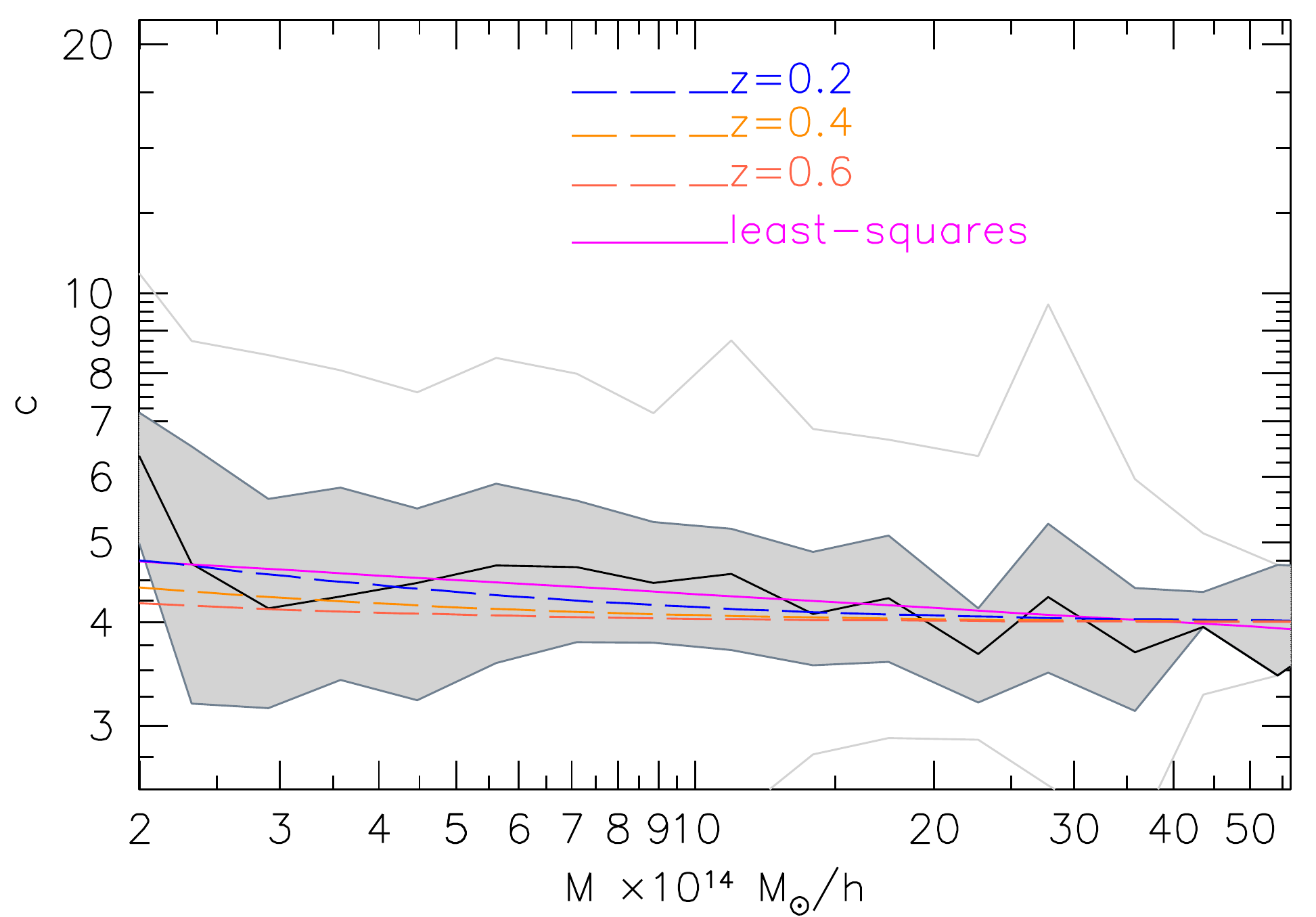}
\includegraphics[width=8.cm]{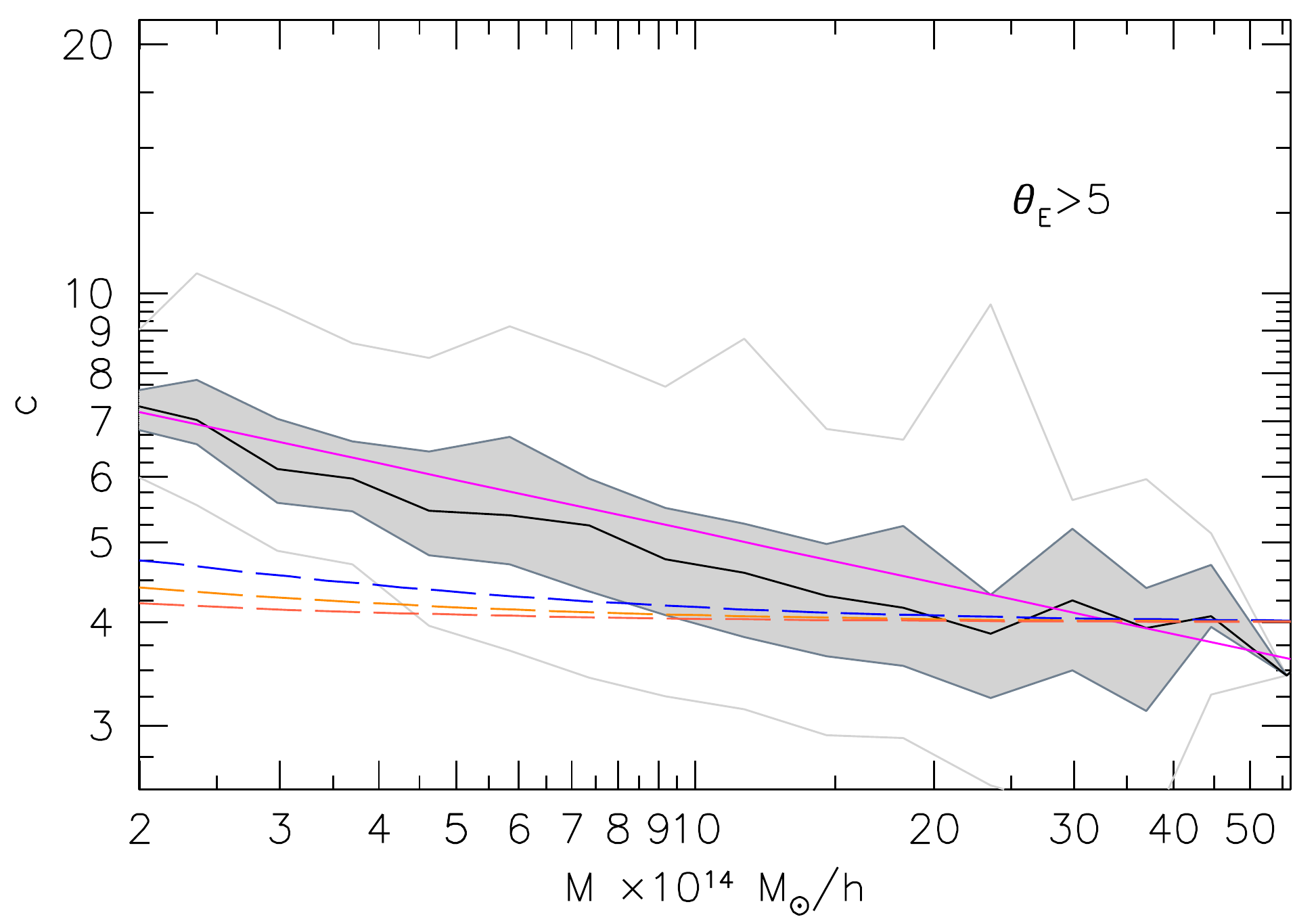}
\includegraphics[width=8.cm]{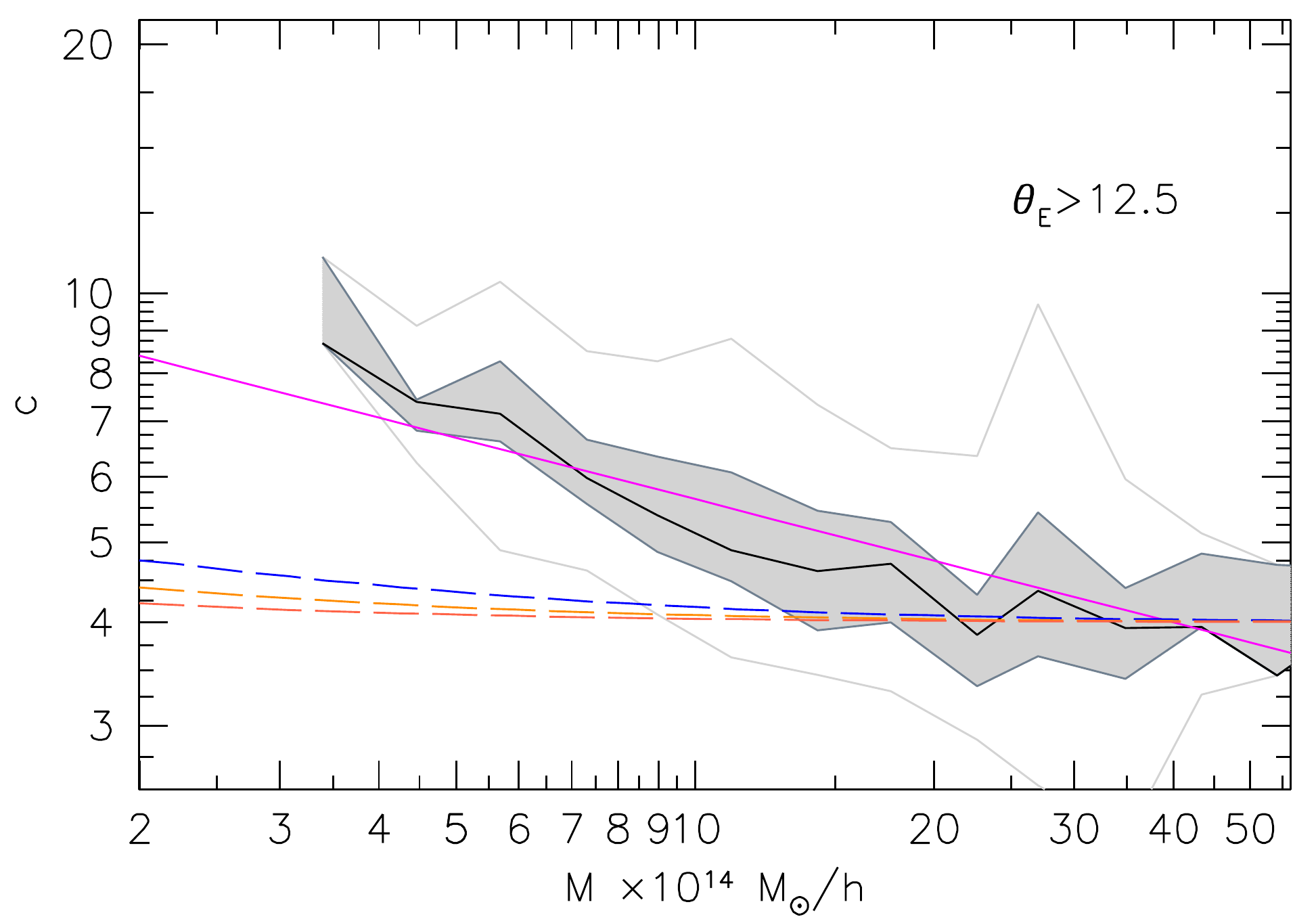}
\includegraphics[width=8.cm]{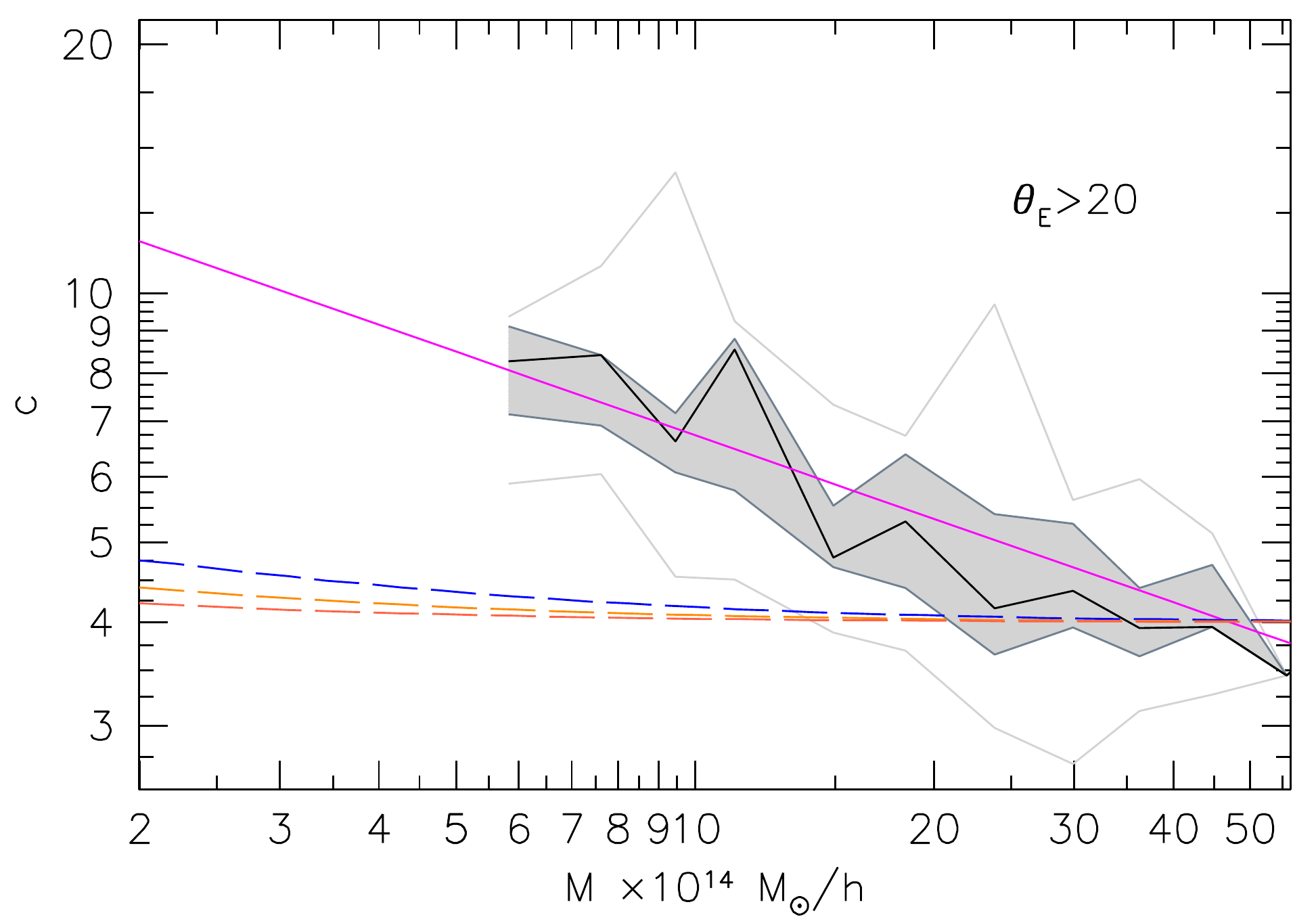}
\caption{Median concentration-mass relation for 10,000 realizations of
  $25$  clusters selected  in  potential ellipticity,  where mass  and
  concentration are  estimated using WL+SL constraints  considering an
  NFW  profile  as  reference  model.  The  shaded  dark  gray  region
  encloses  the first  and the  third quartiles  of the  distribution,
  while the  gray curves are $95\%$  of the data. The  solid (magenta,
  for the colored  version of the figure)  represent the least-squares
  fit to the data.\label{figCLASH}}
\end{figure*}

In  Fig.~\ref{figCLASH} we  show  the  concentration-mass relation  of
these 10,000 realizations.  For each cluster,  we use the mass and the
concentration estimated using a NFW  model as reference combining weak
and strong  lensing constraints. In  each panel the solid  black curve
shows the median of the sample  and the gray region encloses the first
and the third  quartiles. The two external gray  curves enclose $95\%$
of  the  data.  The  different  dashed  curves  show  the  theoretical
concentration-mass relations at three redshifts. The four panels show:
on the  top left the  case in which each  sample does not  contain any
constrain on  the minimum size  of the  Einstein radius, while  in the
other threes the clusters are  randomly selected with the condition of
possessing  an Einstein  radius of  at least  5, 12.5  and 20  arcsec,
respectively. The c-M relation built  from the relaxed samples when no
strong lensing selection is present is in very good agreement with the
theoretical  input models,  while it  increasingly steepens  deviating
from the theoretical  expectations when a lower limit  to the Einstein
radius  size  is  imposed.  In  Table  \ref{tablelsqrelationCLASH}  we
present the slope and the zero point of the least-squares fit relation
to the data, for each panel of the figure.

\begin{table}
  \caption{Least-squares fit to the samples WL+SL selected in potential ellipticity: $\log \mathrm{c} = a \log
    \mathrm{M} + b$ ($M$ is in unit of $M_{\odot}/h$)}
\label{tablelsqrelationCLASH}
\begin{tabular}{@{}lccccc}
\hline
case & a & $\sigma_a$ & | & b & $\sigma_b$  \\ \hline
all clusters & -0.06 & 0.03 & | & 0.69 & 0.03 \\
$\theta_E>5$ & -0.21 & 0.03 & | & 0.92 & 0.03  \\
$\theta_E>12.5$ & -0.25 & 0.04 & | & 1.00 & 0.05 \\
$\theta_E>20$ & -0.34 & 0.06 & | & 1.16 & 0.09 \\
\end{tabular}
\end{table}

\begin{figure}
\includegraphics[width=\hsize]{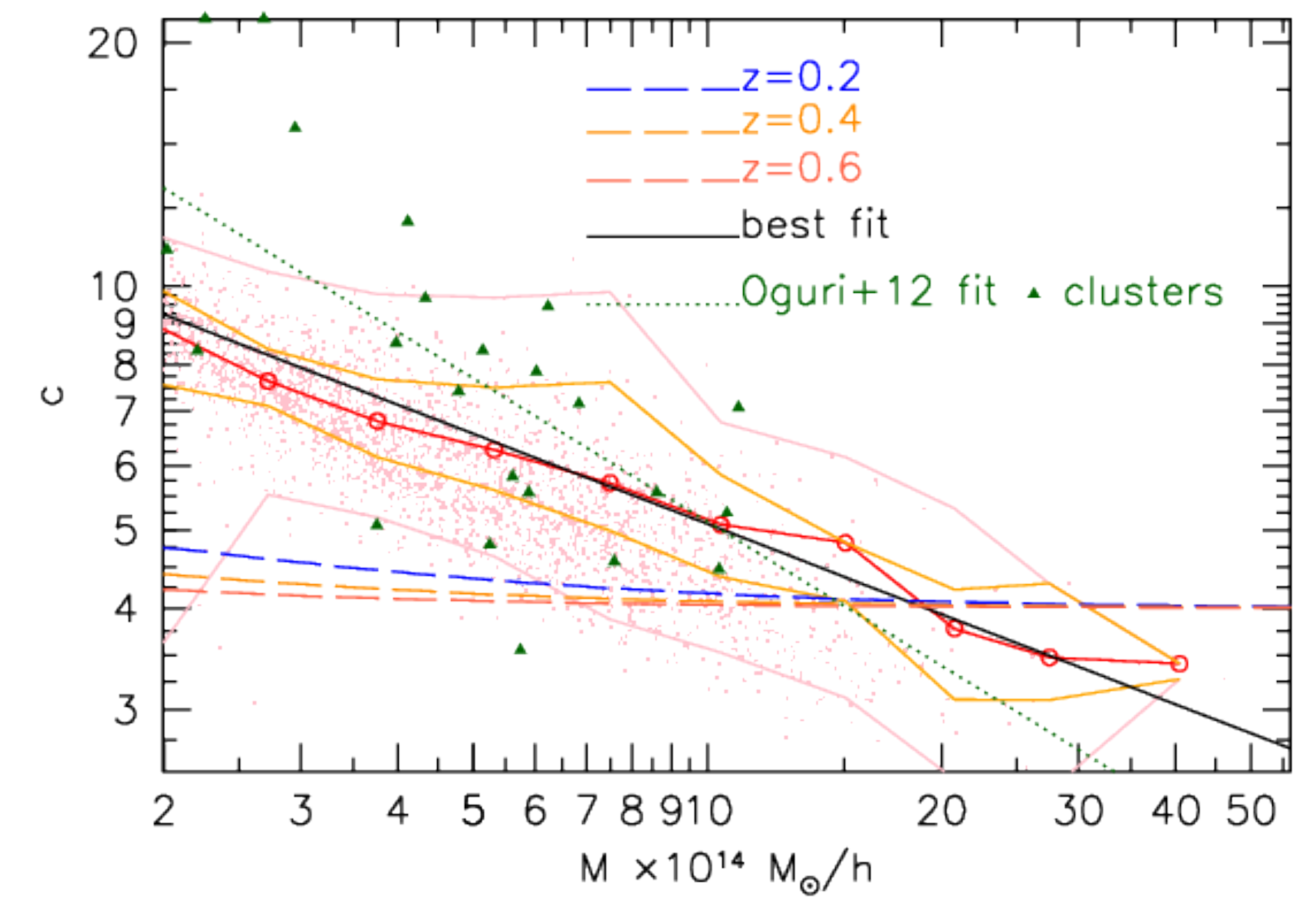}
\caption{Median concentration-mass relation for 10,000 realizations of
  $28$  clusters selected  using the  convergence ellipticity  and the
  Einstein radius  to match  the sample  of \citet{oguri12}.   The red
  circles show  the median  of the distribution  while the  orange and
  pink  lines enclose  the quartiles  and $95\%$  of the  data points,
  respectively.  The black solid line  is the least-squares fit to the
  selected MOKA  clusters. Dashed  curves, as  in Fig.~\ref{figCLASH},
  show  the   theoretical  $c-M$  predictions  at   the  corresponding
  redshifts.\label{fOguriSDSS} }
\end{figure}

As   a   second   case,   in   Fig.~\ref{fOguriSDSS}   we   show   the
concentration-mass  relation  for  10,000 realizations  of  $28$  MOKA
clusters selected  to match the  sample by \citet{oguri12} as  part of
the Sloan Giant Arcs Survey (SGAS).  In this case we have selected the
clusters from our database to  match the redshift, the ellipticity and
the  size  of  Einstein  radius  as  listed  in  Tables  2  and  3  in
\citet{oguri12}.  The  red circles represent  the median of  the $c-M$
relation for  those clusters while  the orange and pink  lines enclose
the quartiles  and $95\%$ of  the data, respectively.  The  black line
shows the least squares fit, that can be read as:
\begin{equation}
c_{vir} = 11.9 \pm 0.1 \; \left( \dfrac{M_{vir}}{10^{14}}\right)^{-0.37\pm0.02}\,.
\end{equation}
The green triangles show the location of the SGAS-SDSS clusters in the
figure when mass  and concentration are estimated using  both weak and
strong  lensing  analyses, while  the  dotted  line  is the  best  fit
relation (equation~26 from \citet{oguri12}). From the figure we notice
that applying a  realistic selection function to  the simulated sample
--  in  which  multiple  mass  components, presence  of  the  BCG  and
adiabatic contraction are considered  -- clarifies the tension between
the $c-M$ relation predicted in numerical simulations and the best-fit
on the observed dataset.

\begin{figure*}
\includegraphics[width=5.8cm]{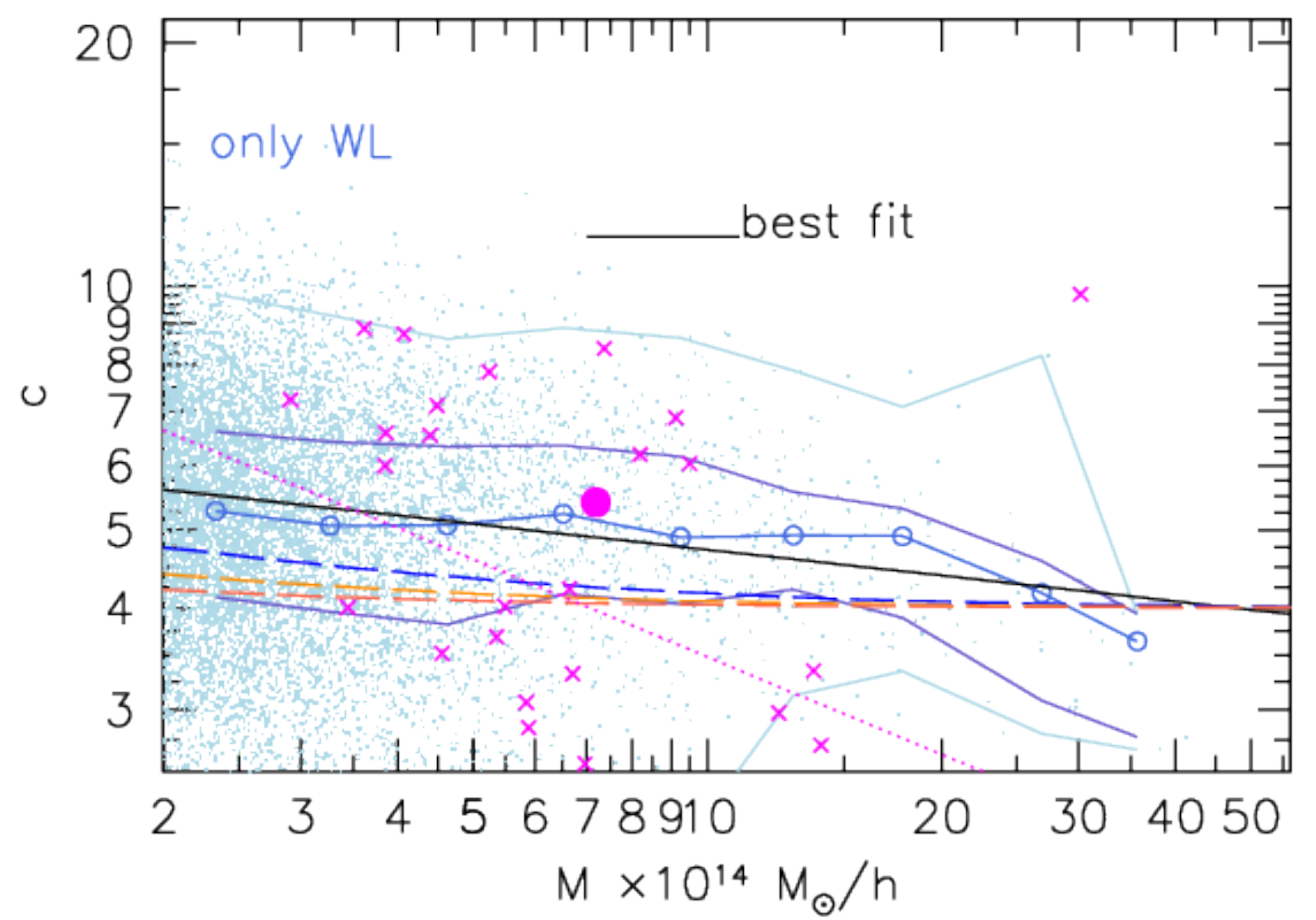}
\includegraphics[width=5.8cm]{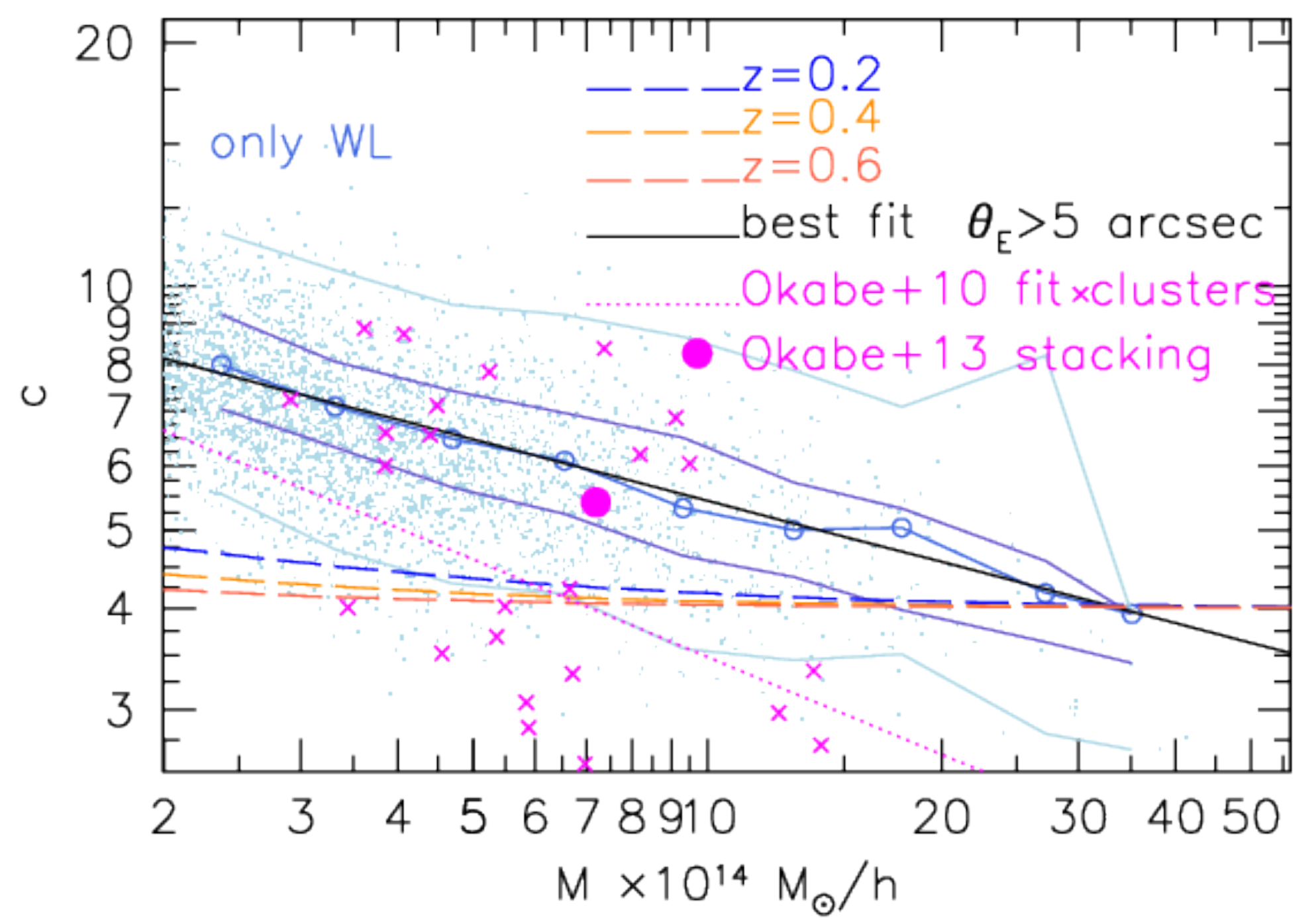}
\includegraphics[width=5.8cm]{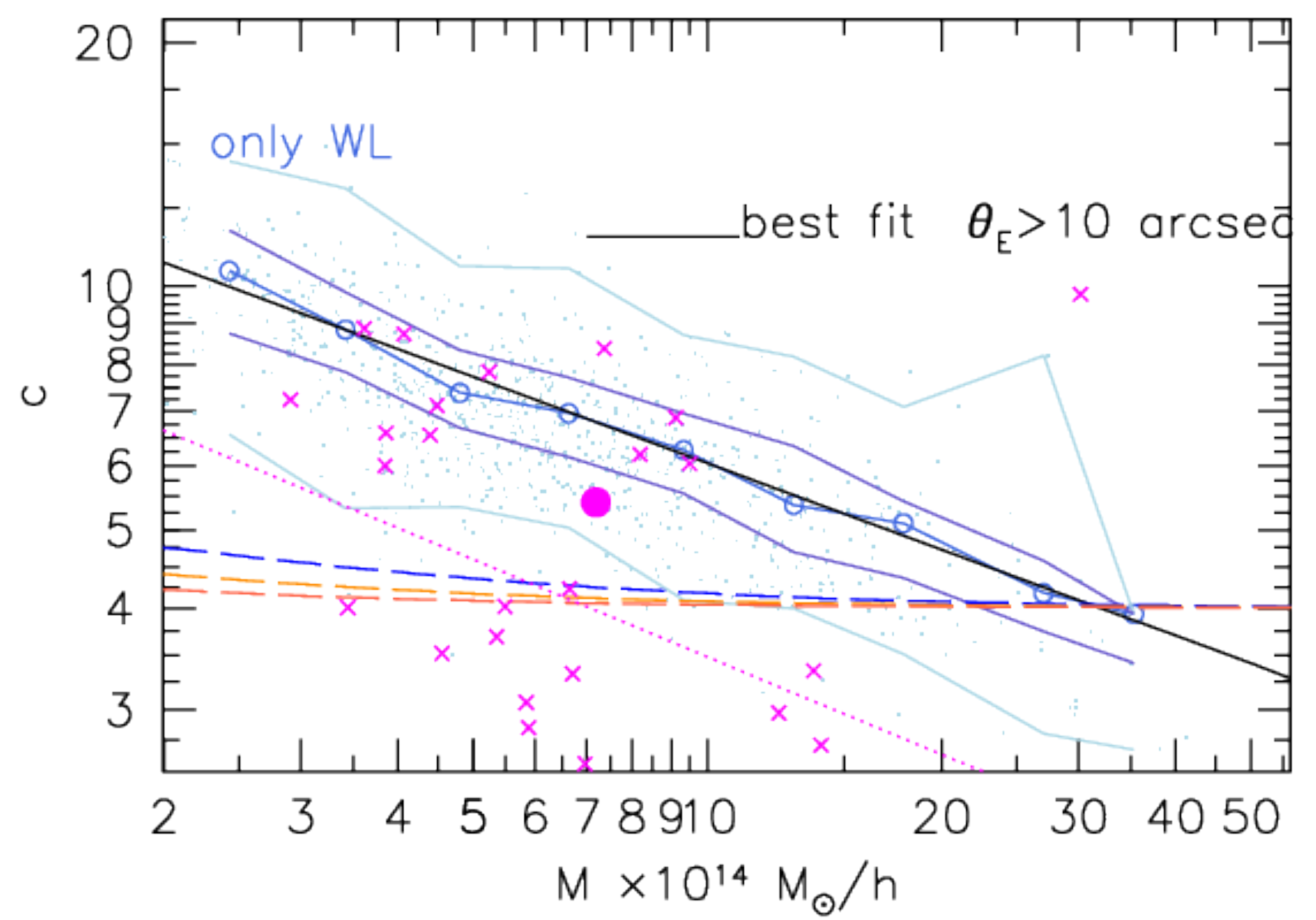}
\caption{Median  concentration-mass  relation  for  10,000
    realizations of $26$  clusters selected in redshift  as the LoCuSS
    sample  by  \citet{okabe10b},  where mass  and  concentration  are
    measured using only WL data through the reduced tangential shear profile.  
    In  the left, central and right panel  we show the
    relation obtained from our  MOKA sample, considering all clusters,
    those  with an  Einstein radius  larger than  $5$ arcsec  and $10$
    arcsec, respectively. In each panel, the blue circles represent the median of the 
    distribution, while the blue and the light-blue lines enclose the quartiles and $95\%$ of the data, respectively.
    \label{fOkabe}}
\end{figure*}

As  a last  case,  we compare  our predictions  to  those obtained  by
analyzing  the weak  lensing data  of  $26$ clusters  from the  LoCuSS
survey, listed  in Table 6 of  \citet{okabe10b}.  In this case,  to be
consistent with their analysis, we  have considered only the situation
in which the  mass and the concentration are estimated  by fitting the
reduced tangential  shear profile. The  MOKA sample has  been selected
from our  database to match the  redshift of the LoCuSS  clusters (see
Table 1 by \citet{okabe10b}.  In  the left panel of Fig.~\ref{fOkabe},
we show the median  estimated concentration-mass relation obtained for
10,000 realizations  of the sample.   The solid black  line represents
the  least-squares fit  to our  data.  In  the central  panel we  have
considered  only clusters  with  an Einstein  radius  larger than  $5$
arcsec, while  in the right  those with $\theta_E>10$ arcsec.   In all
panels   the   magenta  crosses   represent   the   location  in   the
mass-concentration diagram  of LoCuSS  clusters and the  filled circle
the  values obtained  by performing  a  stacking analysis  of them  as
described  by  \citet{okabe13}.   As  already  discussed,  the  strong
lensing selection  tends to  increase both  the normalization  and the
slope of  the concentration-mass relation. In  Table~\ref{tabOkabe} we
summarize the slope  and the zero point of best  fitting the recovered
$c-M$ relations for the three cases.

\begin{table}
  \caption{Best fit parameters of the $c-M$ for 10,000 realizations of
    the  LoCuSS   galaxy  cluster   sample:  $c_{vir}  =   c_N  \left(
    M_{vir}/10^{14} \right)^{-\alpha}$}
\label{tabOkabe}
\begin{tabular}{@{}lccc}
\hline
 $ $ &  $c_N$ &  $\alpha$  \\ \hline
all clusters & $6.03\pm 0.08$ & $0.10\pm 0.03$ \\ 
$\theta_E>5$ & $9.69\pm 0.05$ & $0.25\pm 0.01$ \\
$\theta_E>10$ & $13.68\pm 0.08$ & $0.35\pm 0.01$ \\
\end{tabular}
\end{table}

\section{Summary and Conclusion} \label{secsc}

In  this  work  we have  studied  how  well  galaxy cluster  masses  and
concentrations are recovered using strong and weak lensing signals. Using an
NFW halo model  as reference, we recover mass  and concentration using
only weak lensing data or combining  them with a measurement of the size
of the Einstein  radius.  In addition, for the case  in which we combine
both WL  and SL measurements  we have made  estimates of the  mass and
concentration using  a generalized NFW model, which  reduces both mass
and  concentration  biases while  introducing  a  new free  parameter
(the  inner slope of the density  profile).  We summarize
our study and main results as follows.
\begin{itemize}
\item [-] On average, lensing analysis provides biases on the cluster mass 
that depend on the host halo mass. Small systems typically present  a mass bias
  of about  $15\%$ while  for the  more massive  ones the  bias almost
  vanishes.   The  scatter  has   a  log-normal  distribution  with  a
  $\sigma_{\log M} \approx 0.25$. For the most massive systems, adding
  the constraint on  the size of the Einstein radius  reduces the bias
  to a few percents.

\item [-]The estimated concentration is slightly positively biased and
  decreasing with  the halo mass.  This behavior can be  attributed to
  the presence  of the  BCG at the  center of the  cluster and  to the
  adiabatic  contraction (here  included unlike  in the  previous work
  \citep{giocoli12c}).

\item [-]Adopting a generalized NFW  model for fitting weak and strong
  lensing data  reduces both  mass and concentration  biases. However,
  this introduces an additional free parameter, the inner slope of the
  density profile, $\beta$.
 
\item [-]The  bias and the  scatter in  the estimated mass  modify the
  shape  of  the  mass  function   with  respect  to  the  theoretical
  prediction, from  which the cosmological sample  is drawn.  However,
  for mass bins in which cluster count has a $S/N\geq 5$ the residuals
  between the input  and the recovered mass function are  smaller by a
  factor of $5-7$.

\item [-]The uncertainties on the  cluster mass and concentration also
  change  the mass-concentration  relation.  On  average, when  an NFW
  model  is  used  to  fit  the clusters,  the  normalization  of  the
  recovered  $c-M$ relation  has  a normalization  higher by  $5-7\%$. 
  The use of a generalized NFW helps to recover a $c-M$ relation in better agreement 
  with the theoretical expectations.

\item [-]The biases in the concentration-mass relation, as reported by
  the analyses of different galaxy cluster surveys, are clarified when
  selecting from our MOKA sample a relaxed sub-sample of systems.

\end{itemize}

To conclude, in this work we  have presented a detailed and systematic
analysis of the estimation of mass and concentration of clusters using
lensing data (weak and weak+strong).  We have studied how the mass and
concentration biases  depend on  the halo mass  and redshift.  We have
presented  how the  bias and  the scatter  in the  estimated mass  and
concentration influence the halo mass function and the $c-M$ relation.
In  particular, we  have  discussed how  different selection  criteria
affect the concentration-mass  relation and we have  found that strong
lens clusters may have a concentration  as high as $20-30\%$ above the
average, at the fixed mass.

\section*{Acknowledgements}
CG and RBM's research is part  of the project GLENCO, funded under the
European  Seventh  Framework  Programme,  Ideas,  Grant  Agreement  n.
259349.   We   acknowledge  financial  contributions   from  contracts
ASI/INAF I/023/12/0 and by the PRIN MIUR 2010-2011 ``The dark Universe
and the cosmic evolution of baryons: from current surveys to Euclid''.
CG and LM also acknowledge the financial contribution by the PRIN INAF
2012  ``The Universe  in  the box:  multiscale  simulations of  cosmic
structure''.    SE  acknowledges   the  financial   contribution  from
contracts ASI-INAF  I/088/06/0 and PRIN-INAF  2012.  CG would  like to
thank Giuseppe Tormen and Vincenzo Mezzalira  to have host part of the
computer  jobs run  to produce  the simulated  galaxy cluster  sample.
Part of the simulations of this project have been run during the Class
C Project-HP10CMXLBH  (MOKALEN3) CG  would like to  particularly thank
Matthias Bartelmann  for useful  and stimulating discussions.   We are
also grateful  to Maru\v{s}a Brada\v{c}  Anja Von der  Linden, Stefano
Borgani, Stefano  Andreon and Mauro  Sereno for the  conversations had
during the conference  in Madonna di Campiglio in March  2013.  We are
also  grateful  to the  anonymous  referee  for his/her  comments  and
suggestions that helped to improve the presentation of our results.

\appendix
\section{2D Masses}
\label{app2D} 

The measurement of gravitational  lensing gives an estimate of the
projected  mass  which is  causing  the  distortion  of the  shape  of
background galaxies and  the creation of multiple  images, without any
assumption  about the  dynamical  state of  the  system. However  some
assumption is  needed when the mass reconstruction is deprojected
from 2D to 3D.

\begin{figure}
\includegraphics[width=\hsize]{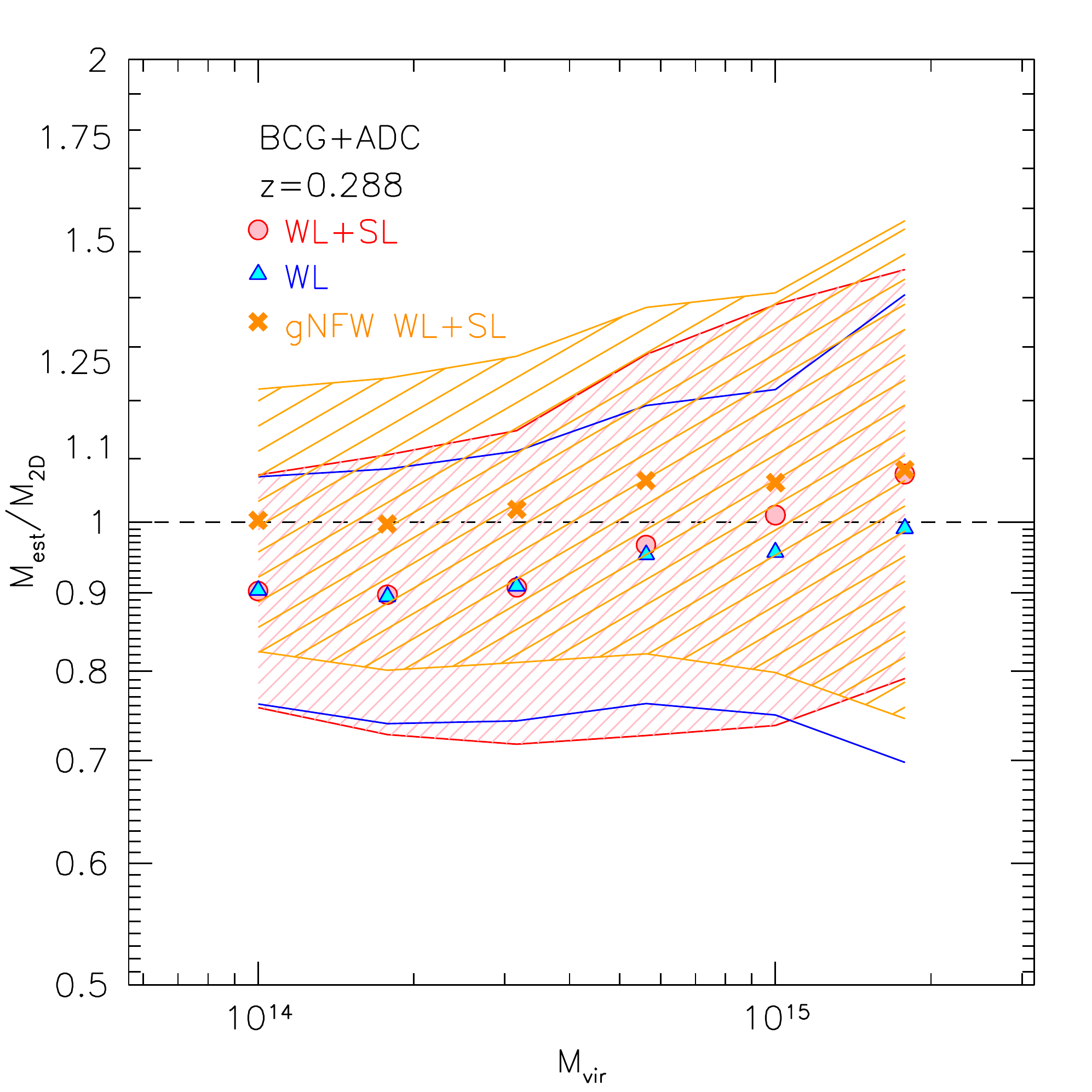}
\caption{Average of  the rescaled --  with respect  to the 2D  mass --
  estimated mass as a function of the 2D cluster mass, for the systems
  at  redshift $z=0.288$.   We  show the  case in  which  the mass  is
  estimated using WL+SL (circles)  and WL only (triangles) information
  using  an  NFW  model  as  a reference  when  minimizing  the  total
  $\chi^2$.  The upper and lower  curves enclose the $1\sigma$ scatter
  of  the  distribution.   Each   mass  bin  contains  $2048$  cluster
  realizations.  \label{2Dfigmassest} The orange crosses show the same
  quantity when  using gNFW  model as  a reference  for both  weak and
  strong lensing.}
\end{figure}

In Fig.~\ref{2Dfigmassest}  we show the  estimated mass derived from  weak and
weak +  strong lensing  rescaled with  respect to the  2D one  as a
function of the  true cluster mass for redshift $z=0.288$.   We do not
show the  results for different redshifts  since they are extremely 
similar this case.  From the figure, when we assume  an NFW halo as  a reference model, 
we  notice that for smaller masses the bias with respect to the 2D mass is of about
ten percent, while  at larger masses it is almost  negligible, well below
$5-7\%$. One more interesting result is that the scatter is
smaller than  the one measured in  the relation $M_{\rm est}/M_{\rm 3D}$.  
The case WL+SL  using generalized NFW model shows on average no 
particular bias for any cluster mass sample.

\section{Convergence and Potential Ellipticity}
\label{CPE} 

\begin{figure*}
\includegraphics[width=8.6cm]{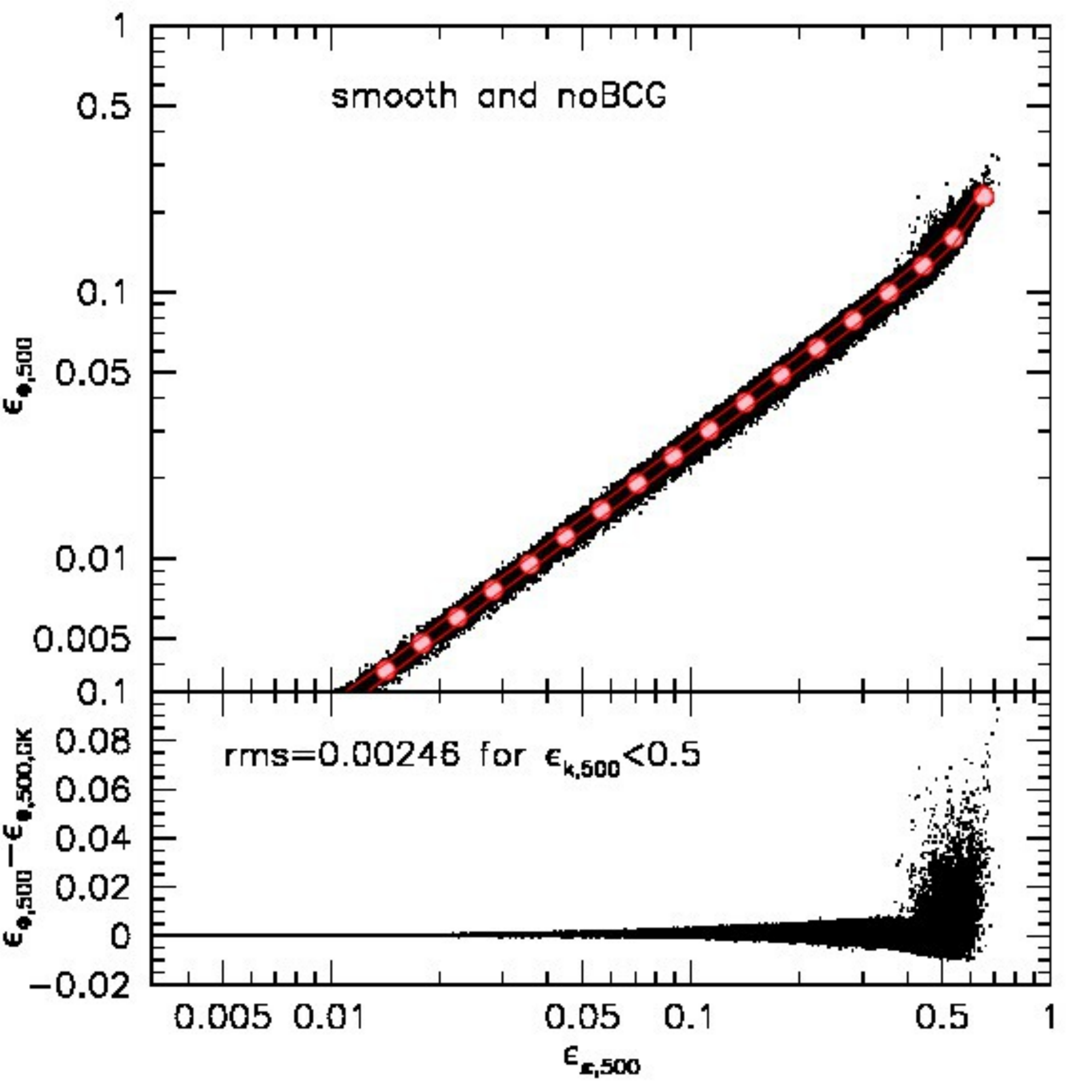}
\includegraphics[width=8.6cm]{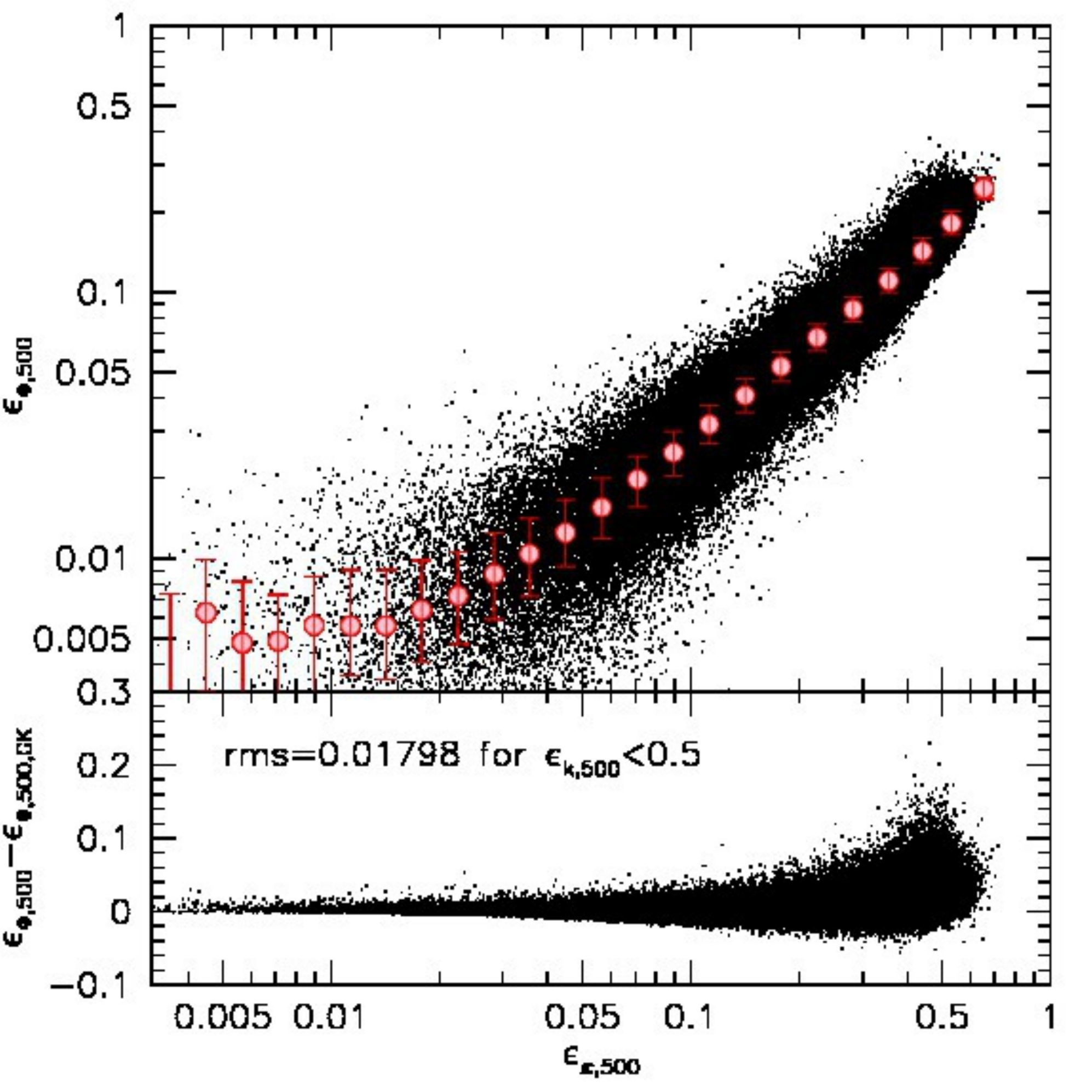}
\caption{Potential vs convergence ellipticity correlation. The dots in
  both panel  show halo by the halo correlation while the  filled circles
  with  the  error   bars correspond to  the  median  and  the   quartiles  of  the
  distribution  in   bins  of   convergence  ellipticity.    Both  the
  convergence and the potential  ellipticities have been measured from
  the respective maps within $R_{500}$. In  the left panel we show the
  correlation  for smooth  haloes  without  substructures and  central
  galaxy.   In the  right panel we  consider more  realistic clusters  where
  substructures  and BCG  are  included. The  bottom  panels show  the
  residuals of the measured potential  ellipticity with respect to the
  one estimated from $\epsilon_{k,500}$ using the \citet{golse02} relation
  (see equation~(\ref{eqgk02})).\label{figellipticities}}
\end{figure*}

The flexibility  of the MOKA code allows  us to study the  relationship between
the convergence and  the potential ellipticity.  In order  to do so we
measure  the ellipticity  from convergence  and lensing  potential maps,
both  within $R_{500}$.   In  this  context, we  use  the relation  of
\citet{golse02}  to link $\epsilon_{\kappa}$  and  $\epsilon_{\phi}$ for
pseudo-elliptical NFW lens models. However, simulated and real clusters
are different from a simple  one-component model and we are interested
in understanding if, and up to what point, their relation is valid for
our MOKA cluster sample. At this aim, we consider two samples of
clusters, one  triaxial without BCG  and satellite population  and the second one
containing them.  The  left panel of Fig.~\ref{figellipticities} shows
the correlation between the  potential and the convergence ellipticity
for the first sample, when considering  together all clusters at all redshifts.
The    filled    circles     represent    the    median    at    fixed
$\epsilon_{\kappa,500}$, while  the solid  lines enclose the  first and
the third  quartiles. In the  bottom panel of  the figure, we  show the
residuals     of     $\epsilon_{\phi,500}$     with     respect     to
$\epsilon_{\phi,500,GK}$  that  represents   the  ellipticity  of  the
potential  directly computed   from $\epsilon_{\kappa,500}$,  using the
\citet{golse02} formalism.

In order to estimate $\epsilon_{\phi,500,GK}$ we solve the equation:
\begin{equation}
\epsilon_{\kappa,500,GK} = a_1 \epsilon_{\Phi,500} + a_2 \epsilon_{\Phi,500}^2,
\label{eqgk02}
\end{equation}
taken from \citet{golse02}, where:
\begin{eqnarray}
a_1 &=& 3.31+0.280\,x_{500} \\ \nonumber
a_2 &=& -2.66-0.512\,x_{500}\,, \nonumber
\end{eqnarray}
and  $x_{500}  =  R_{500}/r_s$.   We recall that  this
relation   has   been  obtained  by applying   the   formalism  to   a
pseudo-elliptical NFW  profile, for a limited range of  the convergence
ellipticity: typically  $\epsilon_{\kappa} \lesssim 0.4$.   We notice,
that  since our  definition of  2D  ellipticity differs  from the  one
adopted  by \citet{golse02},  the  parametrization of  $a_1$ and  $a_2$
differs by a constant factor of  $0.64$. From the bottom left panel we
notice  that  the  \citet{golse02}  formalism  perfectly  captures  the
potential      ellipticity      from     the      convergence      for
$\epsilon_{\kappa,500}<0.5$ with  a very small rms. The
situation  is  different  for  the  case   in  which  we  add  BCG  and
substructures   to   the  convergence   map,   see   right  panel   of
Fig.~\ref{figellipticities}. In  this case, we notice  that on average
the  analytical prediction  is  still  valid but  the  scatter of  the
correlation is much  larger, Moreover the rms  between the ellipticity
measured  in  the  potential  map   and  the  one  inferred  from  the
convergence ellipticity  is larger than  in the previous case  by more
than one order of magnitude.

\bibliographystyle{mn2e}
\bibliography{paper}
\label{lastpage}
\end{document}